\documentclass[10pt,proc]{article} 

\usepackage{fullpage}
\usepackage{booktabs}
\usepackage{graphicx}
\usepackage{xcolor}
\usepackage{multirow}
\usepackage{amsmath}
\usepackage{url}
\usepackage{subcaption}
\usepackage{todonotes}
\usepackage{multirow}
\usepackage{subcaption}
\usepackage[symbol]{footmisc}
\usepackage{paralist}
\usepackage{chngcntr}
\usepackage{xcolor}
\usepackage{rotating}

\usepackage{etoolbox}
\patchcmd{\paragraph}{\@parfont}{\bfseries}{}{}
\patchcmd{\paragraph}{\parindent}{0pt}{}{}

\begin{document}
\title{A Survey of Collection Methods and Cross-Data Set Comparison of Android Unlock Patterns}

\author{
  Adam J. Aviv${}^\bullet$\thanks{Corresponding Author} \quad and \quad Markus D\"urmuth${}^\diamond$\\\\
\small
${}^\bullet$ United States Naval Academy \{{\tt aviv@usna.edu}\}\\
\small
${}^\diamond$ Ruhr-Universit\"at Bochum \{{\tt markus.duermuth@ruhr-uni-bochum.de}\}
}


\maketitle

\begin{abstract}

  Android's graphical password unlock remains one of the most widely used
  schemes for phone unlock authentication, and it is has been studied
  extensively in the last decade since its launch. We have learned that users'
  choice of patterns mimics the poor password choices in other systems, such as
  PIN or text-based passwords. A wide variety of analysis and data collections
  methods was used to reach these conclusions, but what is missing from the
  literature is a systemized comparison of the related work in this space that
  compares both the methodology and the results. In this paper, we take a
  detailed accounting of the different methods applied to data collection and
  analysis for Android unlock patterns. We do so in two dimensions. First we
  systemize prior work into a detailed taxonomy of collection methods, and in
  the second dimension, we perform a detailed analysis of 9 different data sets
  collected using different methods. While this study focuses singularly on the
  collection methods and comparisons of the Android pattern unlock scheme, we
  believe that many of the findings generalize to other graphical password
  schemes, unlock authentication technology, and other knowledge-based
  authentication schemes.


\end{abstract}



\newcommand{\figpatternsample}[0]{
  \begin{figure}[t]
    \begin{center}
      \fbox{\includegraphics[width=0.2\textwidth]{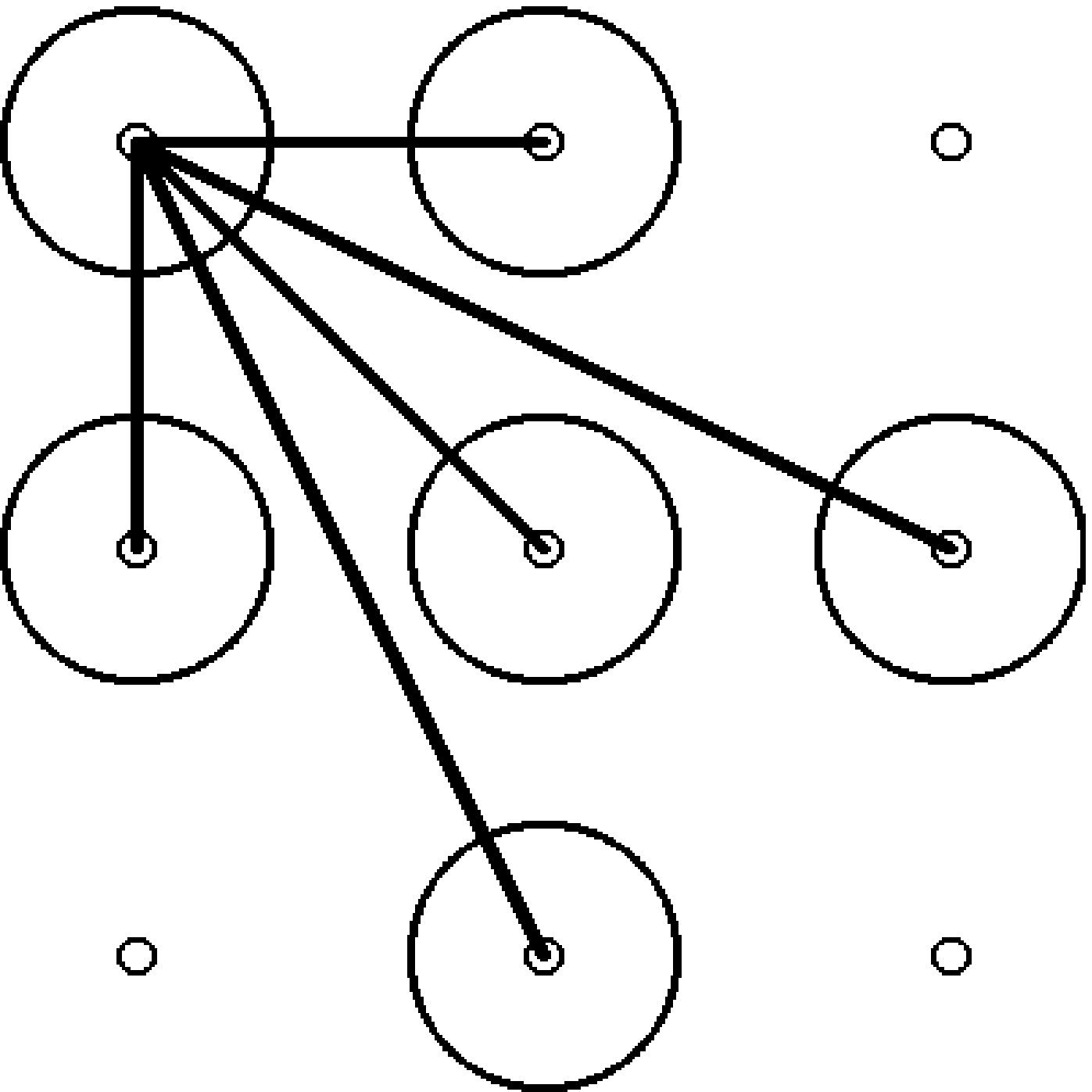}}
        \hspace{.2in}
        \fbox{\includegraphics[width=0.2\textwidth]{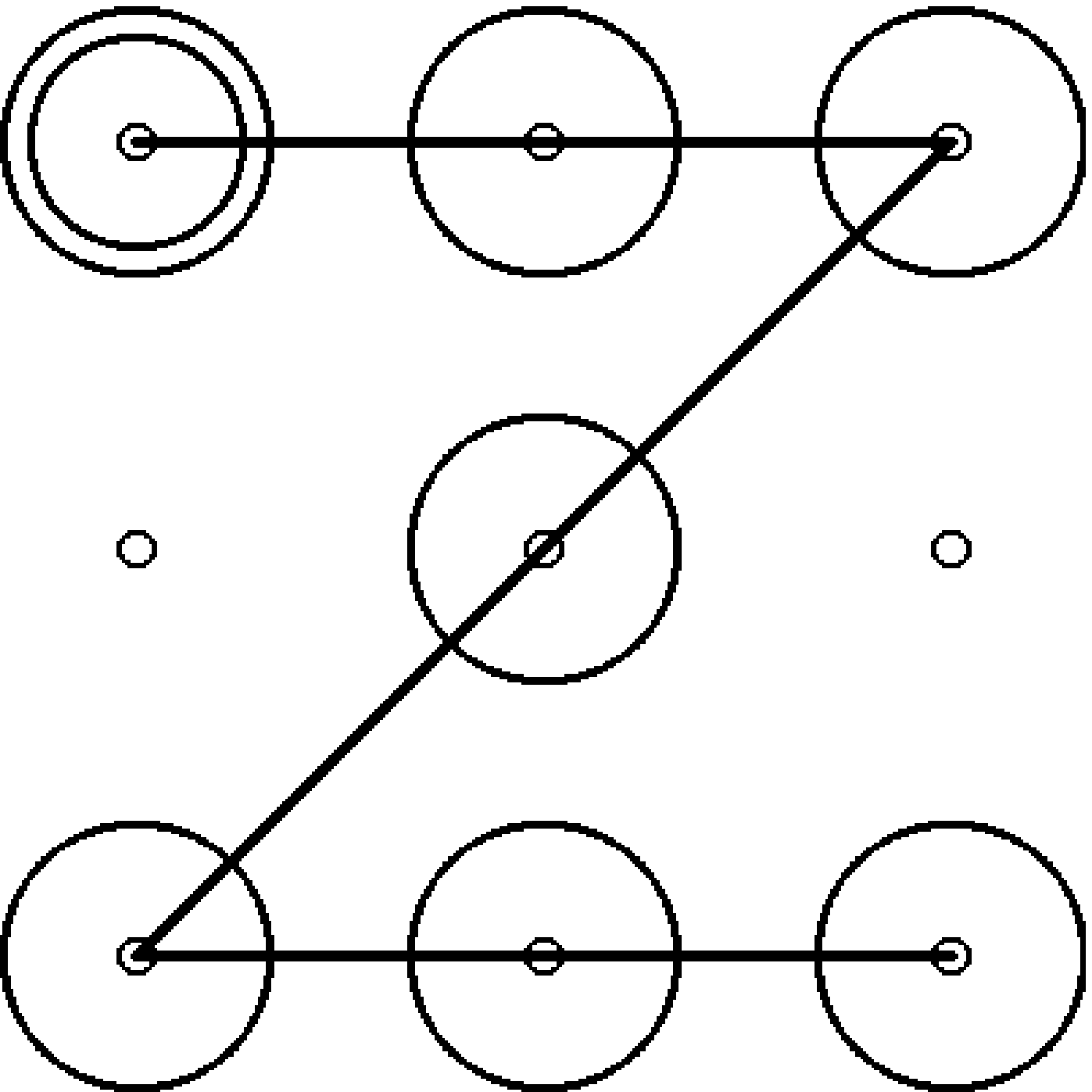}}
      \end{center}
      \caption{({\em left}) Reachable contact points in 3x3 Pattern
        Unlock from the top left contact point, and ({\em right}) a
        common pattern, the {\bf Z}-shaped pattern.}
        \label{fig:patternsample}
\vspace{-.2in}
  \end{figure}
}

\newcommand{\figalternates}[0]{
 \begin{figure*}[tp]
  \centering
  \includegraphics[width=2cm]{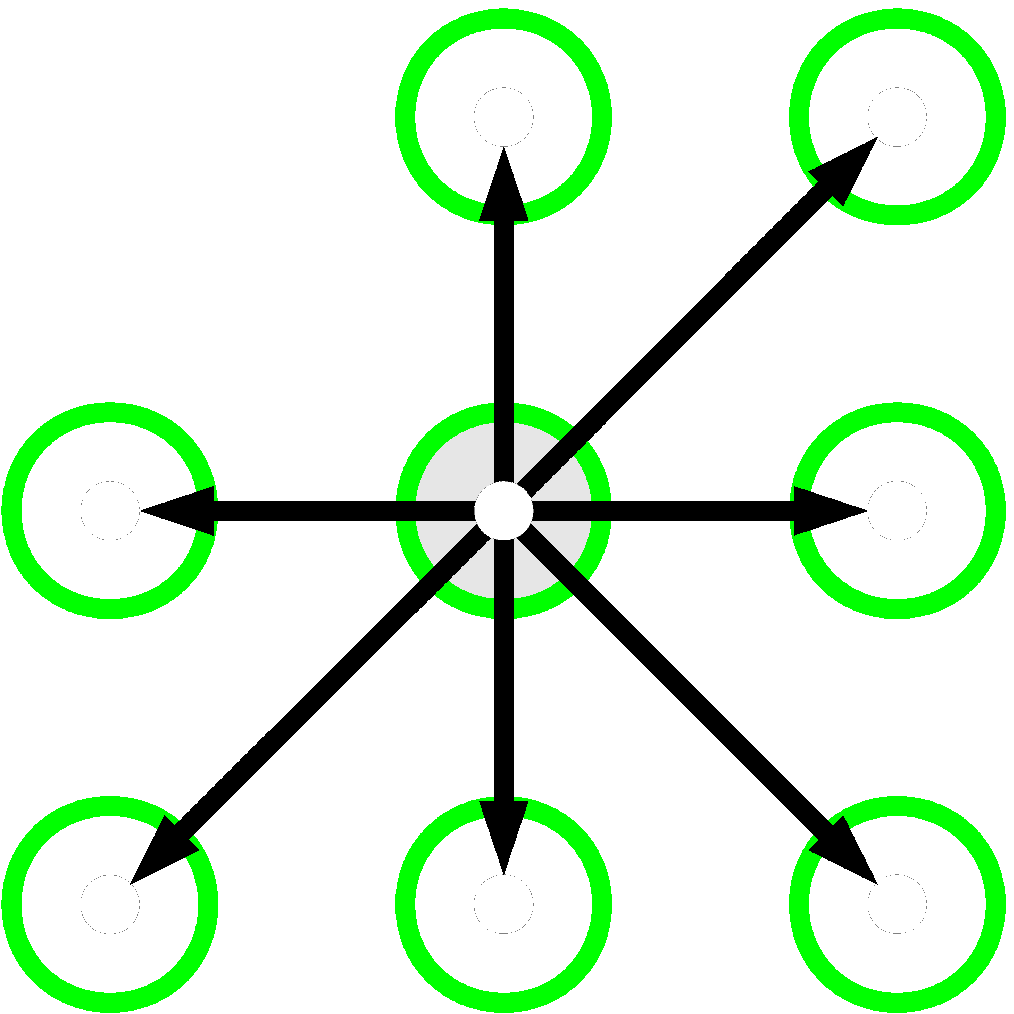}
  \hspace{5mm}
  \includegraphics[width=2cm]{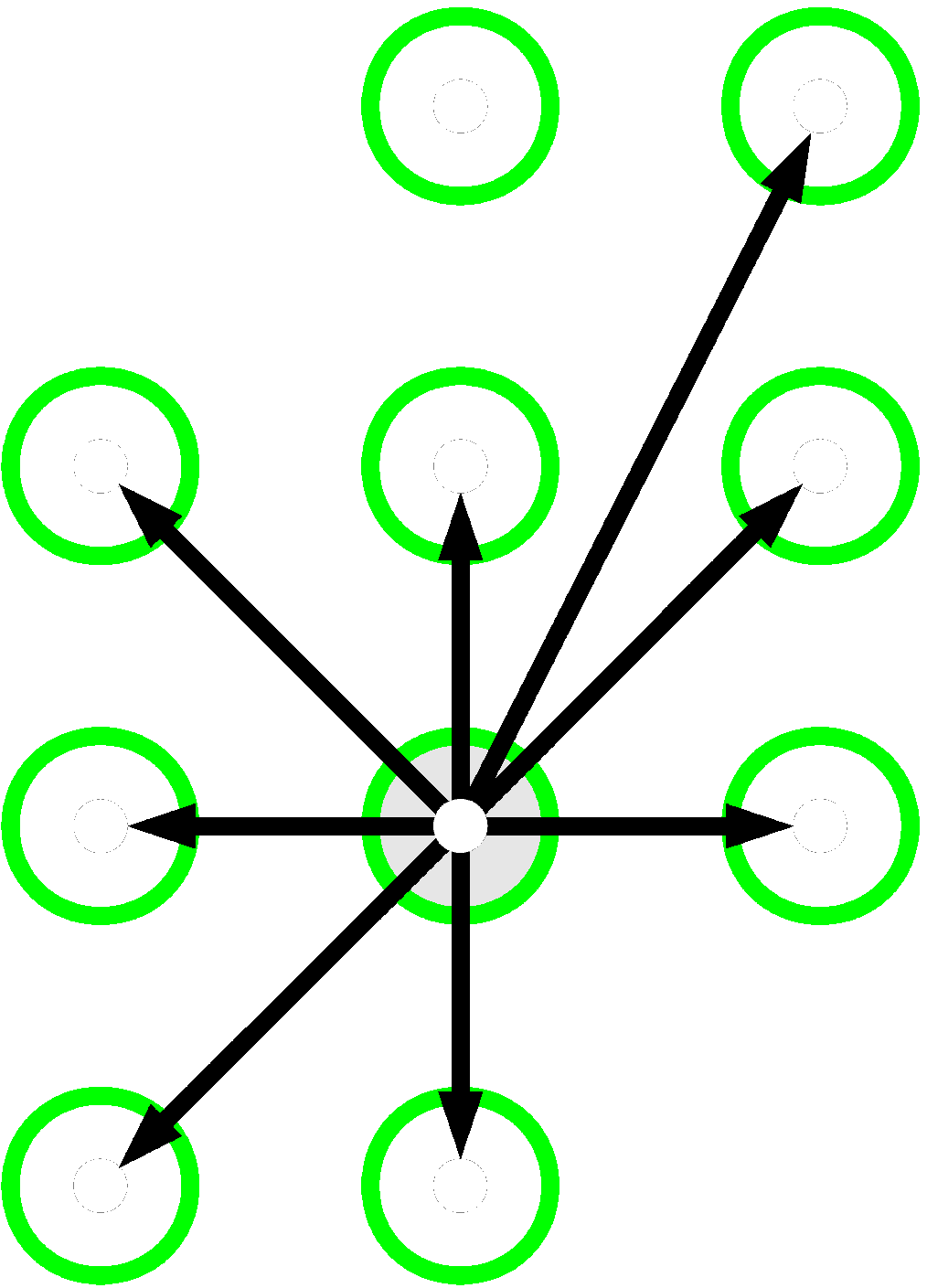}
  \hspace{5mm}
  \includegraphics[width=2cm]{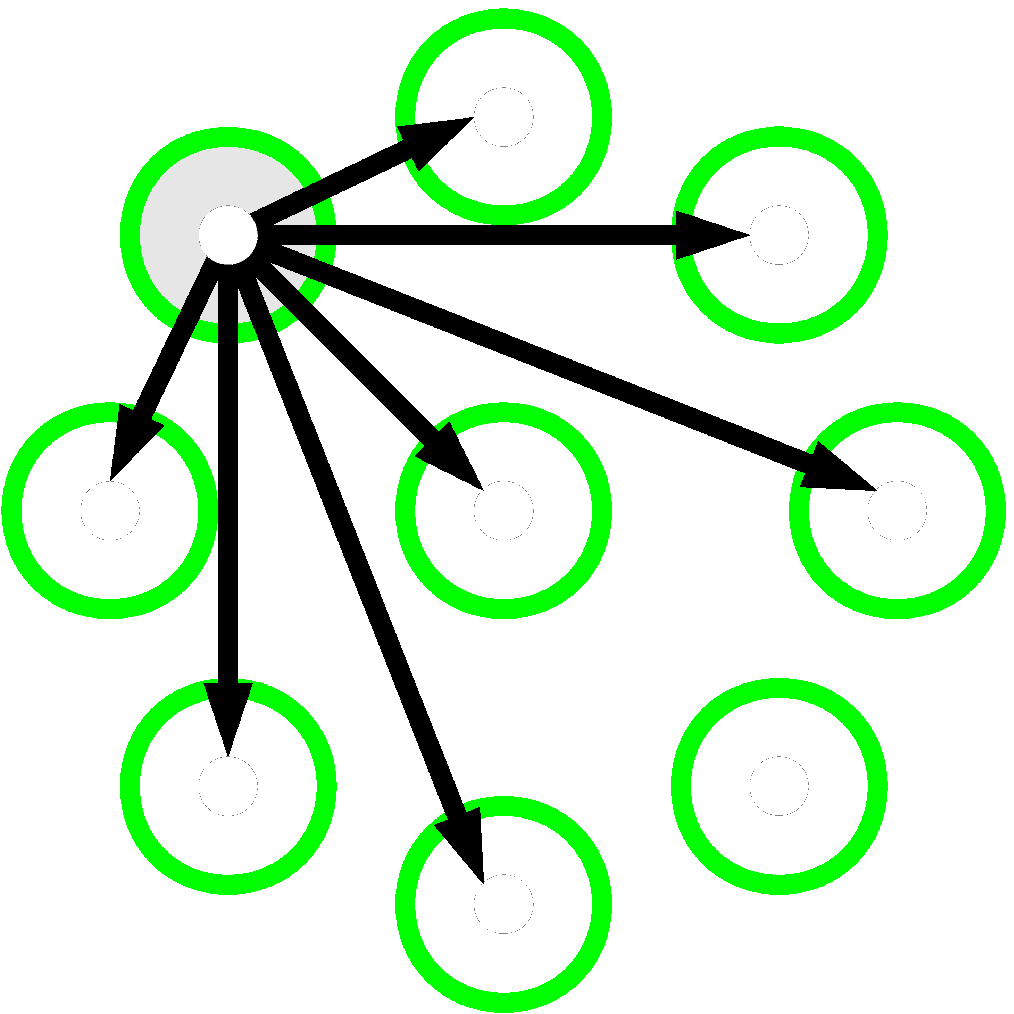}
  \hspace{5mm}
  \includegraphics[width=2cm]{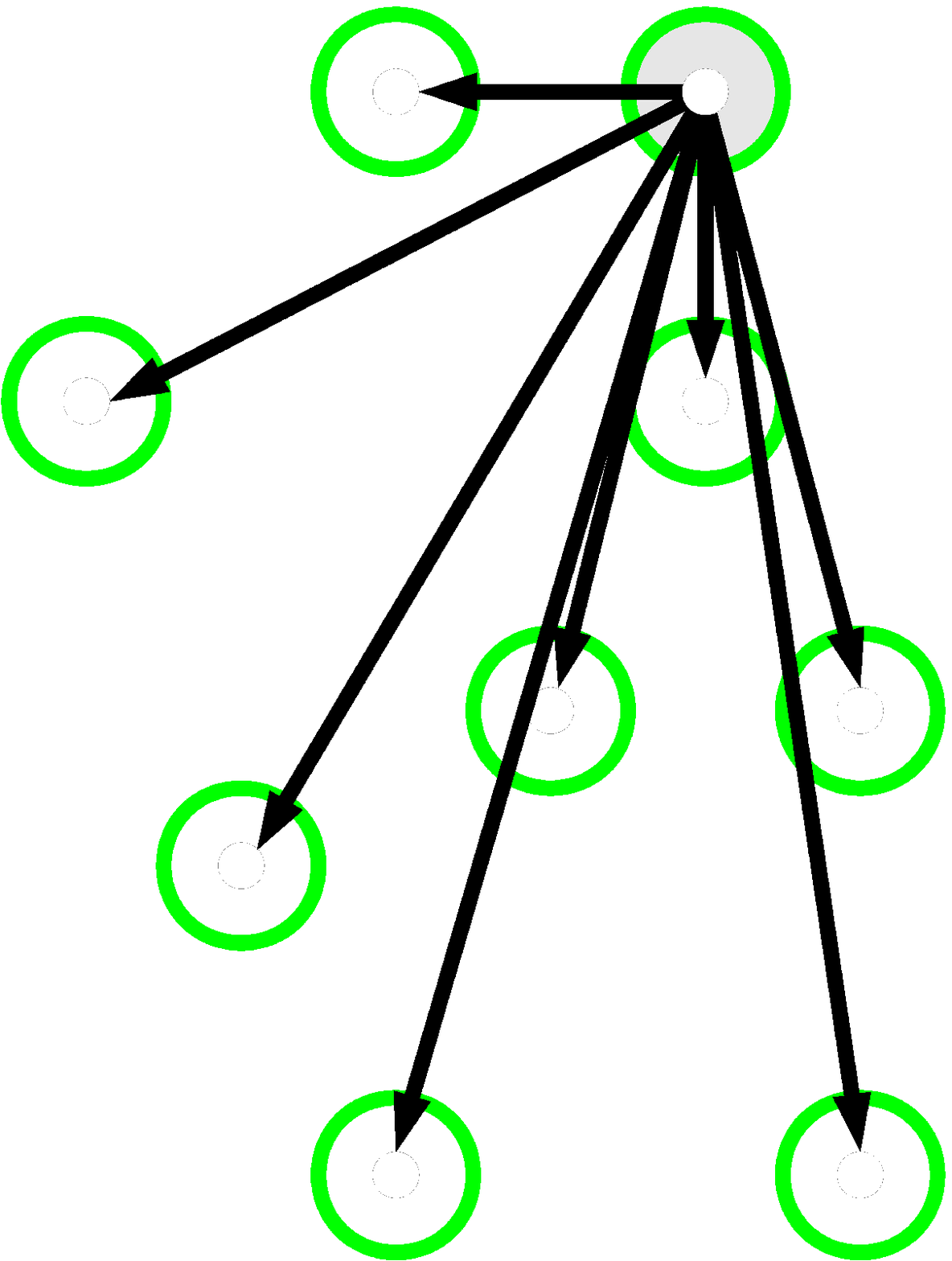}
  \caption{Alternative arrangements tested.}
  \label{fig:alternatives}
\vspace{-.2in}
\end{figure*}
}

\newcommand{\figfreqpatterns}[0]{

\begin{figure*}[!t]
\centering
\small
\resizebox{0.9\linewidth}{!}{
\begin{tabular}{c c c c c c c c c c}

\hline
\multicolumn{10}{c}{ \textbf{ ABK-MP-adv-off } } \\
\fbox{ \includegraphics[width=0.05\textwidth]{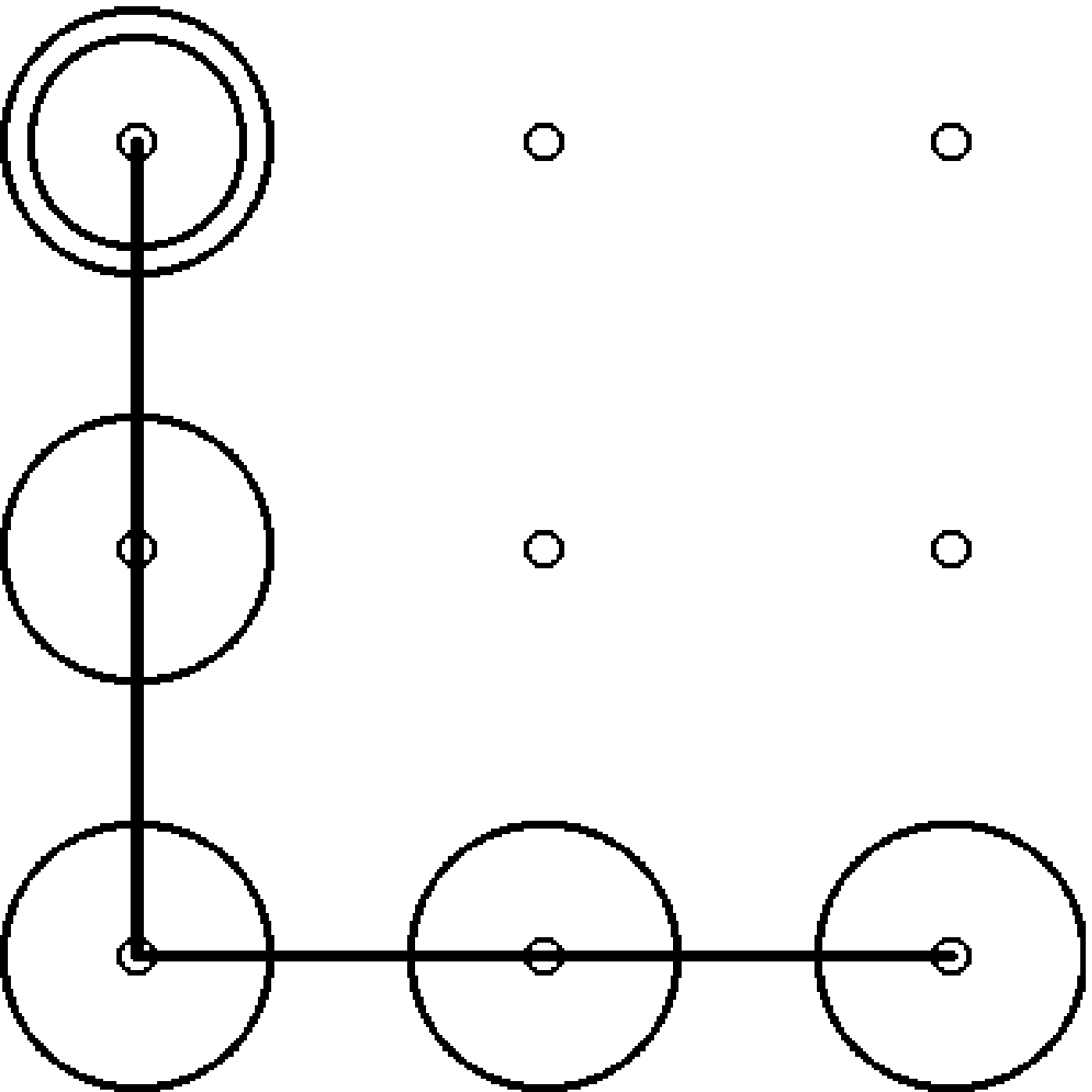} }  & \fbox{ \includegraphics[width=0.05\textwidth]{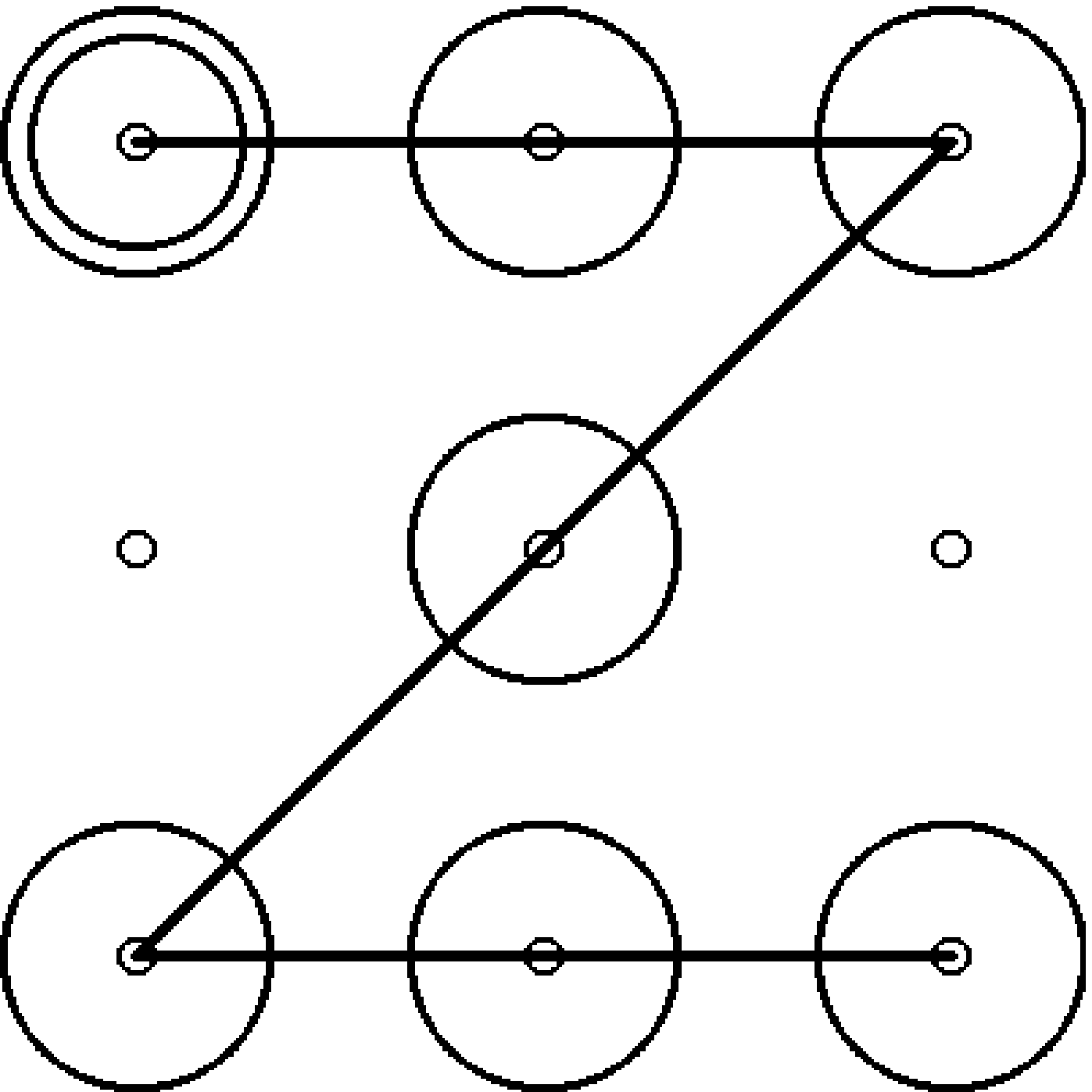} }  & \fbox{ \includegraphics[width=0.05\textwidth]{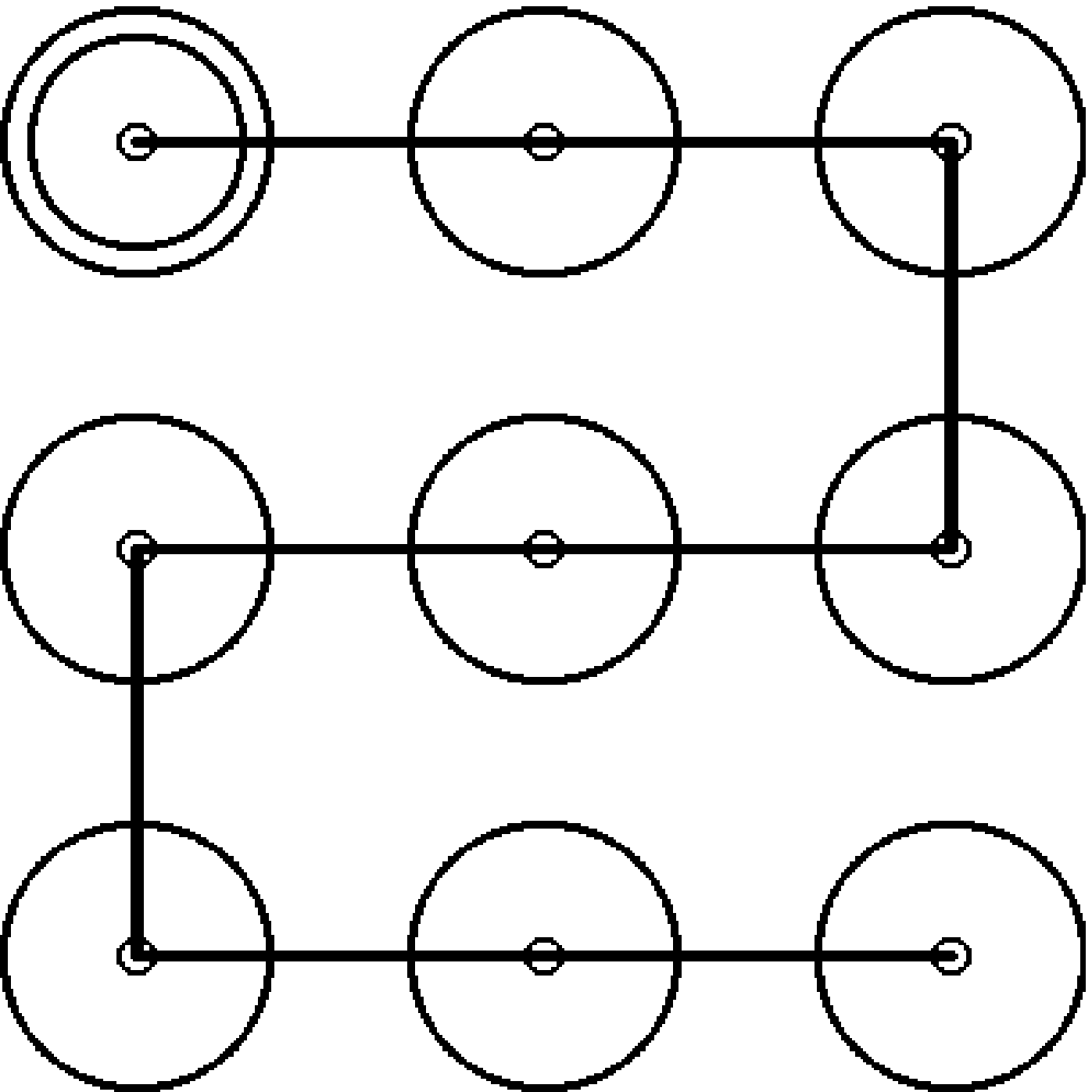} }  & \fbox{ \includegraphics[width=0.05\textwidth]{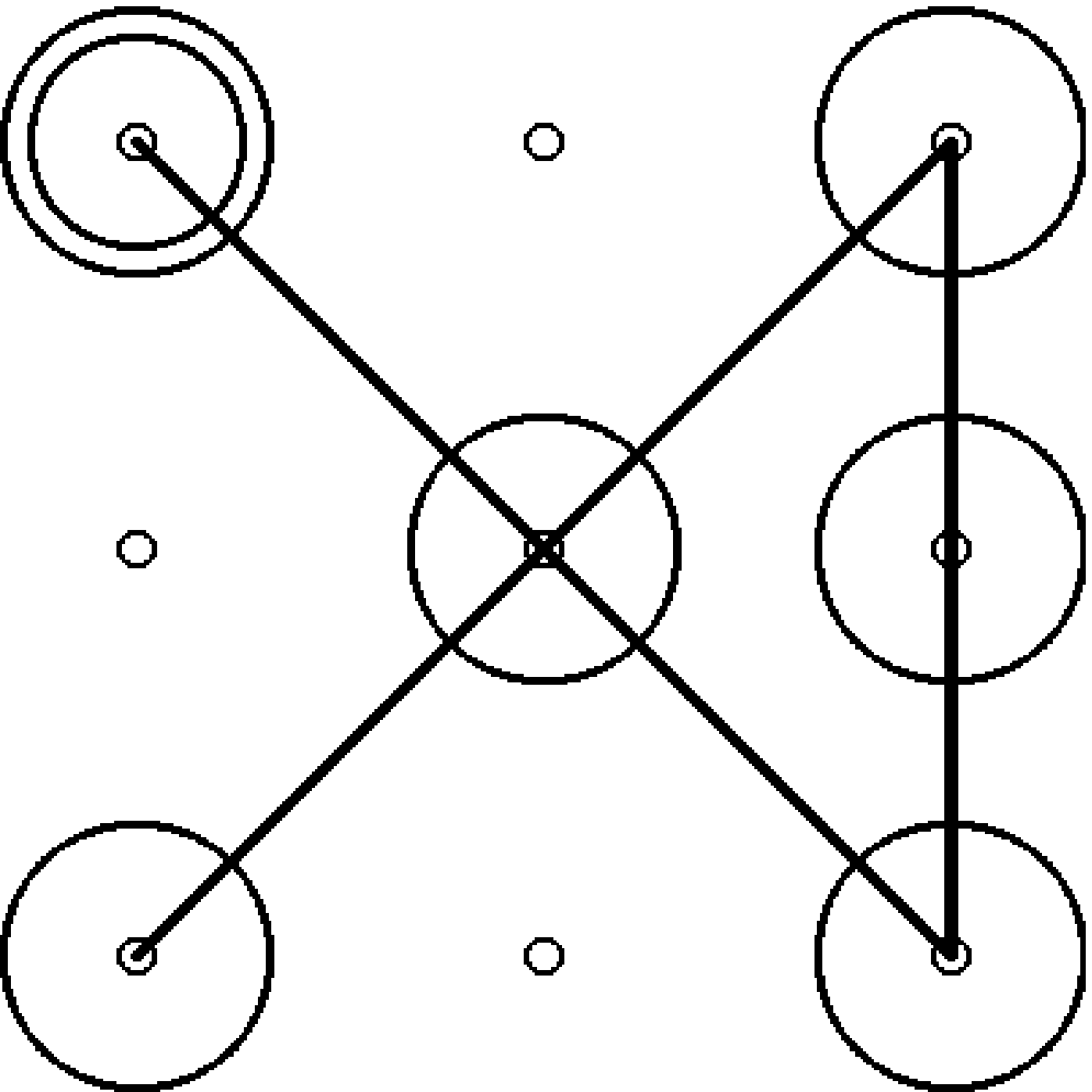} }  & \fbox{ \includegraphics[width=0.05\textwidth]{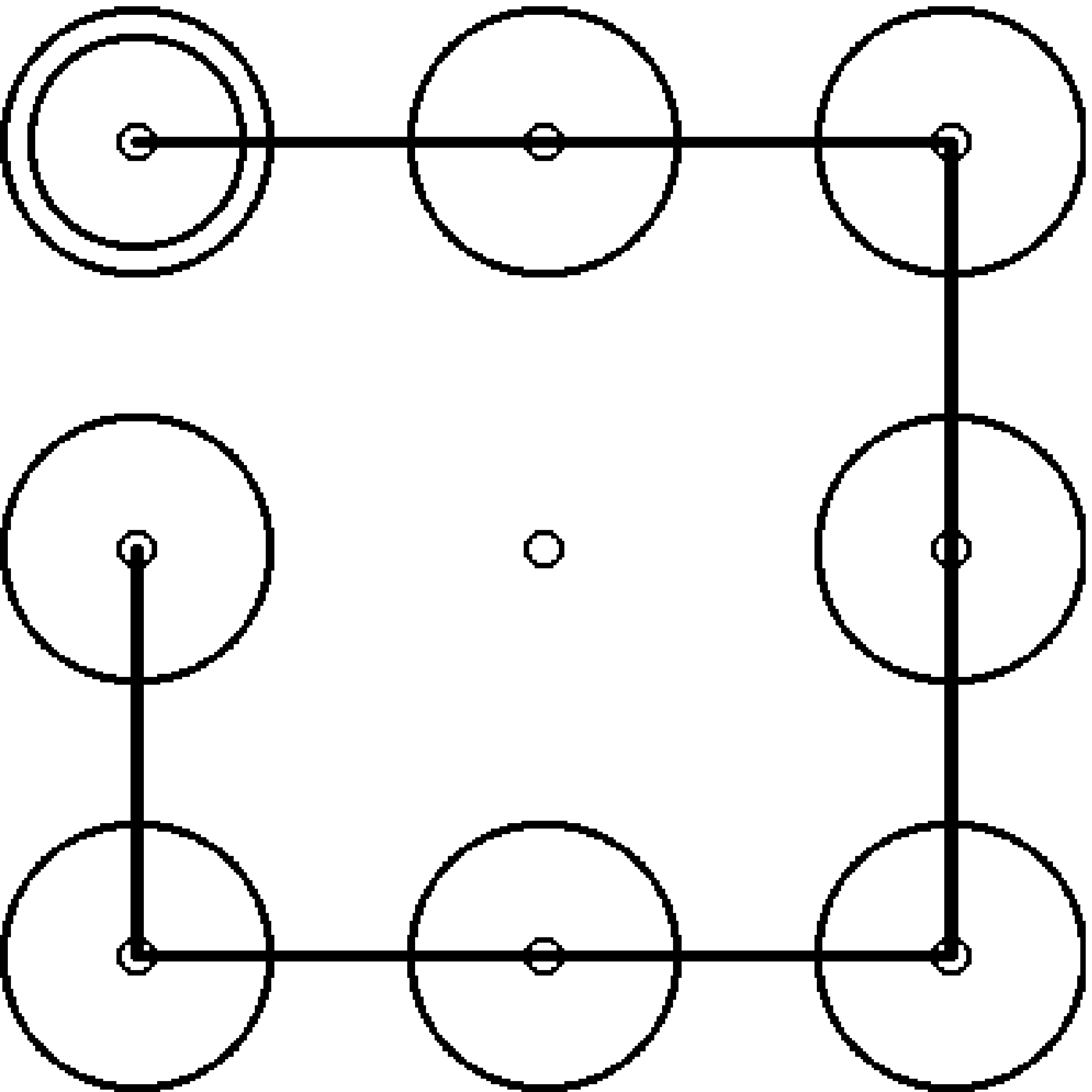} }  & \fbox{ \includegraphics[width=0.05\textwidth]{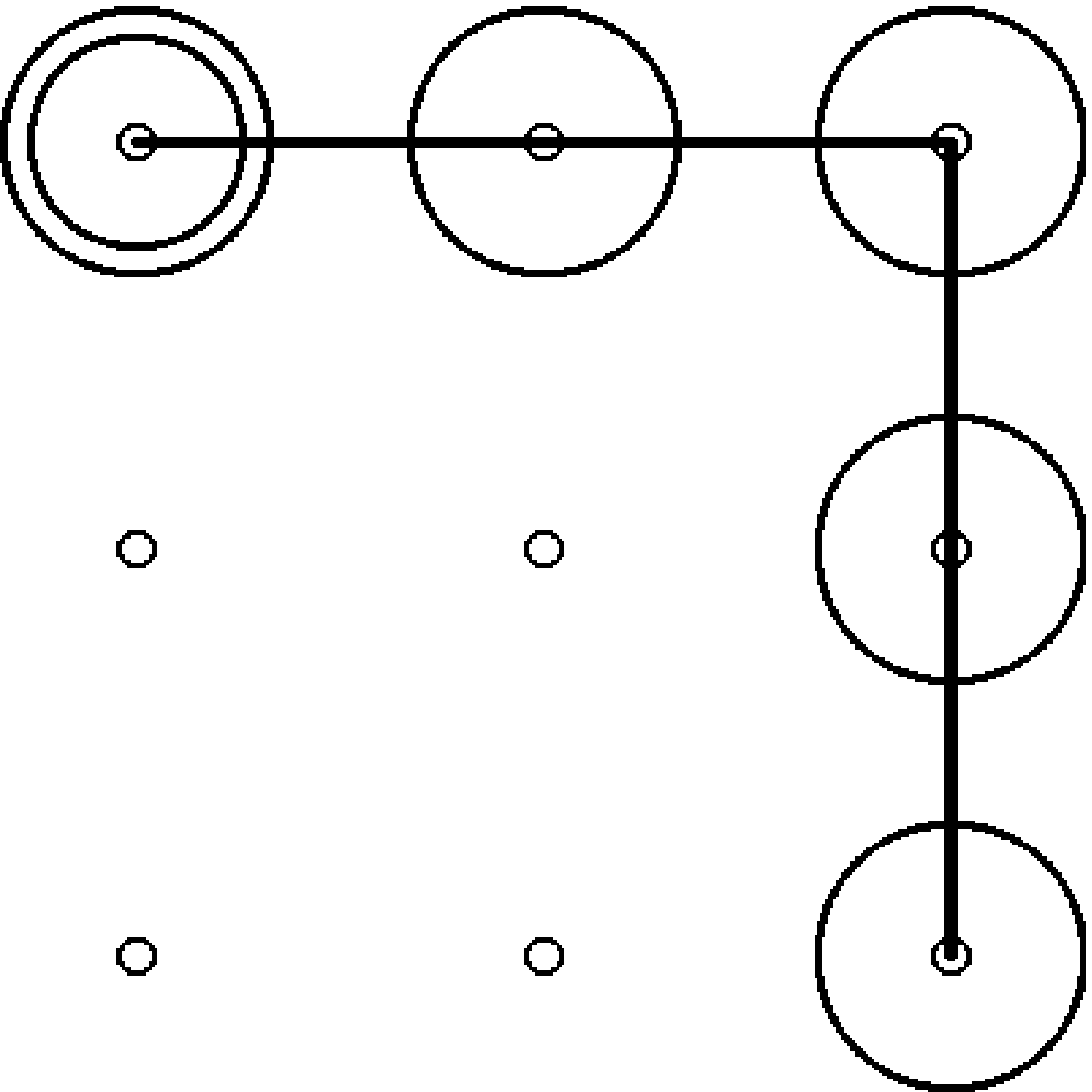} }  & \fbox{ \includegraphics[width=0.05\textwidth]{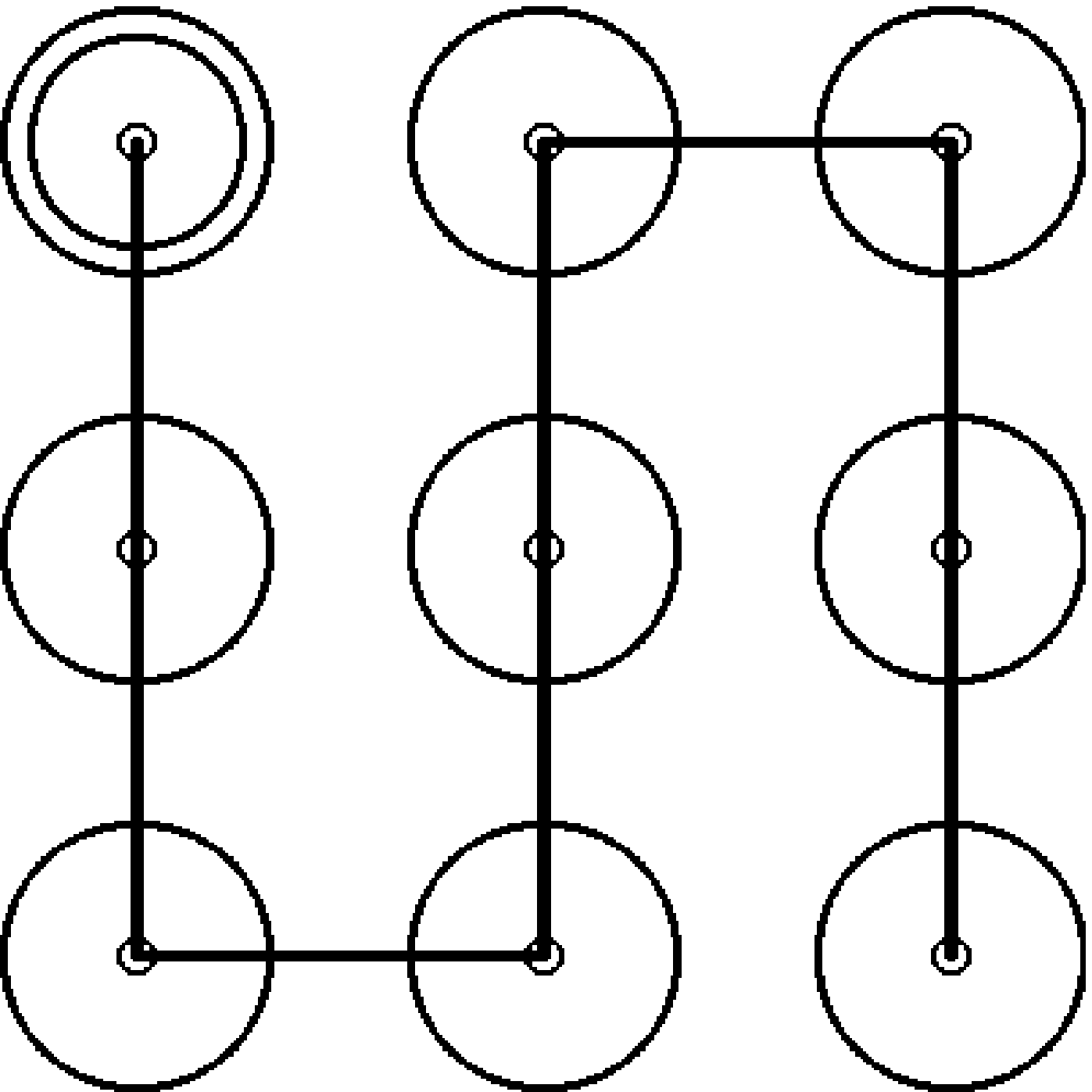} }  & \fbox{ \includegraphics[width=0.05\textwidth]{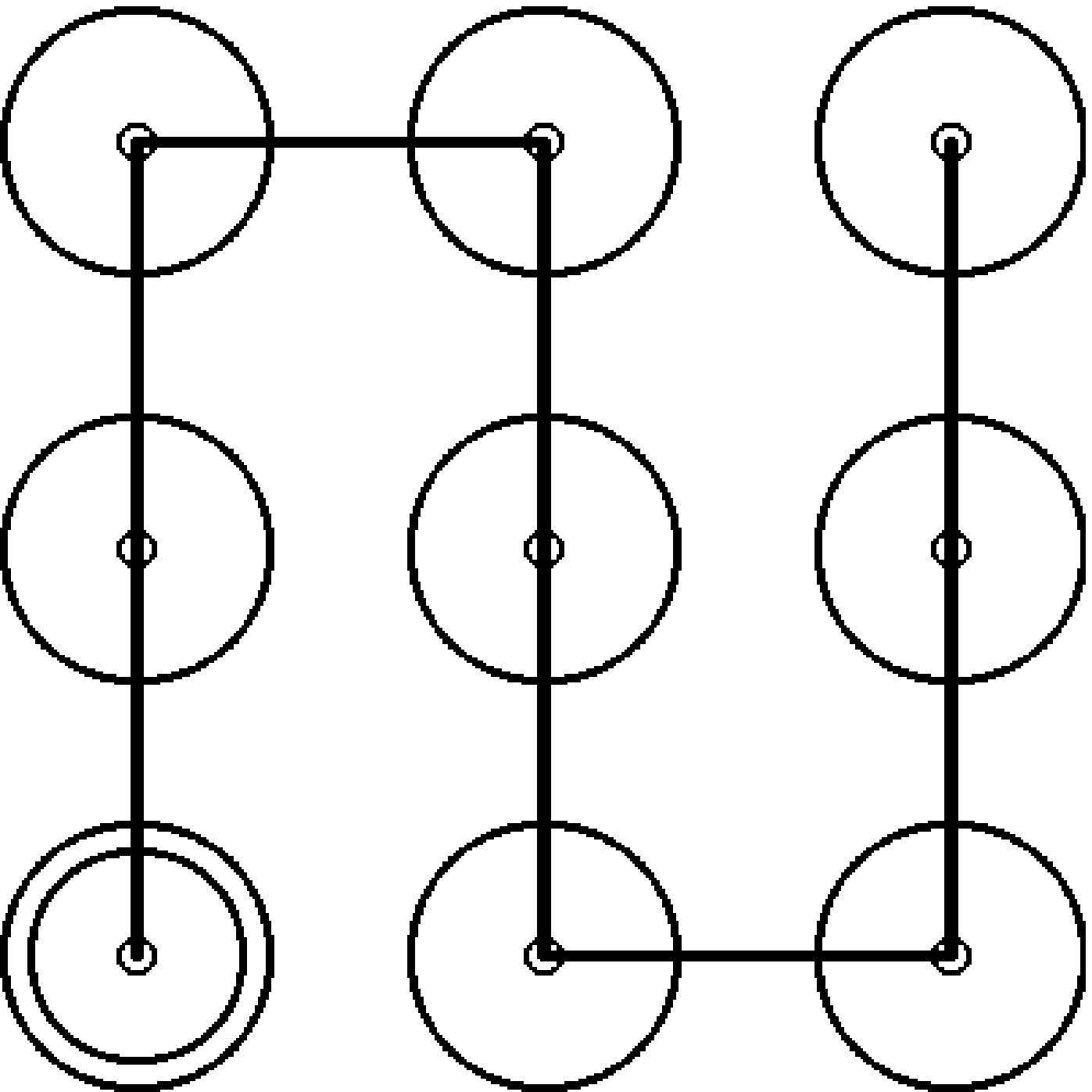} }  & \fbox{ \includegraphics[width=0.05\textwidth]{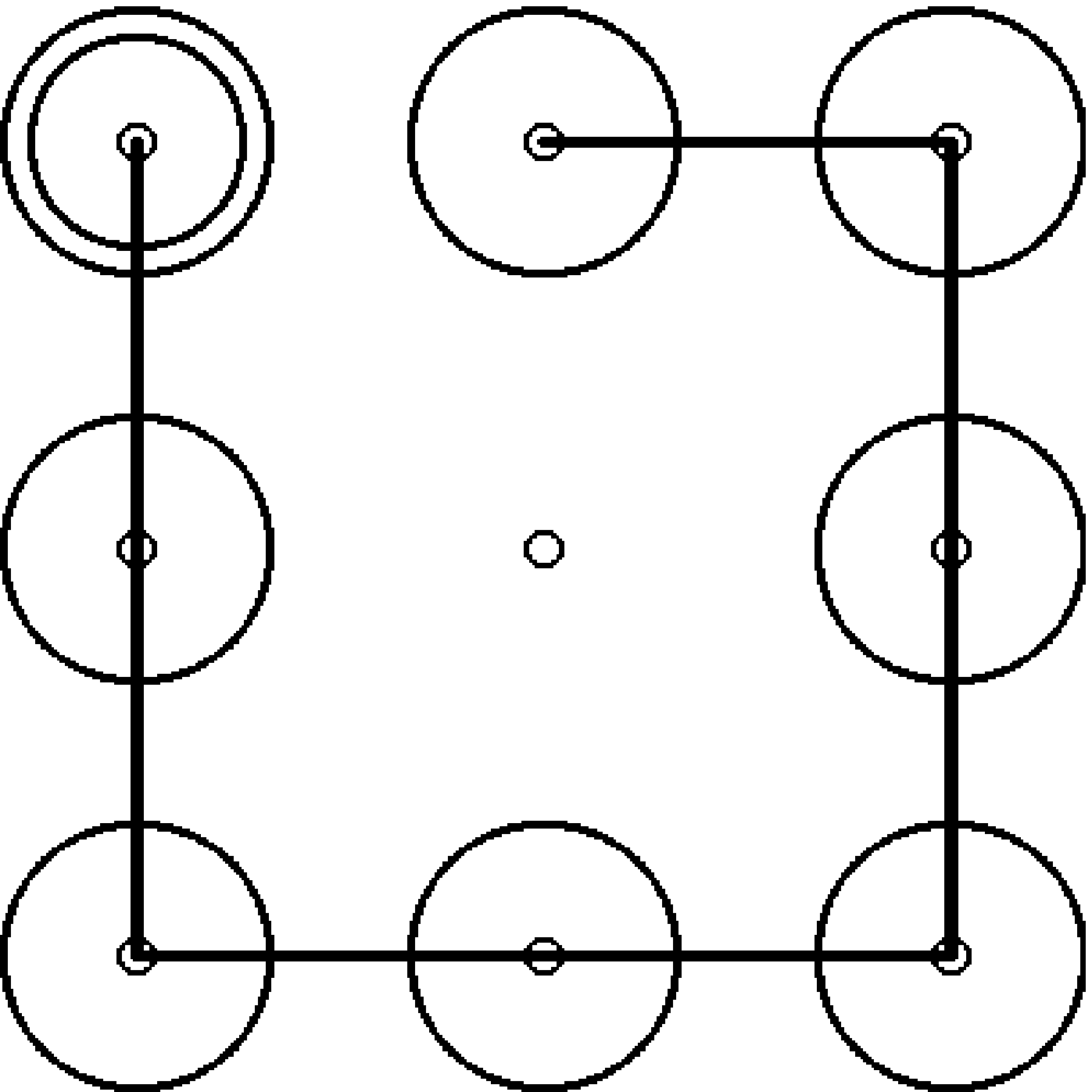} }  & \fbox{ \includegraphics[width=0.05\textwidth]{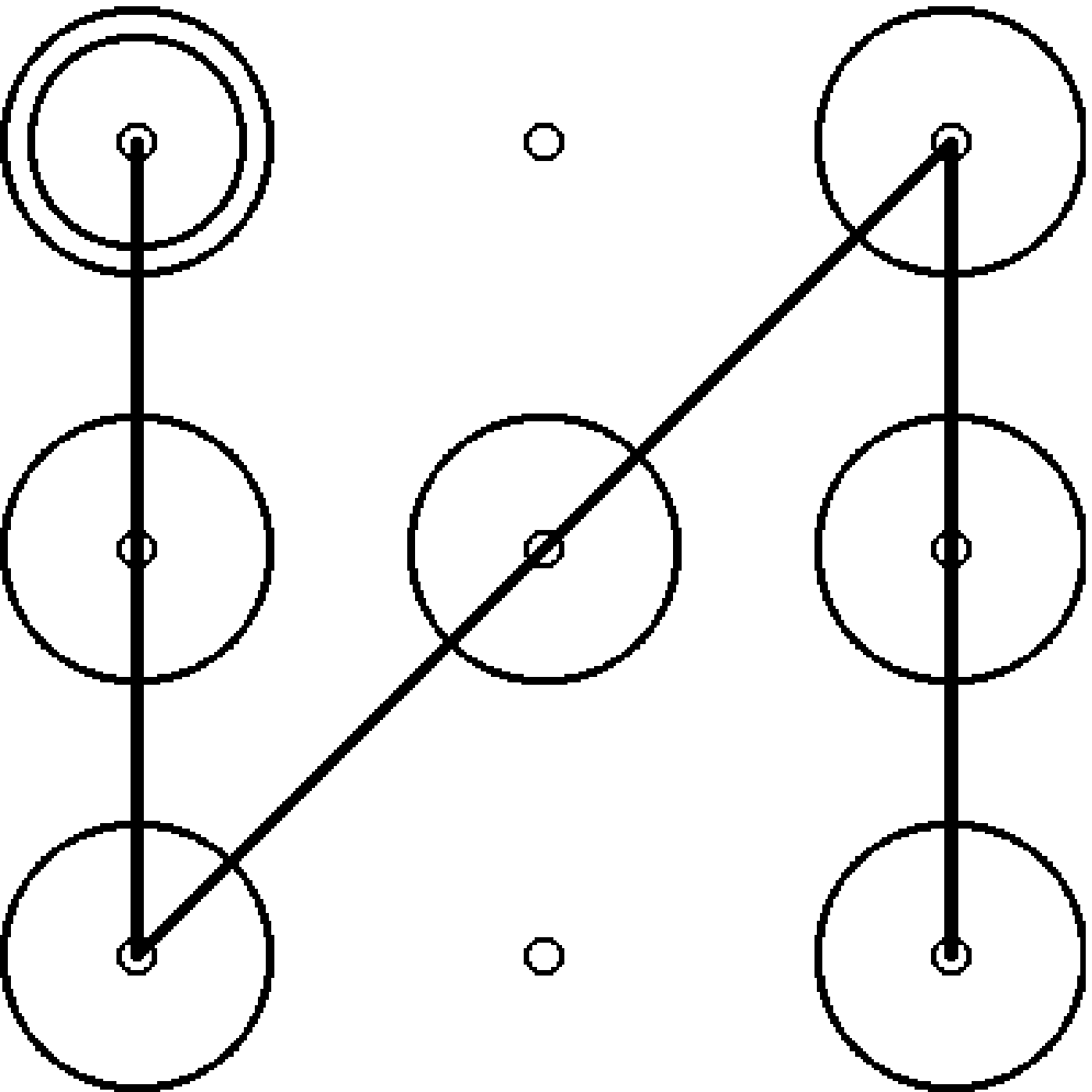} }  \\
\textbf{ freq=9 } & \textbf{ freq=9 } & \textbf{ freq=9 } & \textbf{ freq=7 } & \textbf{ freq=6 } & \textbf{ freq=5 } & \textbf{ freq=5 } & \textbf{ freq=5 } & \textbf{ freq=5 } & \textbf{ freq=5 } \\
\hline
\multicolumn{10}{c}{ \textbf{ABK-MP-self } } \\
\fbox{ \includegraphics[width=0.05\textwidth]{images/patterns/0-1-2-4-6-7-8.eps} }  & \fbox{ \includegraphics[width=0.05\textwidth]{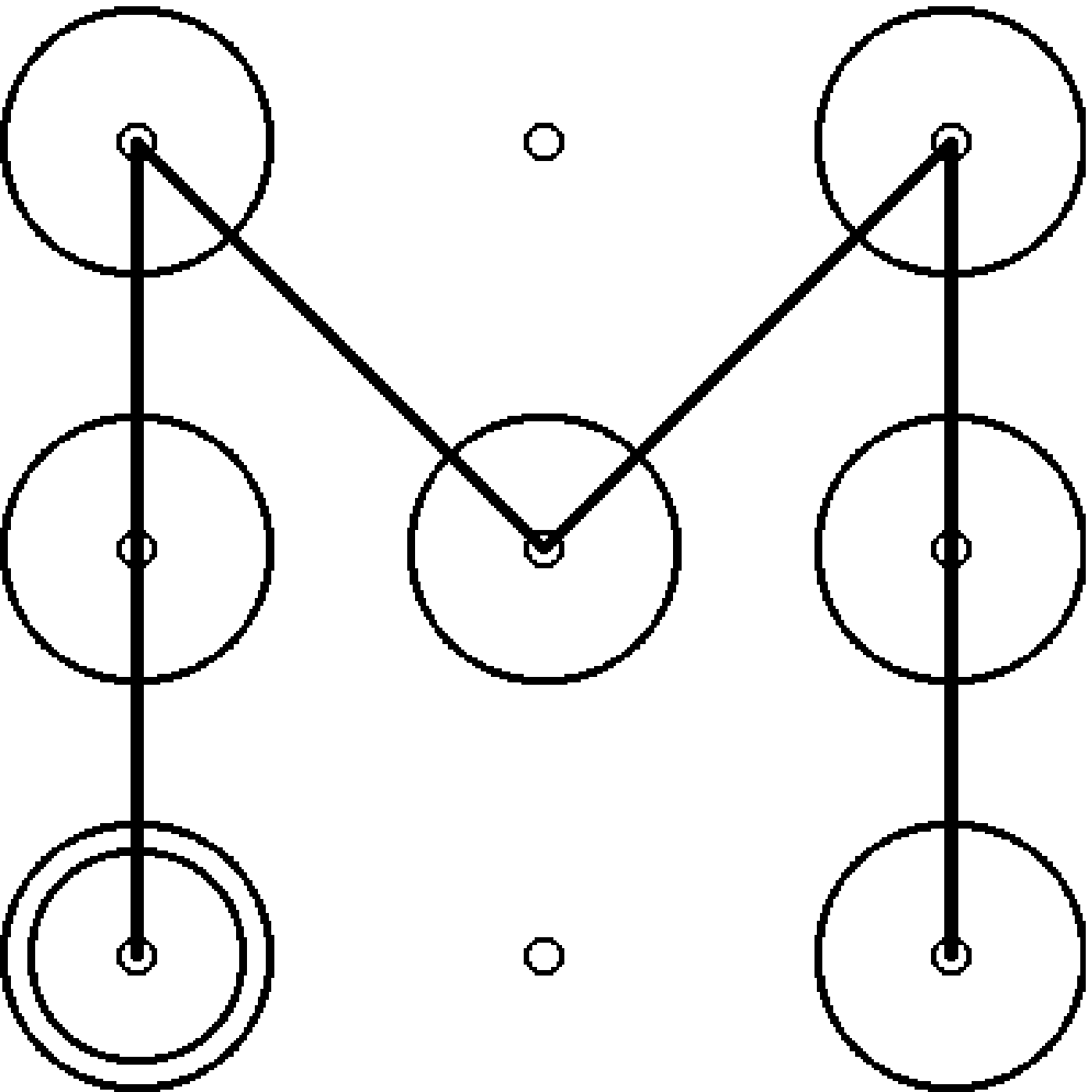} }  & \fbox{ \includegraphics[width=0.05\textwidth]{images/patterns/0-1-2-5-8.eps} }  & \fbox{ \includegraphics[width=0.05\textwidth]{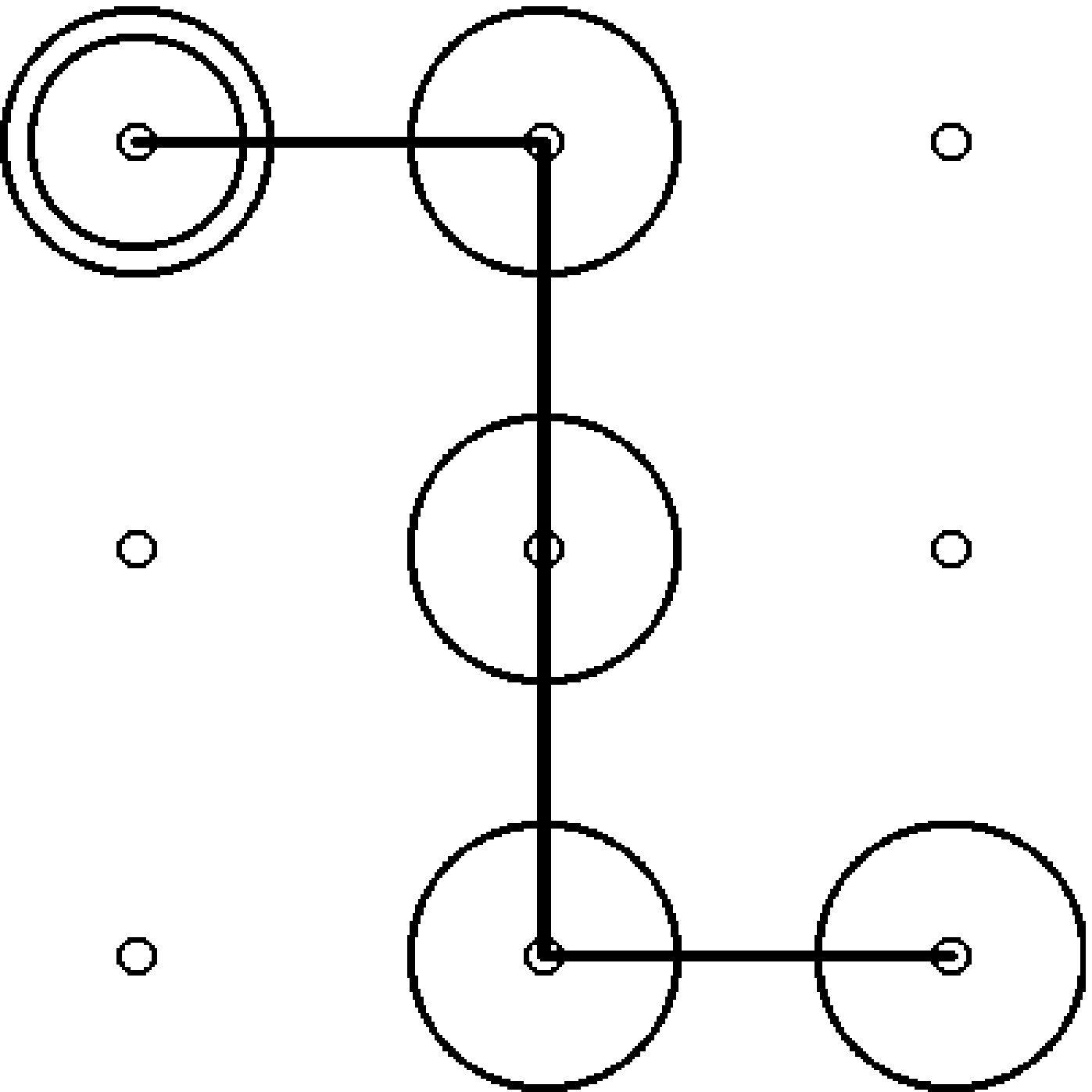} }  & \fbox{ \includegraphics[width=0.05\textwidth]{images/patterns/0-3-6-7-8.eps} }  & \fbox{ \includegraphics[width=0.05\textwidth]{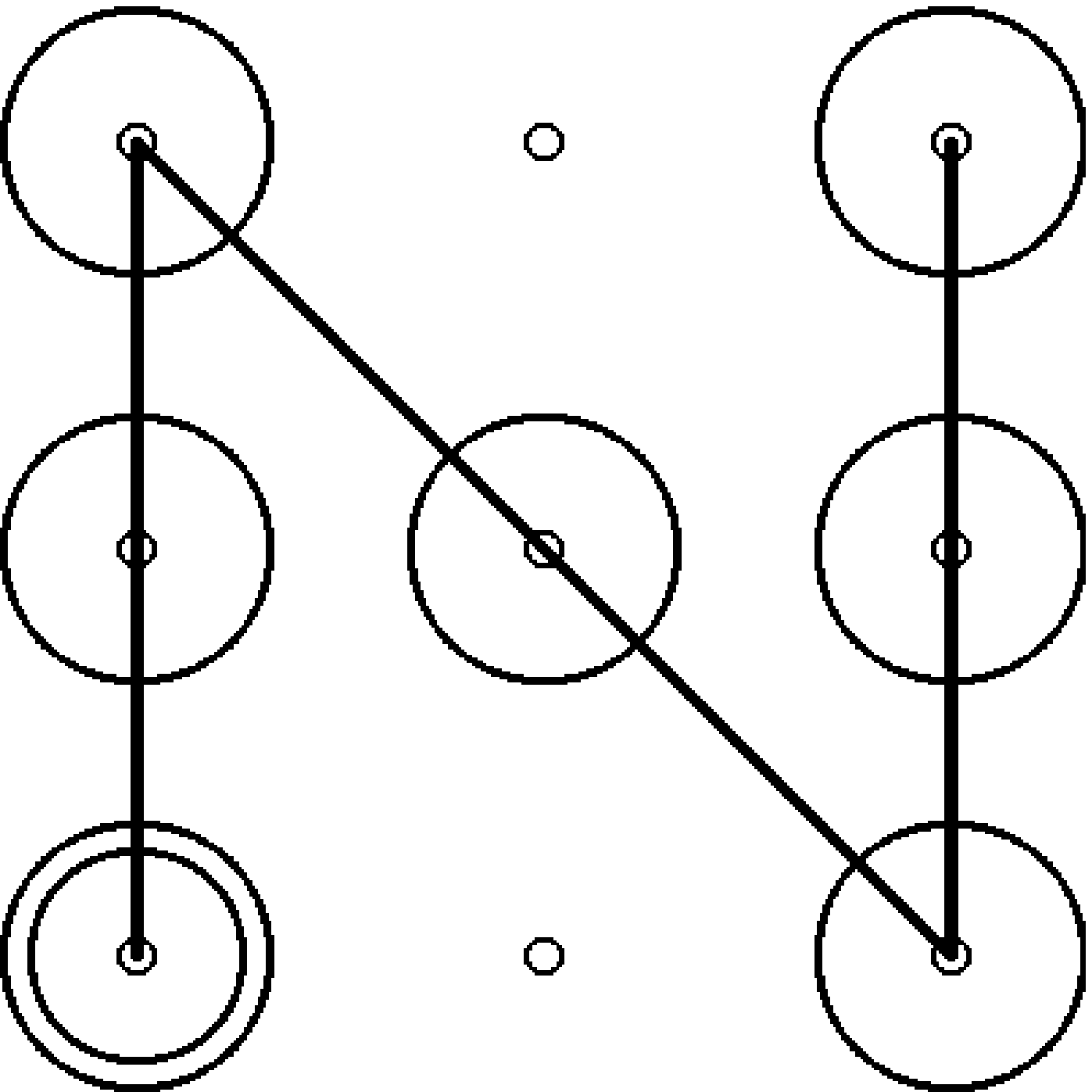} }  & \fbox{ \includegraphics[width=0.05\textwidth]{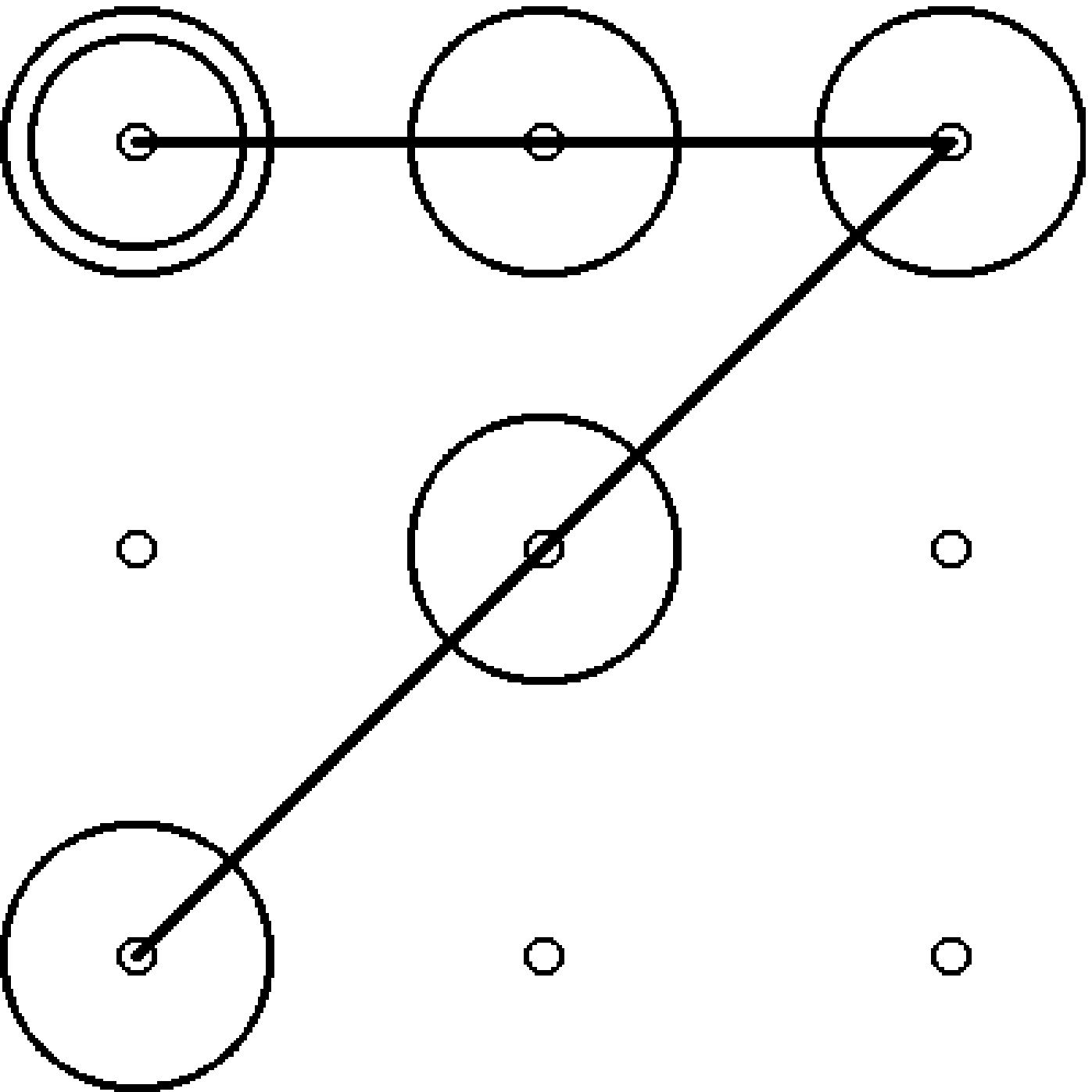} }  & \fbox{ \includegraphics[width=0.05\textwidth]{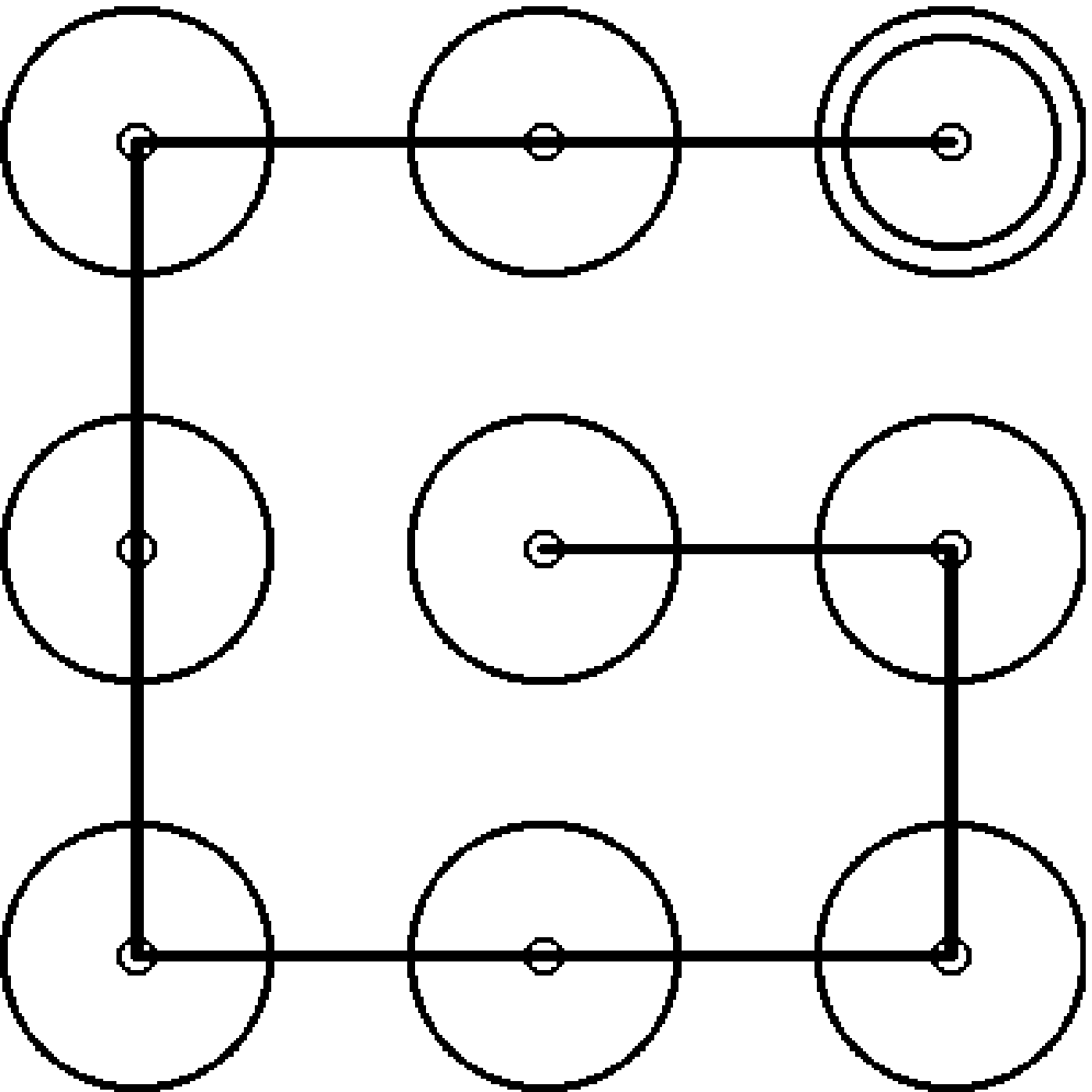} }  & \fbox{ \includegraphics[width=0.05\textwidth]{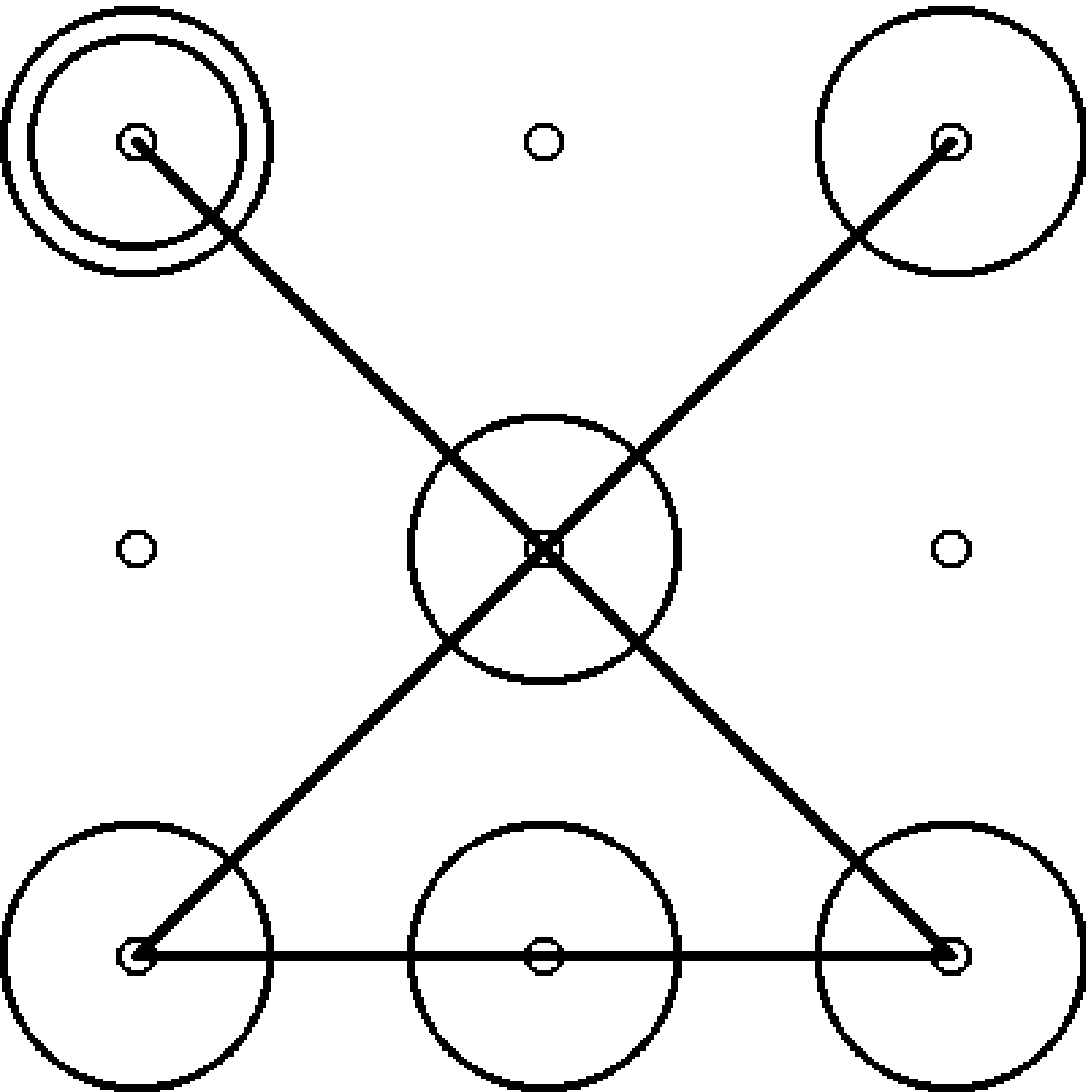} }  & \fbox{ \includegraphics[width=0.05\textwidth]{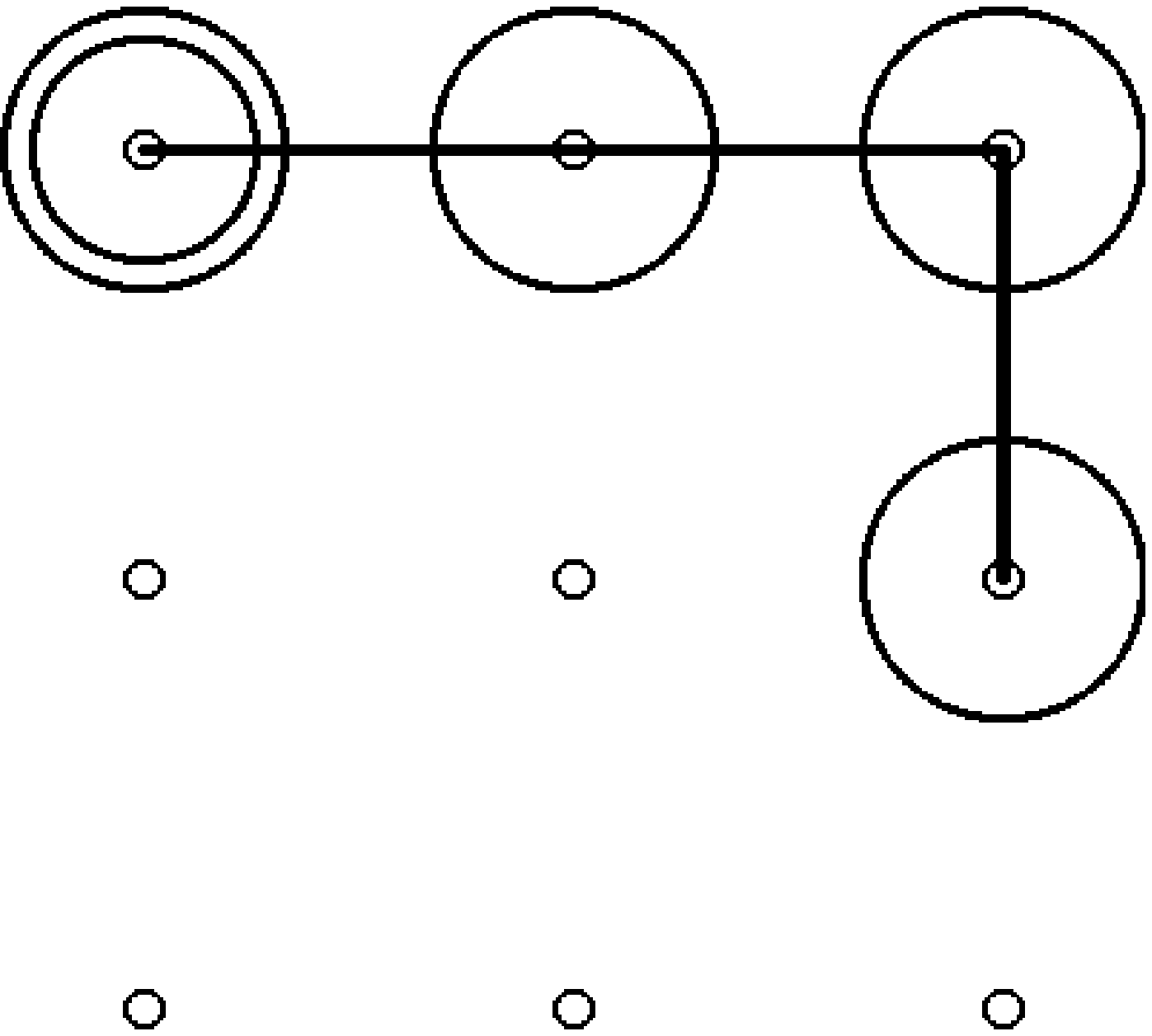} }  \\
\textbf{ freq=17 } & \textbf{ freq=11 } & \textbf{ freq=8 } & \textbf{ freq=8 } & \textbf{ freq=7 } & \textbf{ freq=6 } & \textbf{ freq=6 } & \textbf{ freq=6 } & \textbf{ freq=5 } & \textbf{ freq=4 } \\
\hline
\multicolumn{10}{c}{ \textbf{LDR-banking} } \\
\fbox{ \includegraphics[width=0.05\textwidth]{images/patterns/0-3-6-7-8.eps} }  & \fbox{ \includegraphics[width=0.05\textwidth]{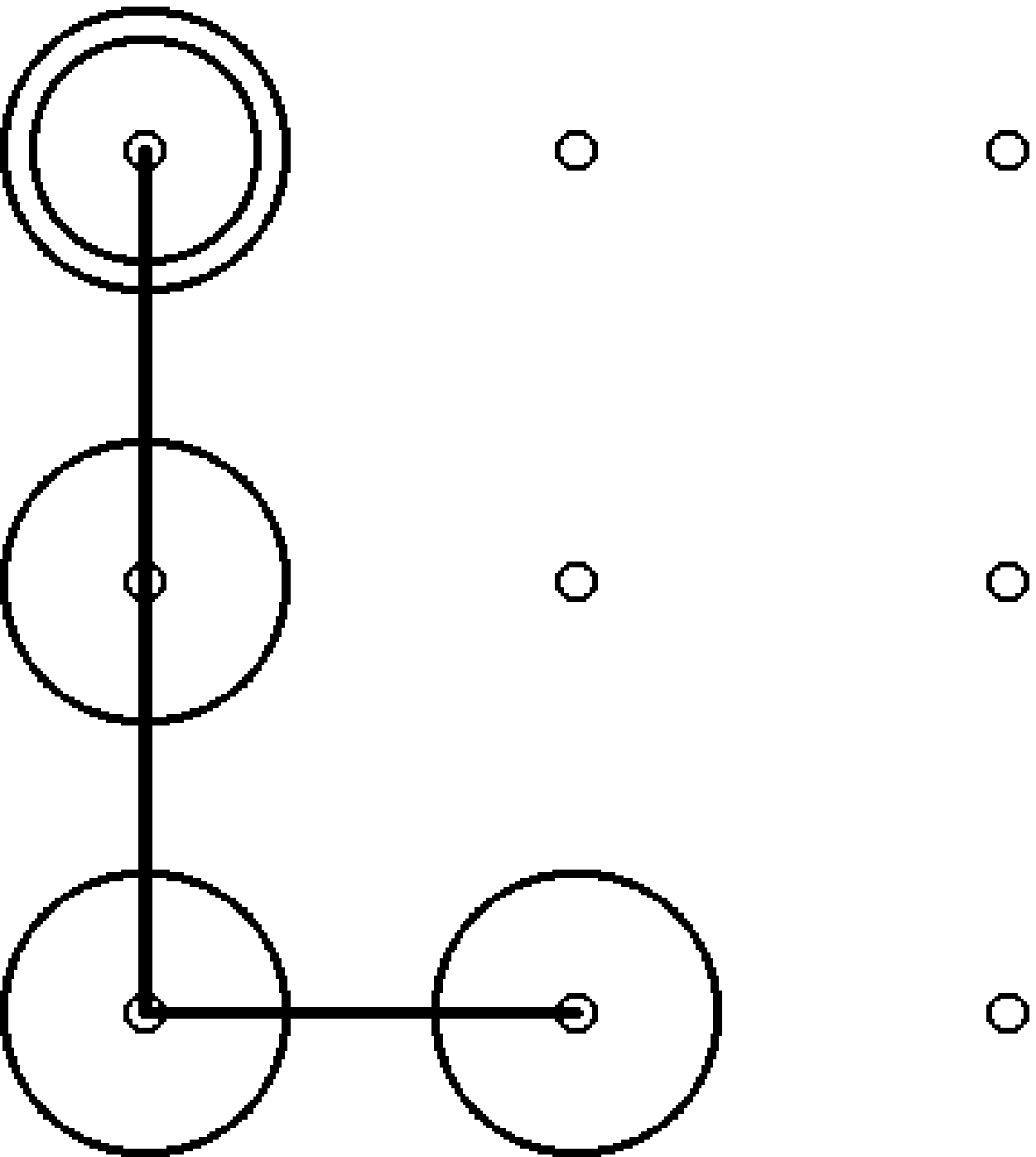} }  & \fbox{ \includegraphics[width=0.05\textwidth]{images/patterns/0-1-2-5.eps} }  & \fbox{ \includegraphics[width=0.05\textwidth]{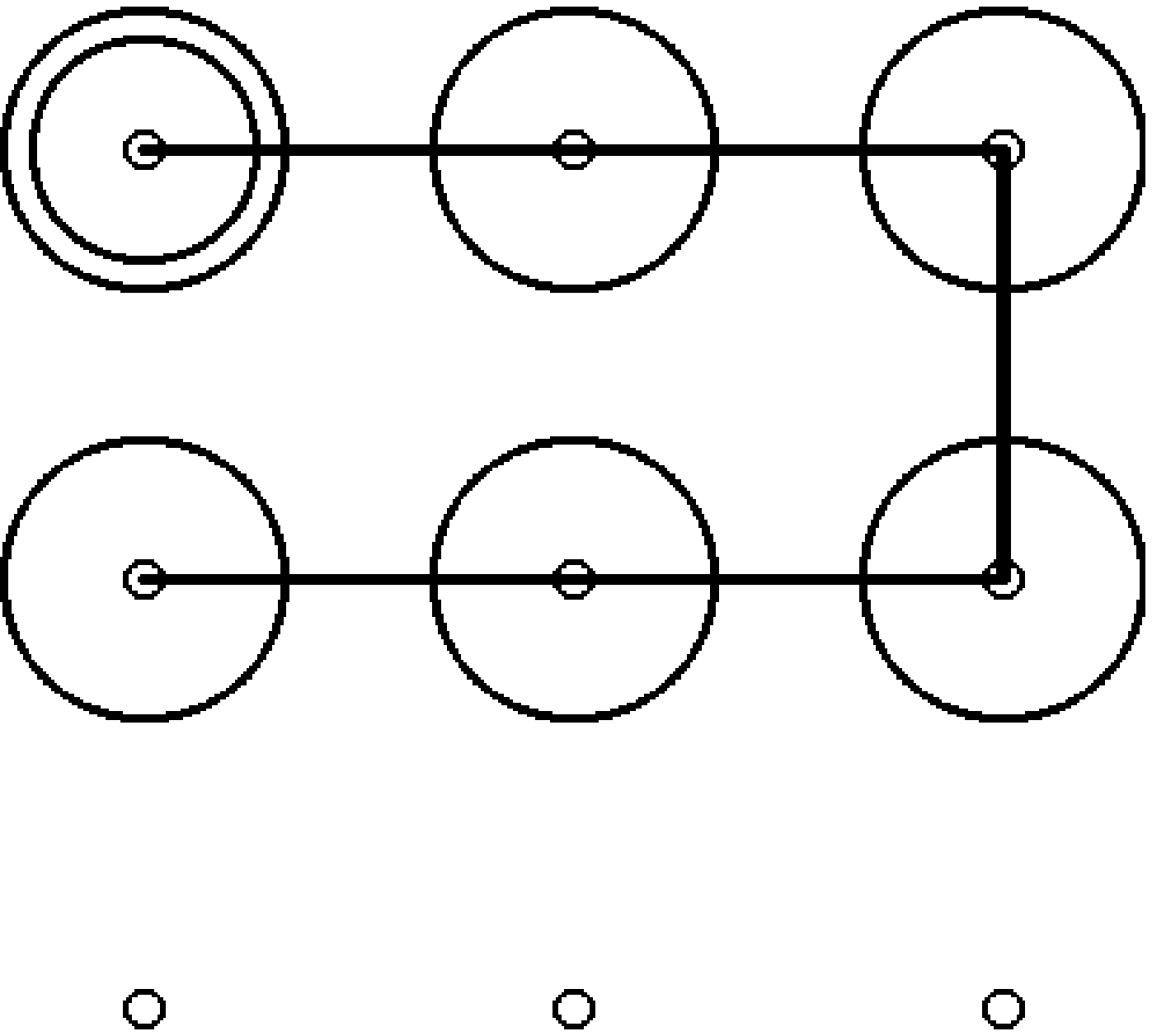} }  & \fbox{ \includegraphics[width=0.05\textwidth]{images/patterns/0-1-2-5-8.eps} }  & \fbox{ \includegraphics[width=0.05\textwidth]{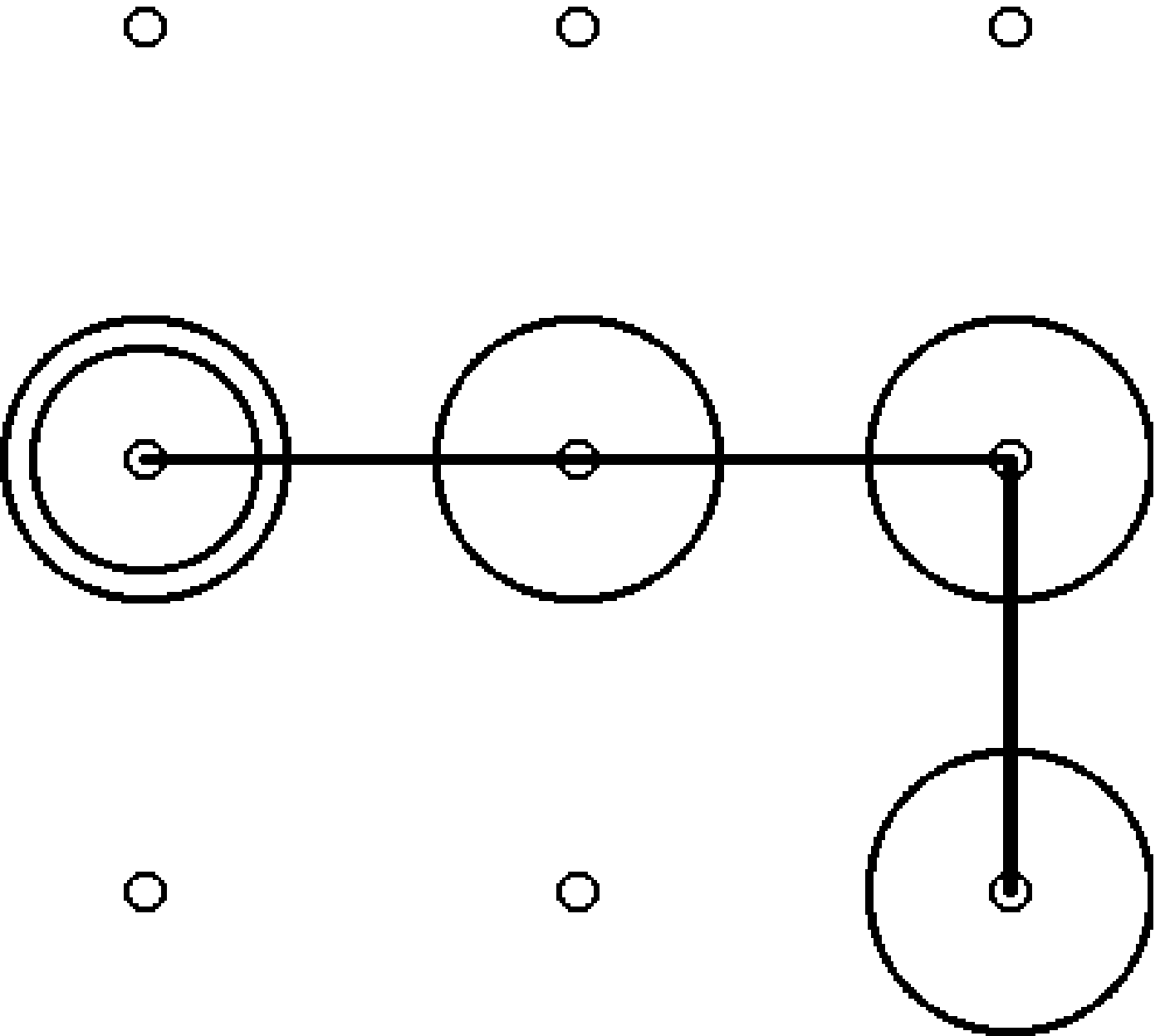} }  & \fbox{ \includegraphics[width=0.05\textwidth]{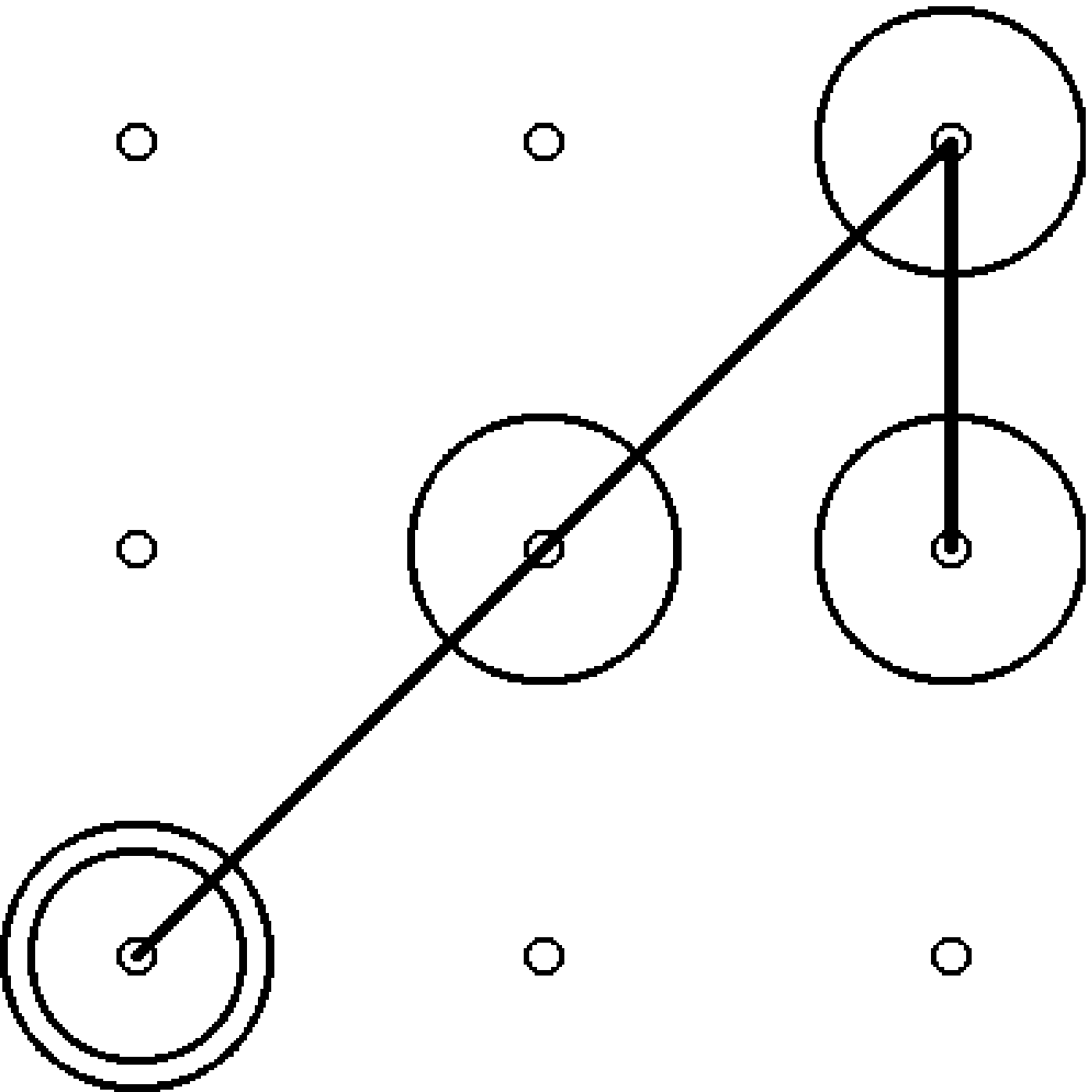} }  & \fbox{ \includegraphics[width=0.05\textwidth]{images/patterns/0-1-2-4-6-7-8.eps} }  & \fbox{ \includegraphics[width=0.05\textwidth]{images/patterns/0-1-2-5-4-3-6-7-8.eps} }  & \fbox{ \includegraphics[width=0.05\textwidth]{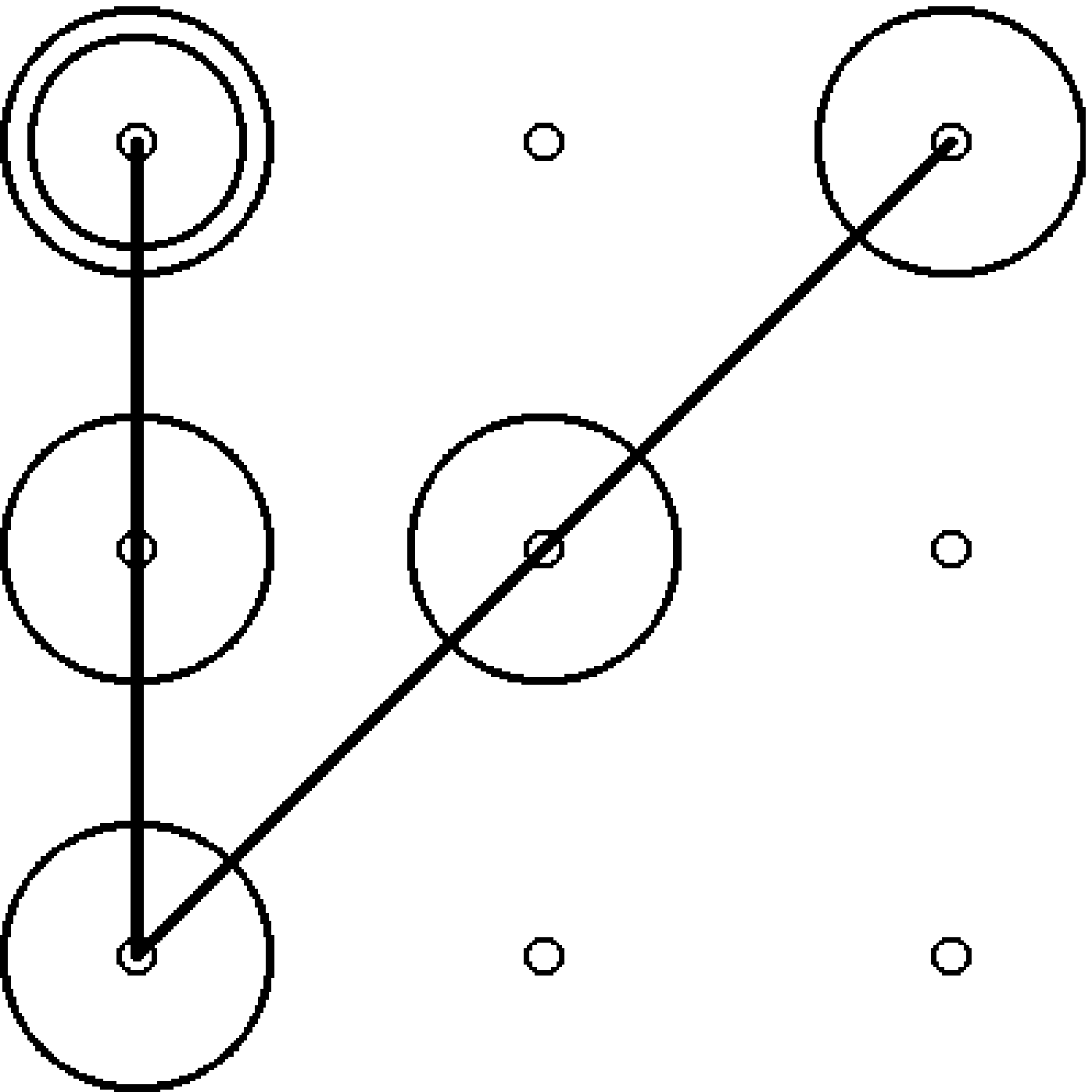} }  \\
\textbf{ freq=17 } & \textbf{ freq=16 } & \textbf{ freq=15 } & \textbf{ freq=9 } & \textbf{ freq=8 } & \textbf{ freq=8 } & \textbf{ freq=6 } & \textbf{ freq=6 } & \textbf{ freq=6 } & \textbf{ freq=6 } \\
\hline
\multicolumn{10}{c}{ \textbf{LDR-shopping} } \\
\fbox{ \includegraphics[width=0.05\textwidth]{images/patterns/0-3-6-7.eps} }  & \fbox{ \includegraphics[width=0.05\textwidth]{images/patterns/0-1-2-5-8.eps} }  & \fbox{ \includegraphics[width=0.05\textwidth]{images/patterns/0-1-2-5.eps} }  & \fbox{ \includegraphics[width=0.05\textwidth]{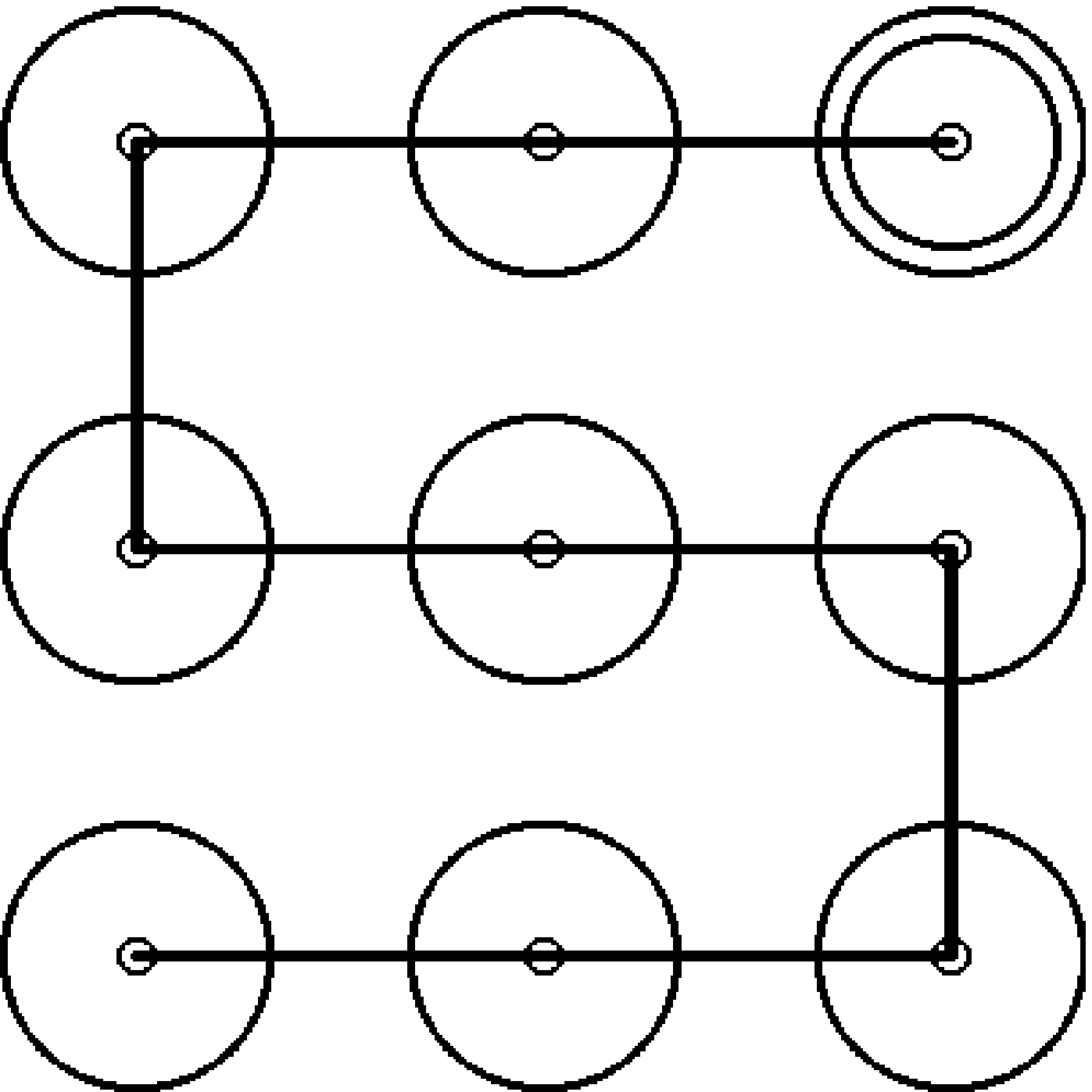} }  & \fbox{ \includegraphics[width=0.05\textwidth]{images/patterns/0-3-6-7-8.eps} }  & \fbox{ \includegraphics[width=0.05\textwidth]{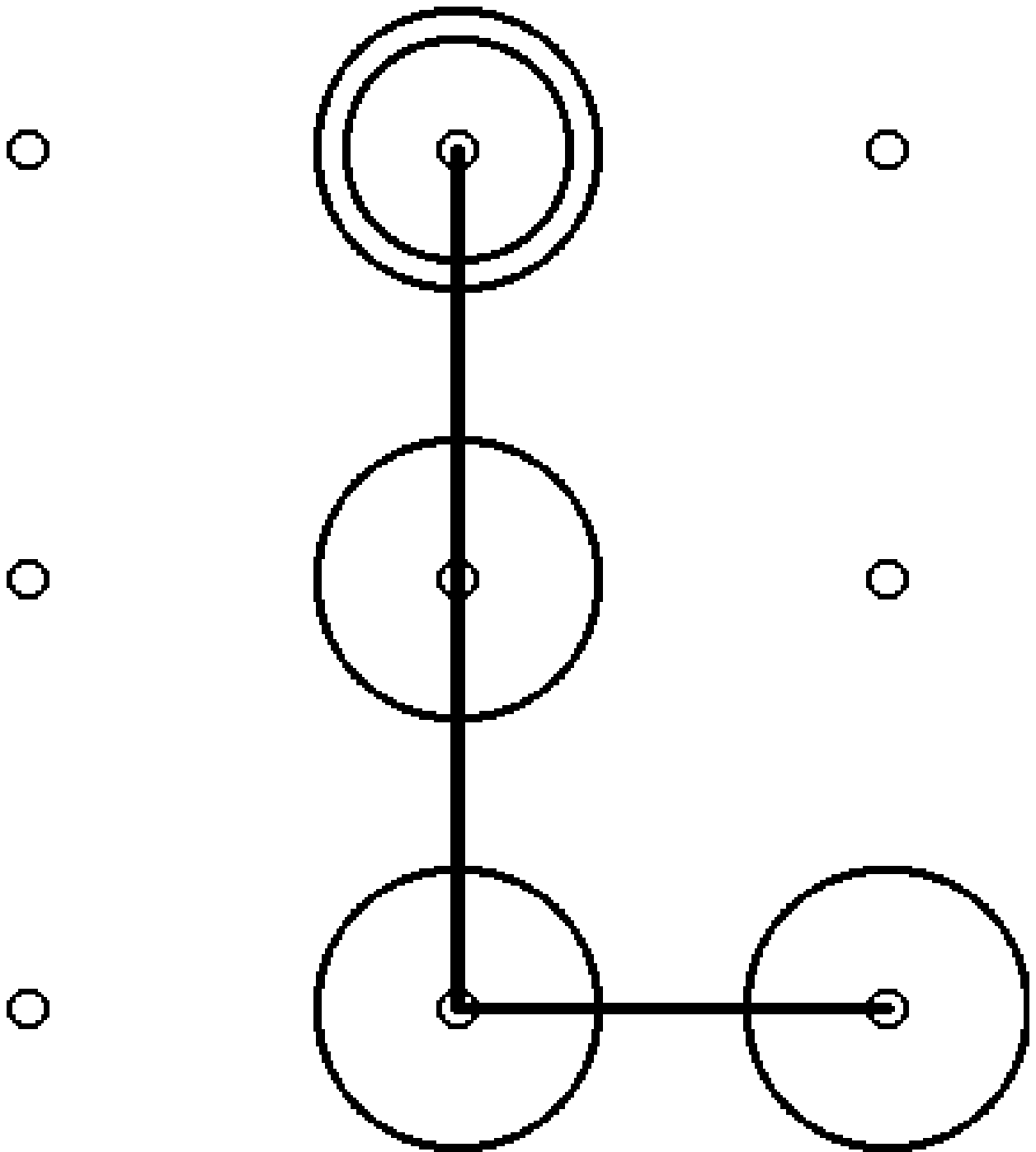} }  & \fbox{ \includegraphics[width=0.05\textwidth]{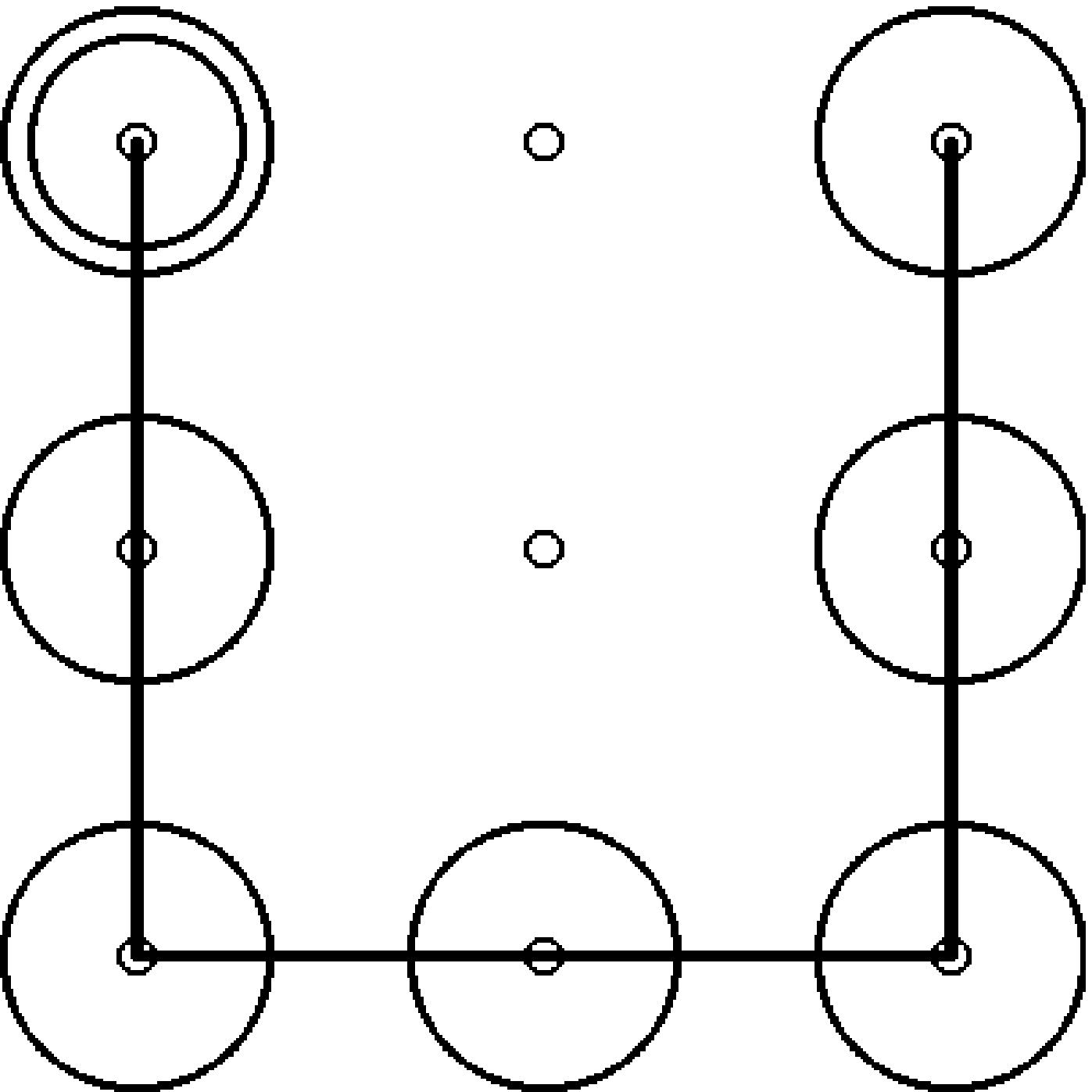} }  & \fbox{ \includegraphics[width=0.05\textwidth]{images/patterns/0-1-2-4-6-7-8.eps} }  & \fbox{ \includegraphics[width=0.05\textwidth]{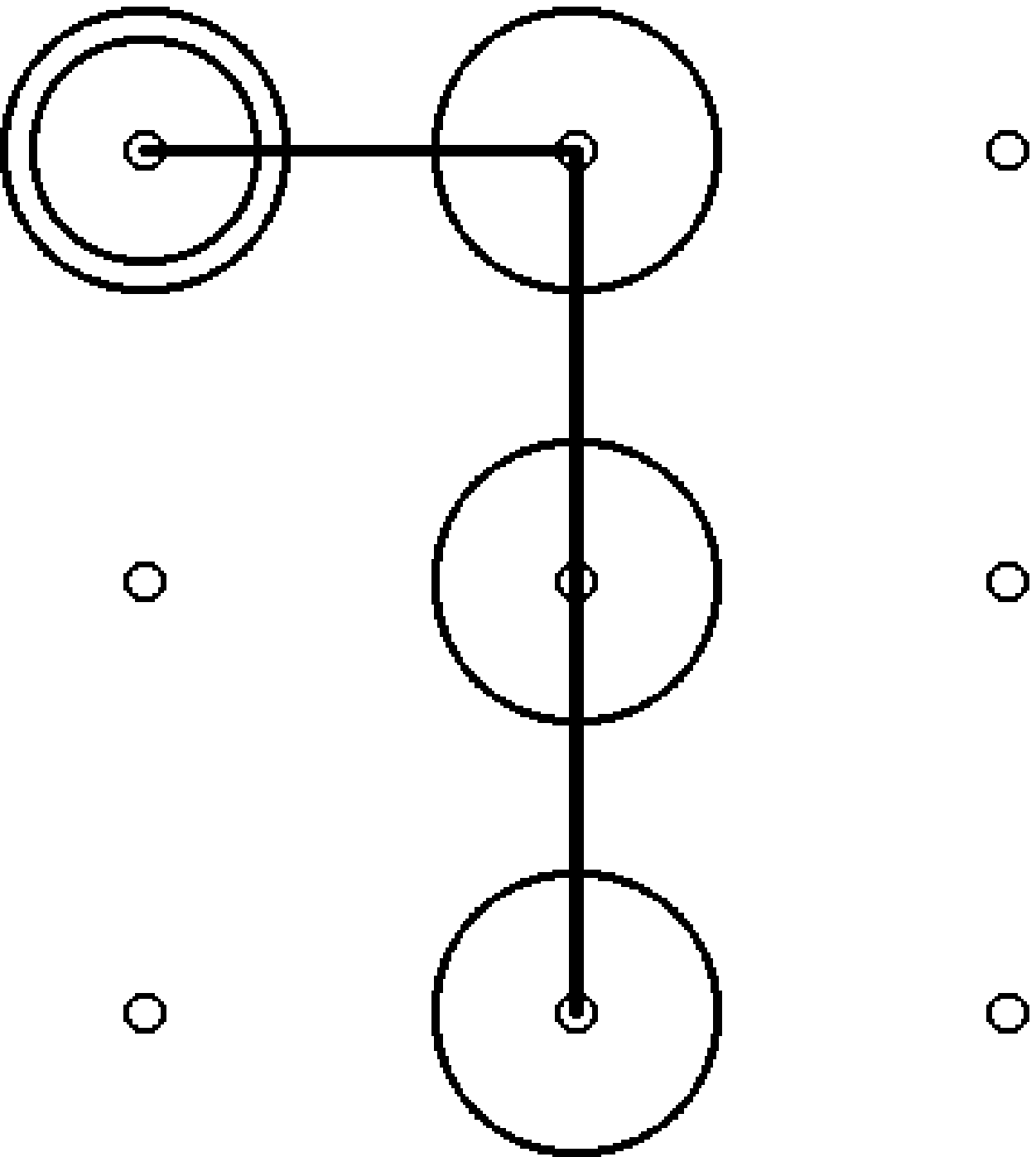} }  & \fbox{ \includegraphics[width=0.05\textwidth]{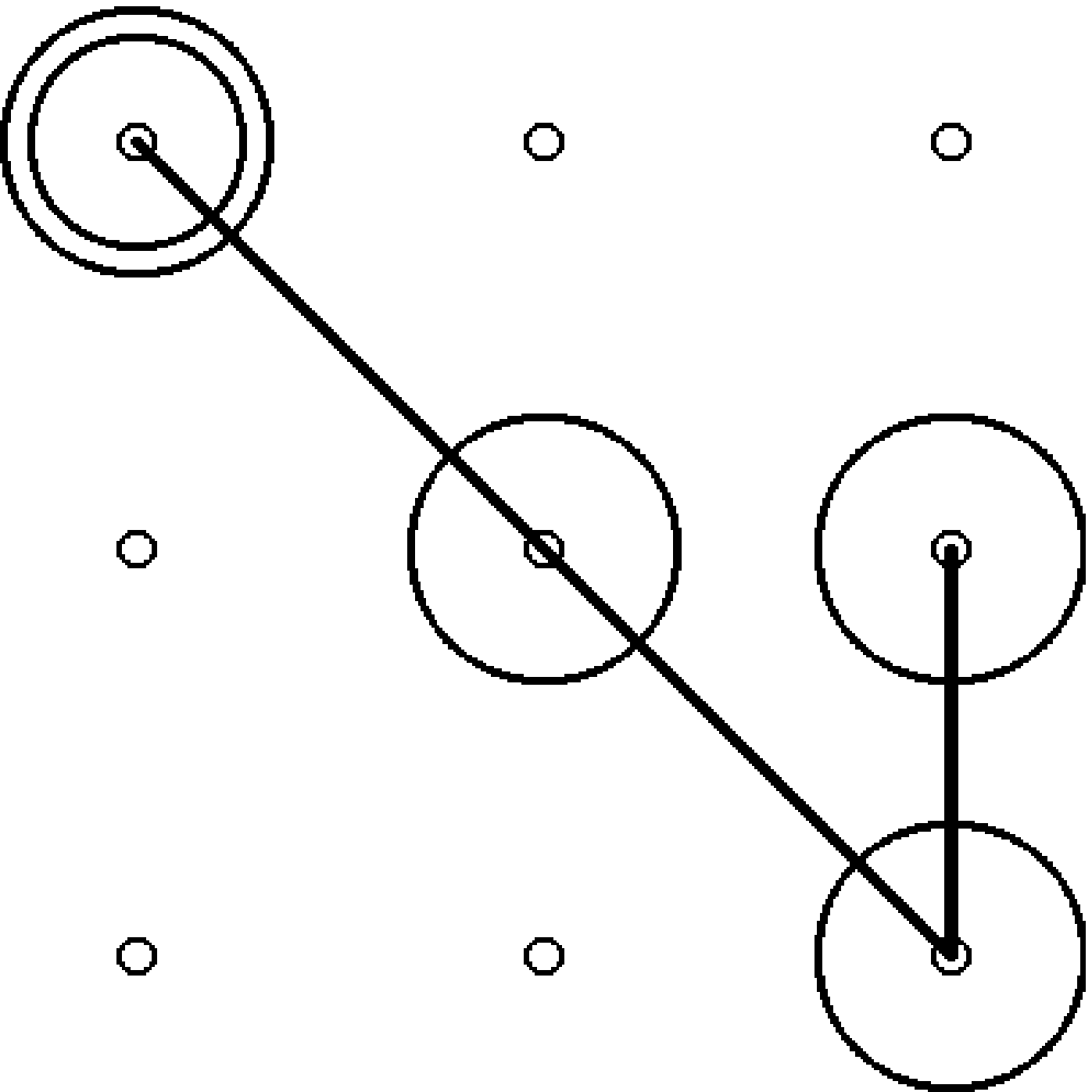} }  \\
\textbf{ freq=26 } & \textbf{ freq=24 } & \textbf{ freq=21 } & \textbf{ freq=19 } & \textbf{ freq=18 } & \textbf{ freq=15 } & \textbf{ freq=14 } & \textbf{ freq=10 } & \textbf{ freq=9 } & \textbf{ freq=9 } \\
\hline
\multicolumn{10}{c}{ \textbf{LDR-unlock} } \\
\fbox{ \includegraphics[width=0.05\textwidth]{images/patterns/0-3-6-7-8.eps} }  & \fbox{ \includegraphics[width=0.05\textwidth]{images/patterns/0-3-6-7.eps} }  & \fbox{ \includegraphics[width=0.05\textwidth]{images/patterns/0-1-2-5-8.eps} }  & \fbox{ \includegraphics[width=0.05\textwidth]{images/patterns/0-1-2-4-6-7-8.eps} }  & \fbox{ \includegraphics[width=0.05\textwidth]{images/patterns/0-1-2-5.eps} }  & \fbox{ \includegraphics[width=0.05\textwidth]{images/patterns/0-4-8-5.eps} }  & \fbox{ \includegraphics[width=0.05\textwidth]{images/patterns/0-1-4-7.eps} }  & \fbox{ \includegraphics[width=0.05\textwidth]{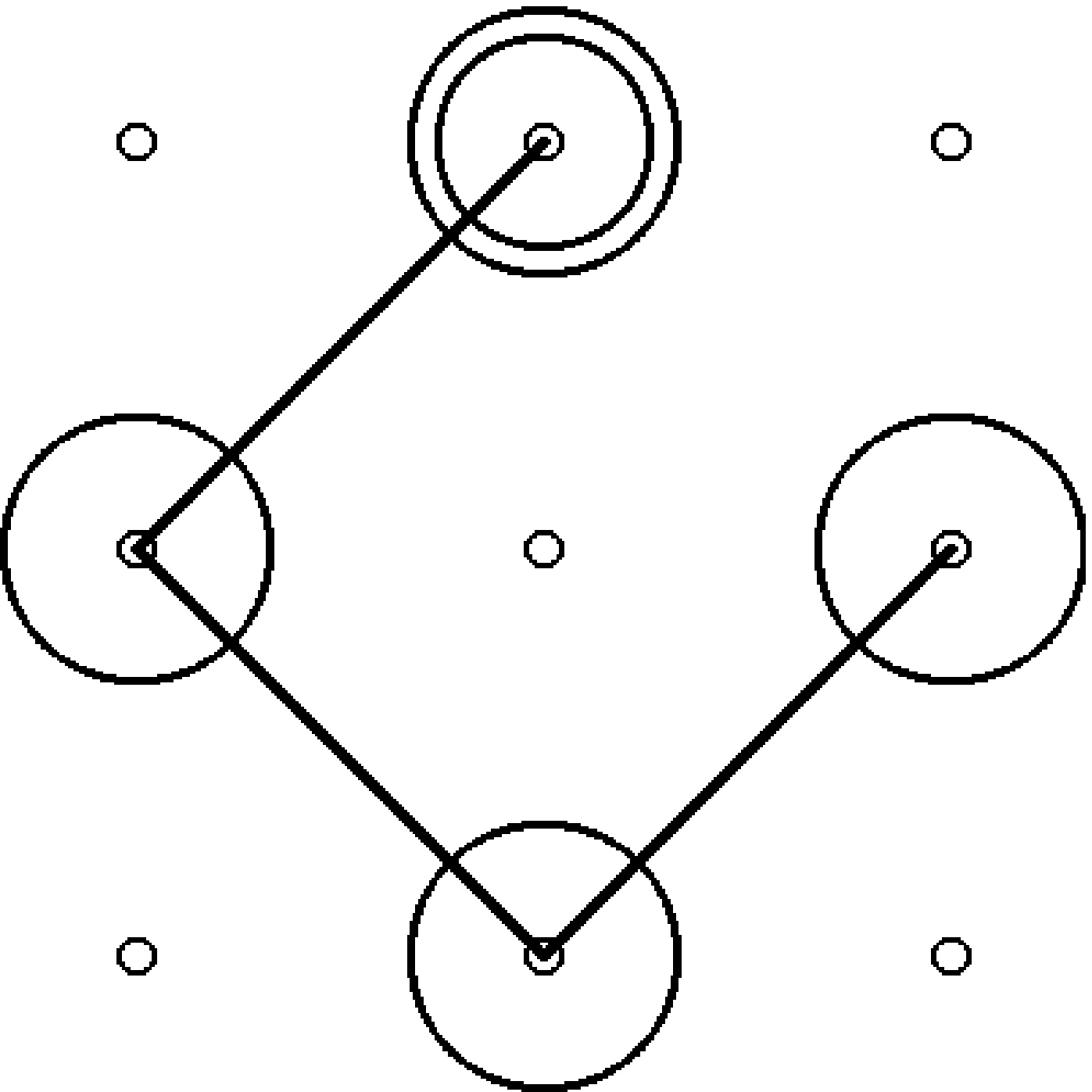} }  & \fbox{ \includegraphics[width=0.05\textwidth]{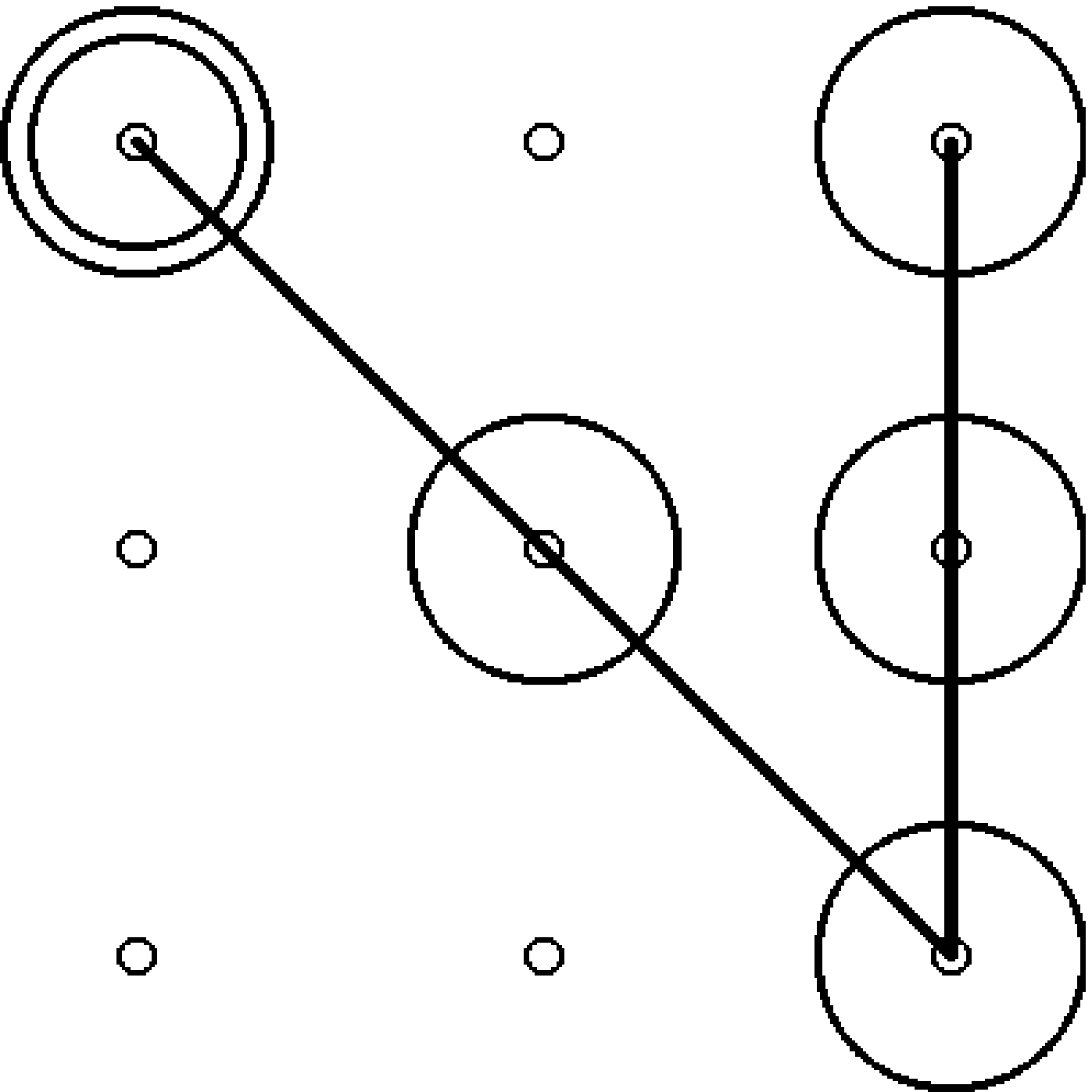} }  & \fbox{ \includegraphics[width=0.05\textwidth]{images/patterns/6-3-0-4-2-5-8.eps} }  \\
\textbf{ freq=27 } & \textbf{ freq=23 } & \textbf{ freq=22 } & \textbf{ freq=16 } & \textbf{ freq=15 } & \textbf{ freq=12 } & \textbf{ freq=8 } & \textbf{ freq=8 } & \textbf{ freq=8 } & \textbf{ freq=8 } \\
\hline
\multicolumn{10}{c}{ \textbf{UDWH-offensive} } \\
\fbox{ \includegraphics[width=0.05\textwidth]{images/patterns/0-1-2-4-6-7-8.eps} }  & \fbox{ \includegraphics[width=0.05\textwidth]{images/patterns/0-1-2-5-8.eps} }  & \fbox{ \includegraphics[width=0.05\textwidth]{images/patterns/6-3-0-4-8-5-2.eps} }  & \fbox{ \includegraphics[width=0.05\textwidth]{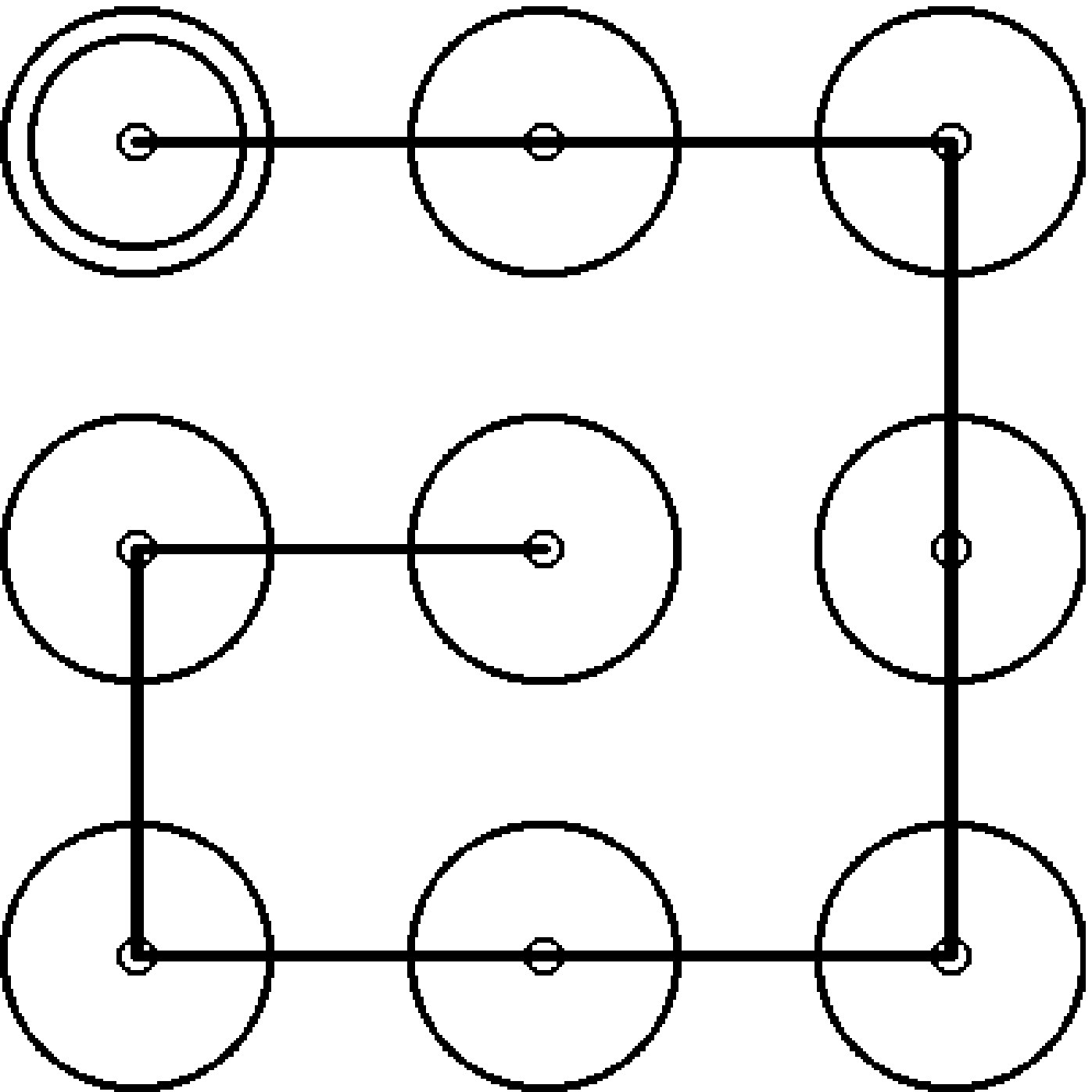} }  & \fbox{ \includegraphics[width=0.05\textwidth]{images/patterns/0-1-2-5-4-3-6-7-8.eps} }  & \fbox{ \includegraphics[width=0.05\textwidth]{images/patterns/0-3-6-7-8.eps} }  & \fbox{ \includegraphics[width=0.05\textwidth]{images/patterns/6-3-0-4-2-5-8.eps} }  & \fbox{ \includegraphics[width=0.05\textwidth]{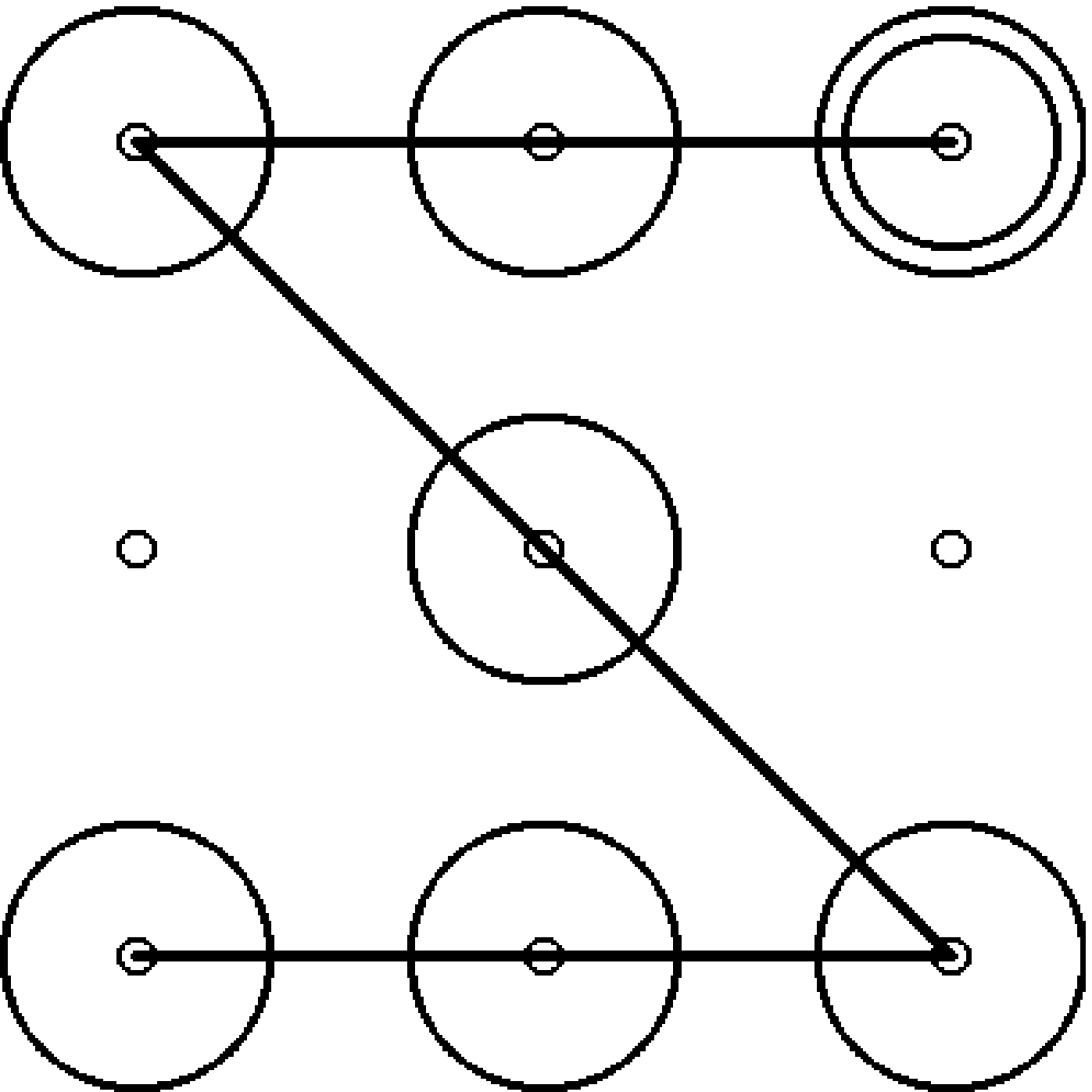} }  & \fbox{ \includegraphics[width=0.05\textwidth]{images/patterns/0-3-6-7-8-5-2.eps} }  & \fbox{ \includegraphics[width=0.05\textwidth]{images/patterns/0-3-6-4-2-5-8.eps} }  \\
\textbf{ freq=25 } & \textbf{ freq=22 } & \textbf{ freq=11 } & \textbf{ freq=9 } & \textbf{ freq=9 } & \textbf{ freq=8 } & \textbf{ freq=7 } & \textbf{ freq=6 } & \textbf{ freq=6 } & \textbf{ freq=6 } \\
\hline
\multicolumn{10}{c}{ \textbf{ZEBOdLAH} } \\
\fbox{ \includegraphics[width=0.05\textwidth]{images/patterns/0-4-8-5.eps} }  & \fbox{ \includegraphics[width=0.05\textwidth]{images/patterns/0-3-6-7.eps} }  & \fbox{ \includegraphics[width=0.05\textwidth]{images/patterns/0-1-2-5.eps} }  & \fbox{ \includegraphics[width=0.05\textwidth]{images/patterns/0-1-2-4-6-7-8.eps} }  & \fbox{ \includegraphics[width=0.05\textwidth]{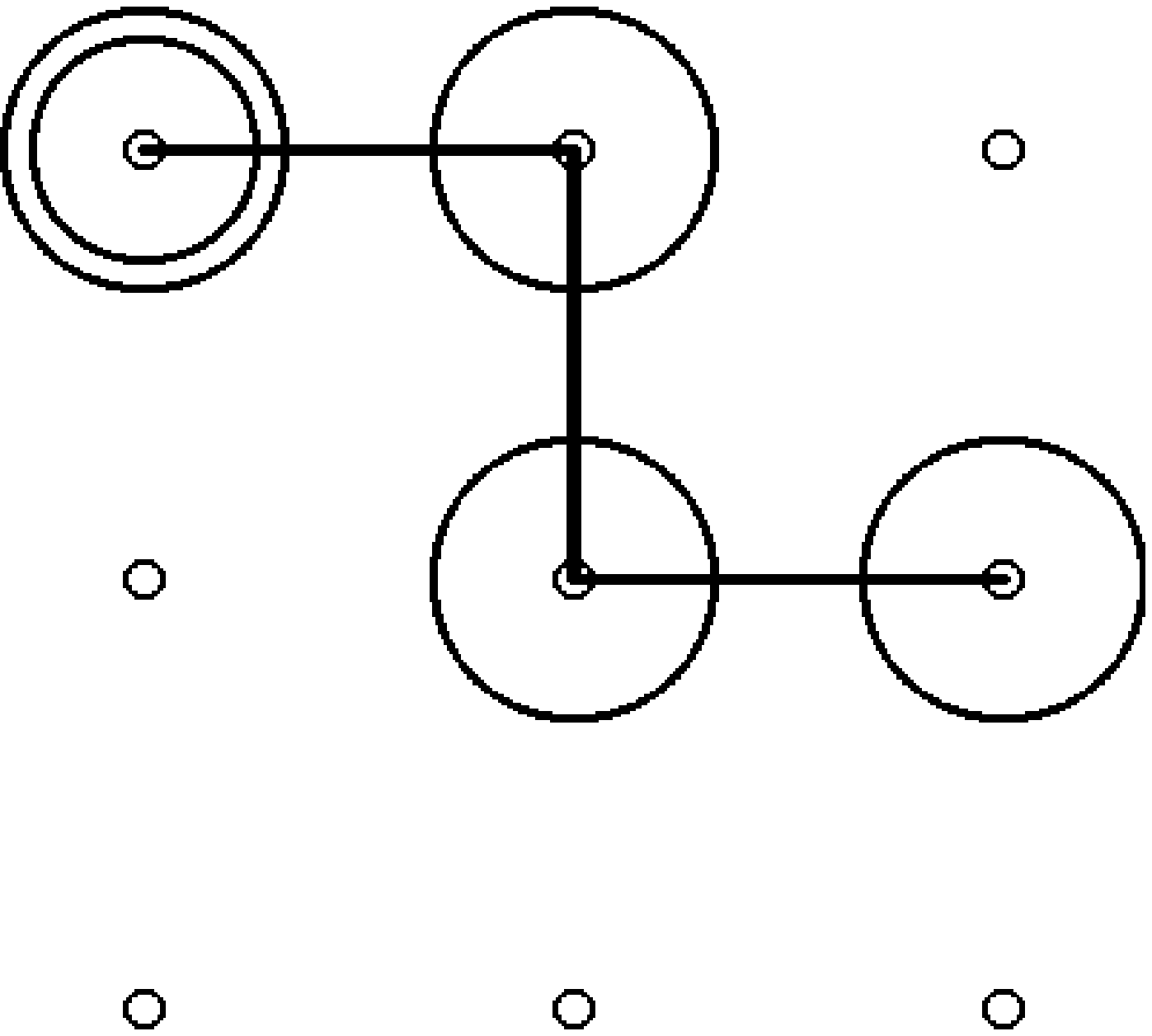} }  & \fbox{ \includegraphics[width=0.05\textwidth]{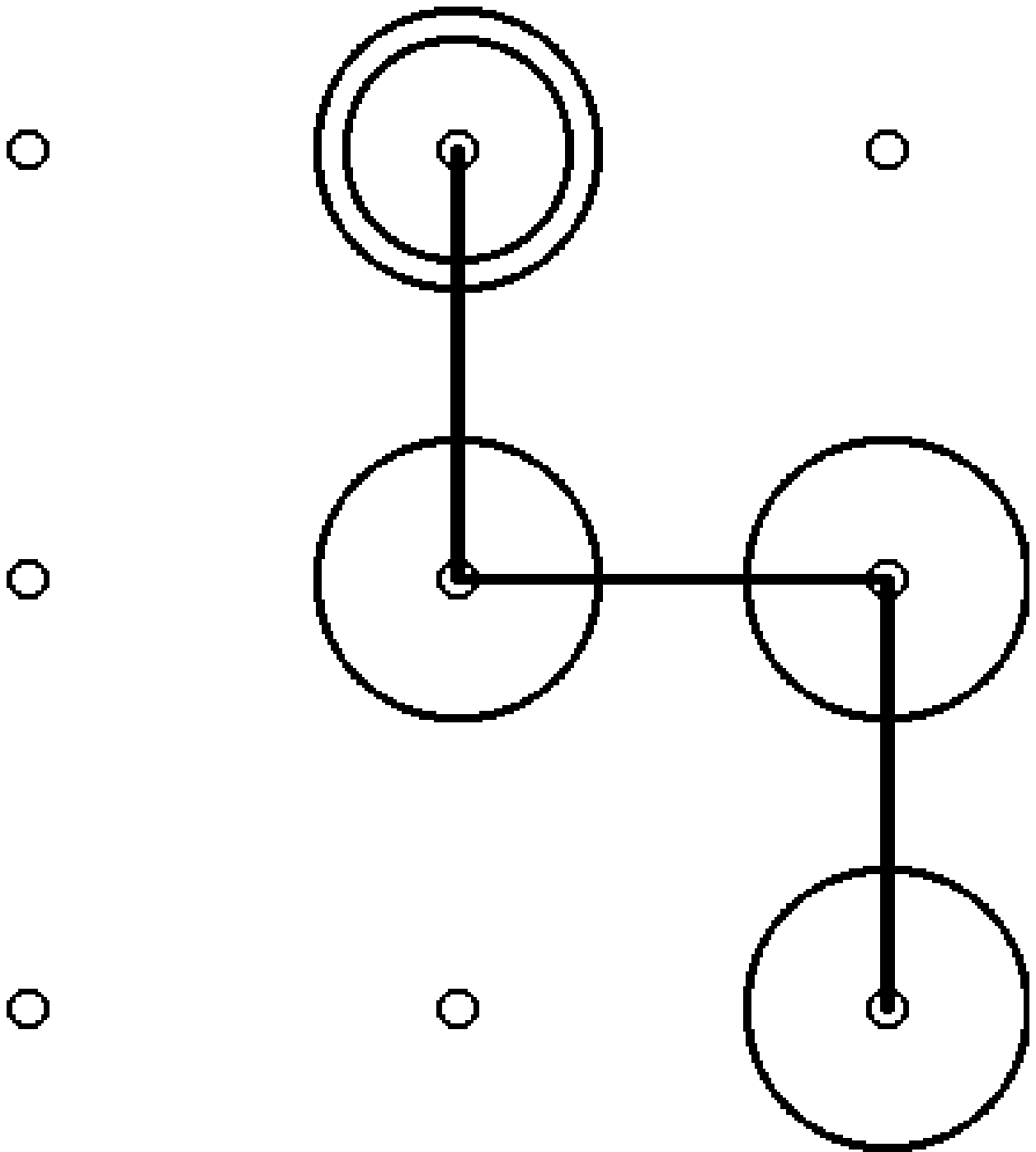} }  & \fbox{ \includegraphics[width=0.05\textwidth]{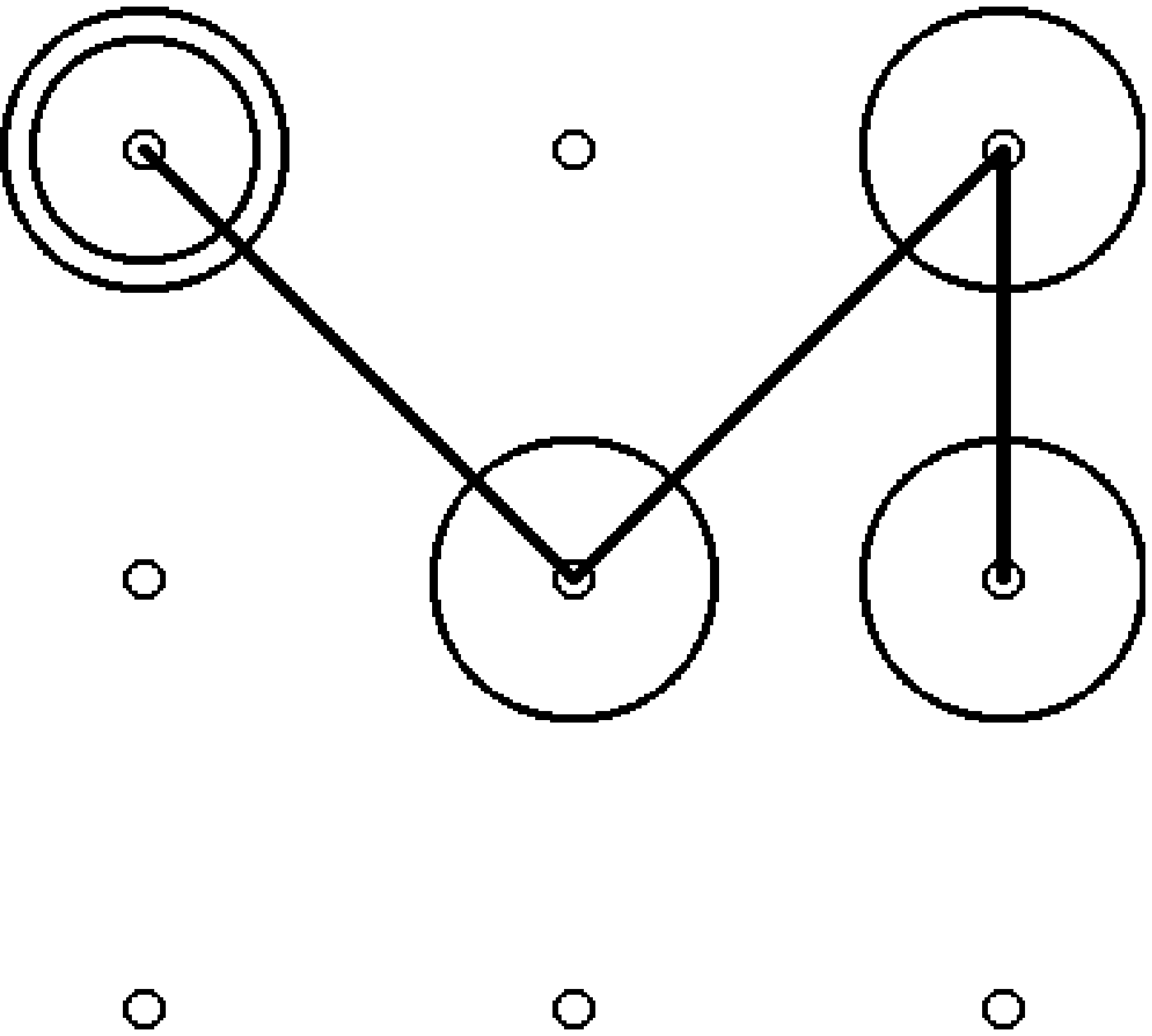} }  & \fbox{ \includegraphics[width=0.05\textwidth]{images/patterns/1-4-7-8.eps} }  & \fbox{ \includegraphics[width=0.05\textwidth]{images/patterns/0-1-2-5-8.eps} }  & \fbox{ \includegraphics[width=0.05\textwidth]{images/patterns/0-1-4-7.eps} }  \\
\textbf{ freq=12 } & \textbf{ freq=11 } & \textbf{ freq=10 } & \textbf{ freq=7 } & \textbf{ freq=7 } & \textbf{ freq=6 } & \textbf{ freq=6 } & \textbf{ freq=6 } & \textbf{ freq=5 } & \textbf{ freq=5 } \\
\hline
\multicolumn{10}{c}{ \textbf{Total} } \\
\fbox{ \includegraphics[width=0.05\textwidth]{images/patterns/0-1-2-5-8.eps} }  & \fbox{ \includegraphics[width=0.05\textwidth]{images/patterns/0-1-2-4-6-7-8.eps} }  & \fbox{ \includegraphics[width=0.05\textwidth]{images/patterns/0-3-6-7-8.eps} }  & \fbox{ \includegraphics[width=0.05\textwidth]{images/patterns/0-3-6-7.eps} }  & \fbox{ \includegraphics[width=0.05\textwidth]{images/patterns/0-1-2-5.eps} }  & \fbox{ \includegraphics[width=0.05\textwidth]{images/patterns/1-4-7-8.eps} }  & \fbox{ \includegraphics[width=0.05\textwidth]{images/patterns/0-1-2-5-4-3-6-7-8.eps} }  & \fbox{ \includegraphics[width=0.05\textwidth]{images/patterns/0-4-8-5.eps} }  & \fbox{ \includegraphics[width=0.05\textwidth]{images/patterns/2-1-0-3-4-5-8-7-6.eps} }  & \fbox{ \includegraphics[width=0.05\textwidth]{images/patterns/6-3-0-4-2-5-8.eps} }  \\
\textbf{ freq=95 } & \textbf{ freq=91 } & \textbf{ freq=90 } & \textbf{ freq=86 } & \textbf{ freq=70 } & \textbf{ freq=41 } & \textbf{ freq=40 } & \textbf{ freq=40 } & \textbf{ freq=39 } & \textbf{ freq=37 } \\
\hline
\end{tabular}}
\caption{Top 10 most frequent patterns in each of the datasets, excluding {\em defensive} patterns due to low frequency numbers for each of the patterns.}
\label{fig:freqpatterns}
\end{figure*}
}

\newcommand{\figvizfeatures}[0]{
\begin{figure*}[t]
\footnotesize
\centering
\begin{tabular}{ c | c c c }
{\bf Spatial Features} & \multicolumn{3}{c}{\textbf{Visual Properties}} \\
\midrule
\fbox{\includegraphics[width=0.12\linewidth]{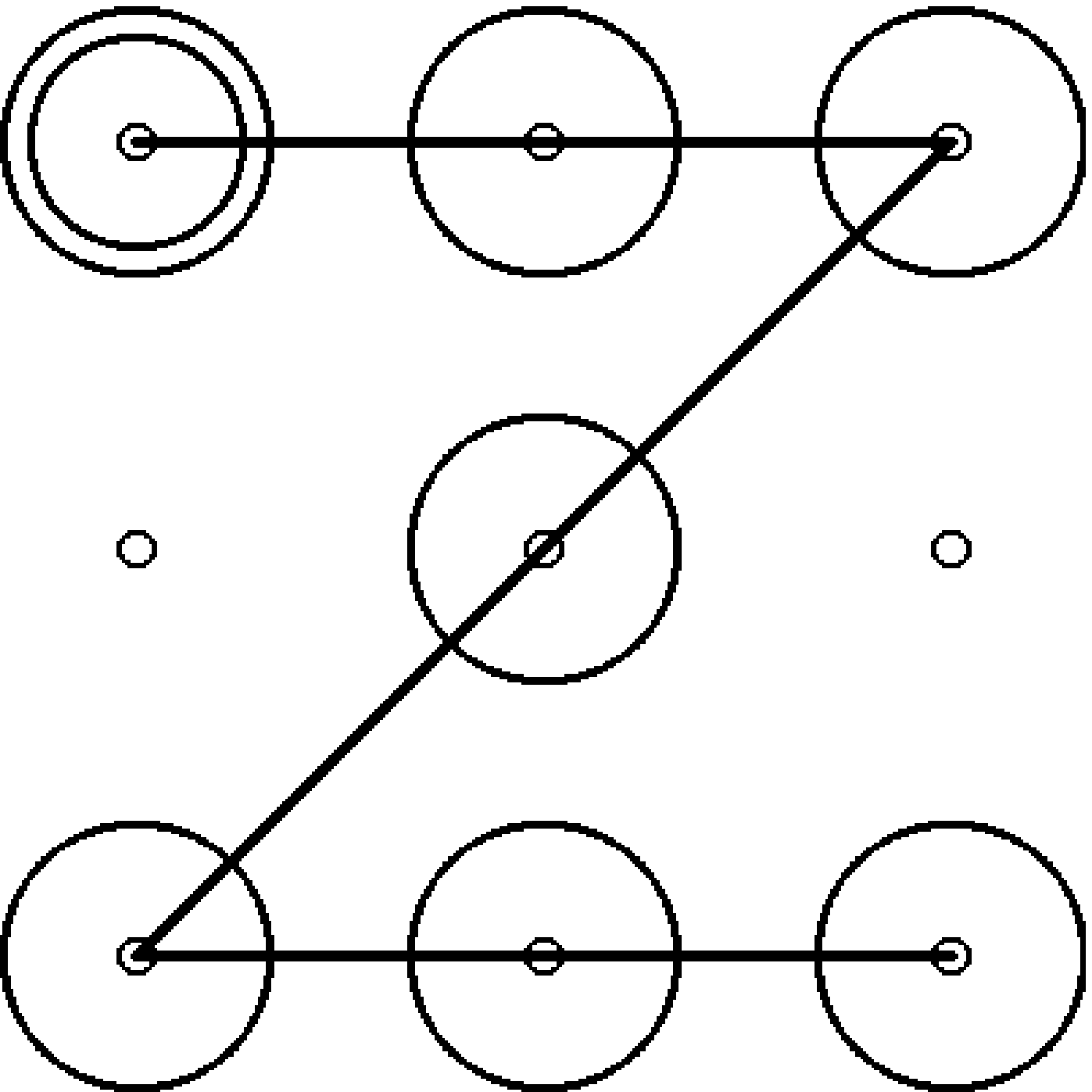}} &
\fbox{\includegraphics[width=0.12\linewidth]{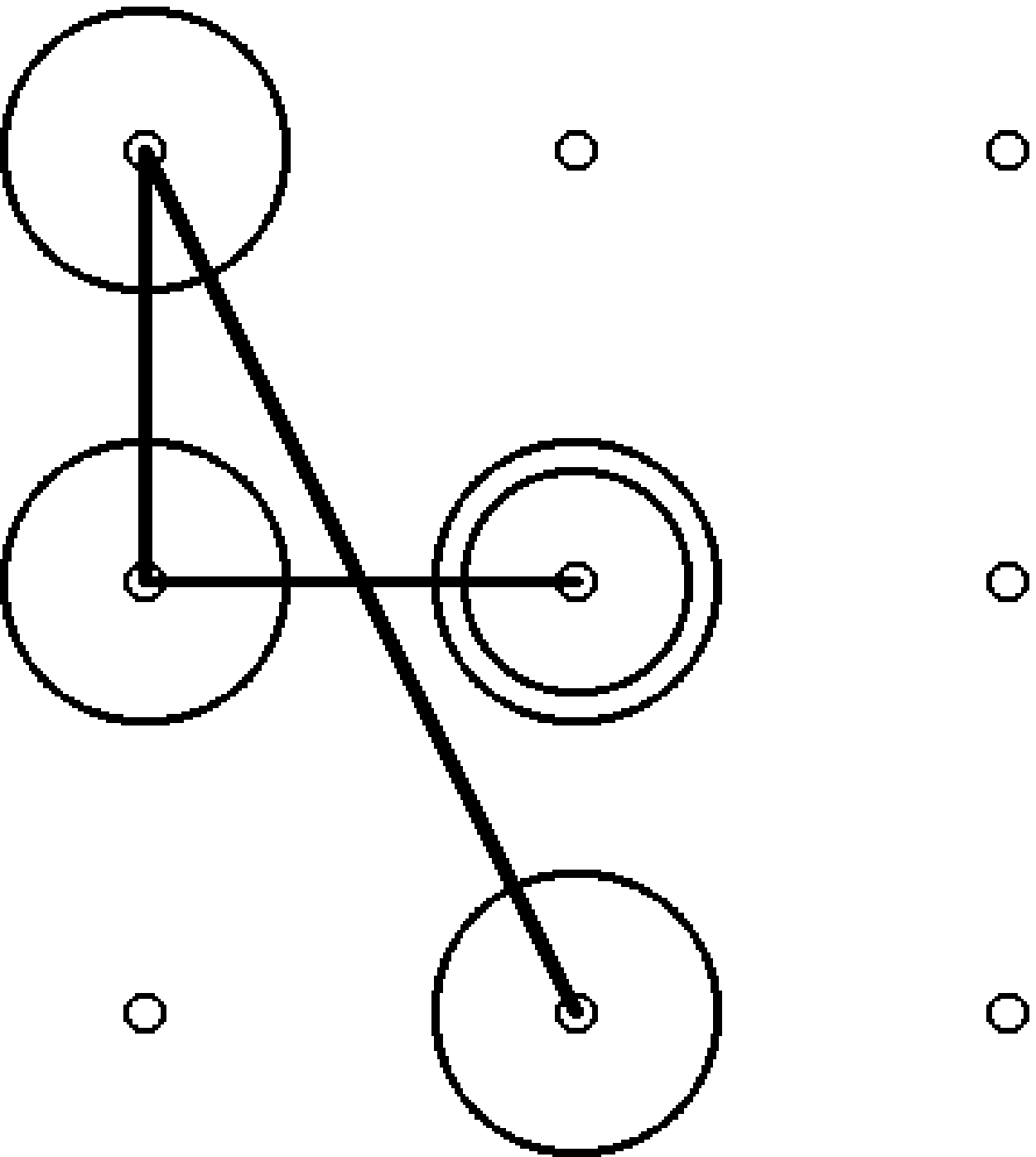}} &
\fbox{\includegraphics[width=0.12\linewidth]{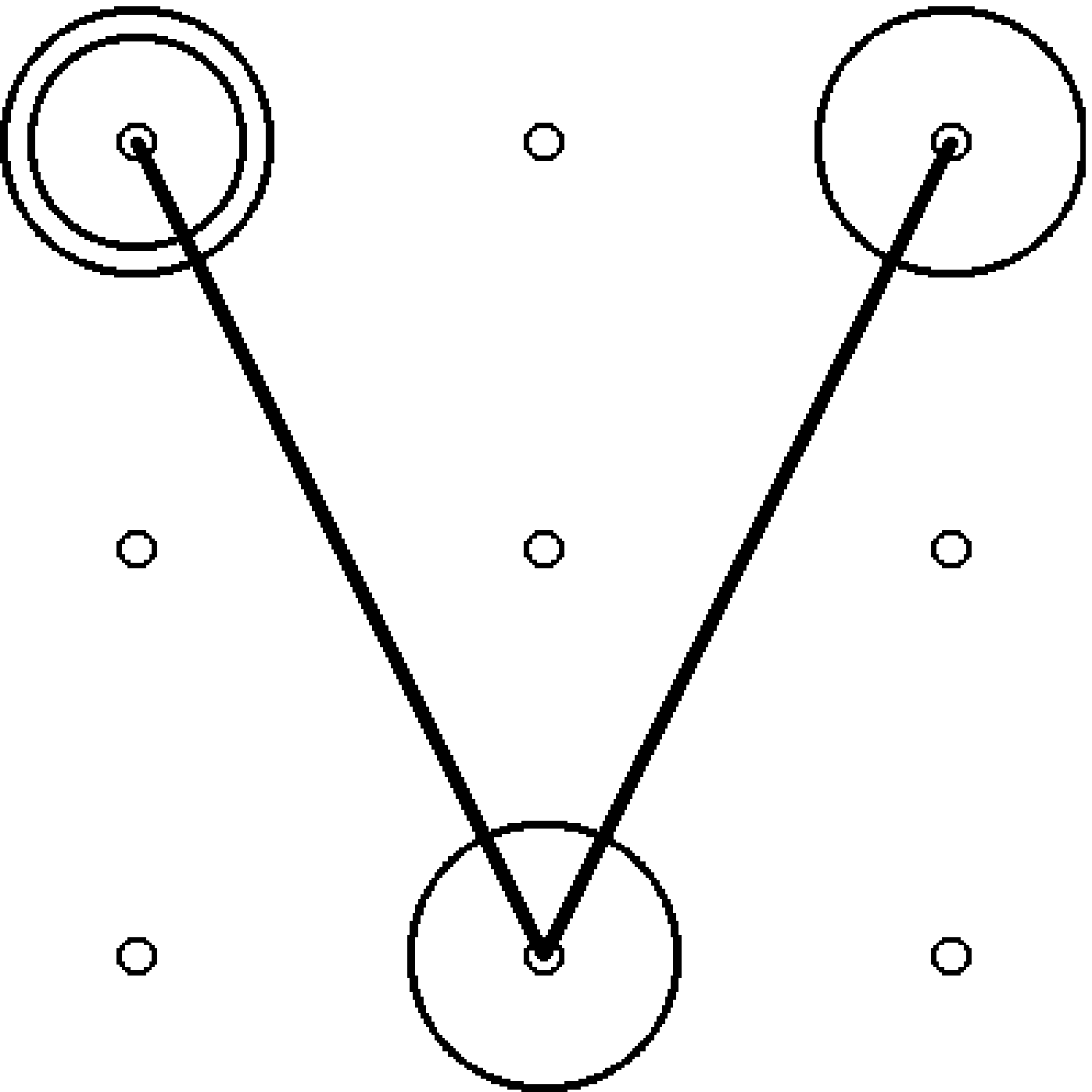}} &
\fbox{\includegraphics[width=0.12\linewidth]{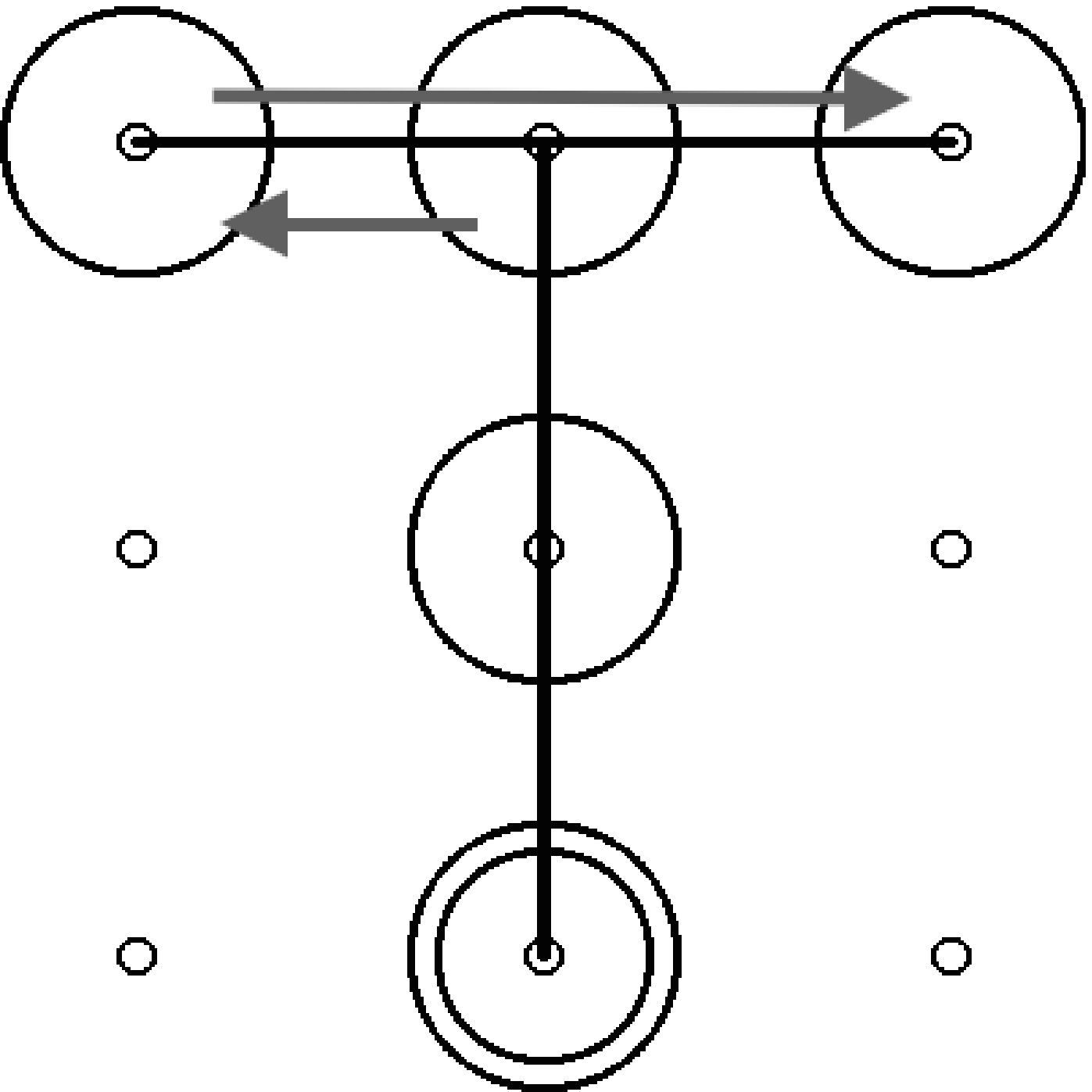}}\\
Length=7 &  Cross 
& KMove & Non-adj\\
Stroke-Length=6.82 & \\
Turns=2 &\\
\end{tabular}
\caption{Pictorial of the spatial and visual features: {\em length} is the
  number of contact points, {\em stroke length} (slength) is the length of the line segments
  with (0,0) mapped to the center point, {\em turns} is the number of direction
  changes, {\em cross} is any cross over,
  {\em kmove} is a
  connection of two contact points over 2 and across 1 (like a knight moves in
  chess), and {\em non-adj} is when a previously contacted point is traced over to
  connect two non-adjacent points.}
\label{fig:vizfeatures}
\end{figure*}}

\newcommand{\figlengths}[0]{
\begin{figure*}[t]
  \centering

  \begin{minipage}{0.48\linewidth}
    \centering
  \includegraphics[width=0.8\linewidth]{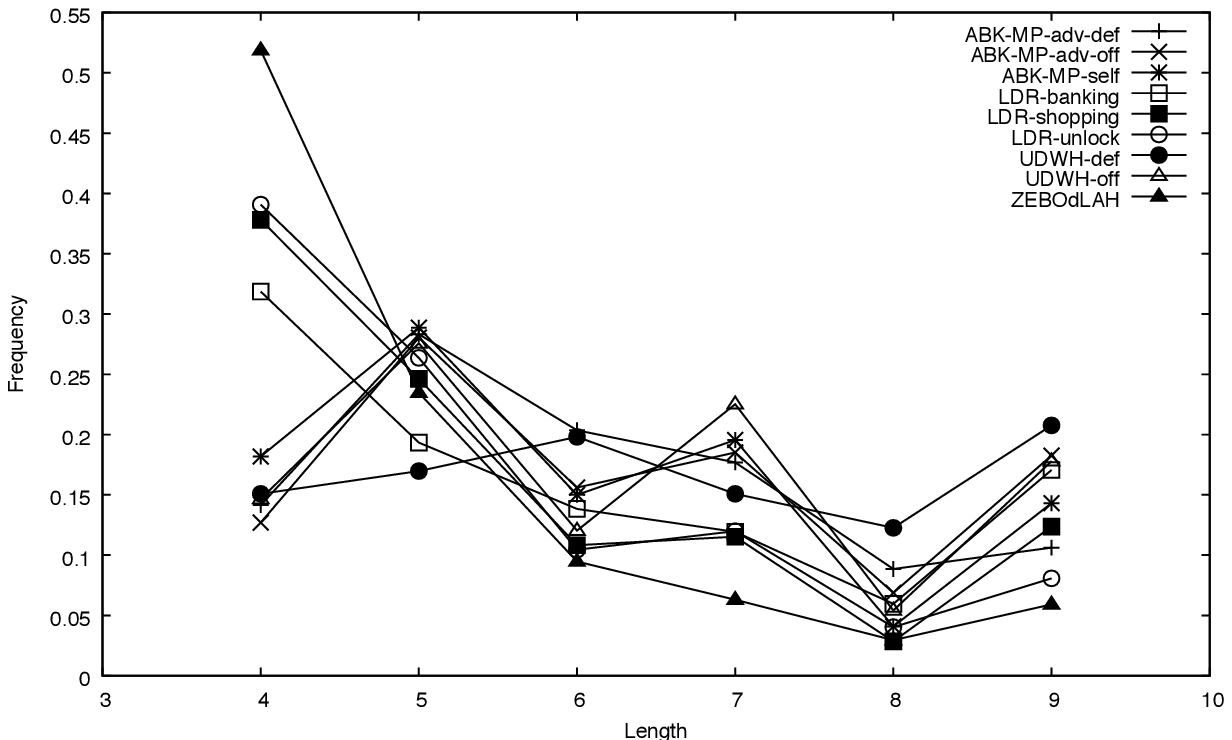}
  \caption{Histogram of lengths in each of the data sets}
  \label{fig:length}
  \end{minipage}
  \hfill
  \begin{minipage}{0.48\linewidth}
    \centering
  \includegraphics[width=0.8\linewidth]{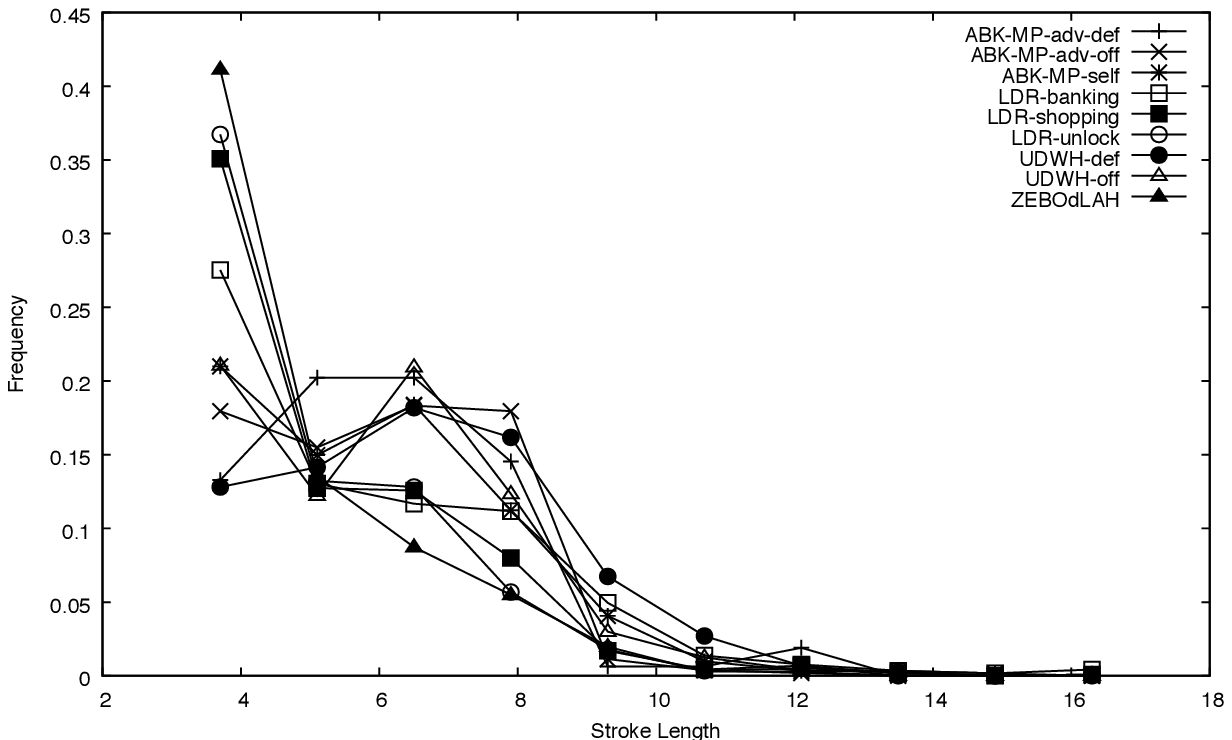}
  \caption{Histogram of {\em stroke} lengths in each of the data sets}
  \label{fig:slength}
  \end{minipage}

\end{figure*}}


\newcommand{\tableterms}[0]{%
  \begin{table}[!t]

    \centering
    \scriptsize
    
  \begin{tabular} { l  p{6cm}  }
    \toprule
        {\em Term} & {\em Description}\\
        \midrule
        Adversarial & Participants are asked to generate patterns that they believe others have used ({\em offensive}) and patterns that others cannot guess ({\em defensive})\\\\
        Self-Report & Participants are asked to honestly (with some checks) report the pattern that they use (or have used) on their mobile device\\\\
        Situational & The survey proposes scenarios underwhich participants should consider generating a pattern, e.g., for a bank vs. just to unlock the device.\\\\
        \hline\\
        Online & The survey occurred via an online form or web application without the oversight of a researcher\\\\
        App & The survey occured using a specialized mobile App, either installed on the participants device or researchers device\\\\
        Local & The survey occurred within the purview of the researcher\\\\
        \hline\\
        On-device & The survey occurred using a mobile device\\\\
        User-device & The survey occurred using the participants own mobile device\\\\
        Pen-and-Paper & The survey occurred using pen-and-paper where participants ``draw'' the patterns\\\\
        Mouse & The survey allowed participants to use a ``mouse'' rather than touch to enter patterns \\\\
        \hline\\
        Altered & The survey measured altered forms of unlock patterns, e.g., background images, expanded contact points, less contact points \\\\

        \bottomrule
  \end{tabular}
  \caption{Overview of Collection Method Techniques and their Descriptions}
  \label{tab:terms}
  \end{table}%

}

\newcommand{\tabletermsanalyze}[0]{%
  \begin{table}[!tp]
    
    \scriptsize
    \centering
    
    
  \begin{tabular} { l  p{5cm}  }
    \toprule
        {\em Term} & {\em Description}\\
        \midrule
        
        Visual & Patterns were analyzed based on visual or structural
        properties, such as the length, crosses, or other aspects that
        affect its visual construction \\\\

        Meter & Patterns were anlayzed based on a password-like meter construction developed by the authors \\\\

        Physical & Patterns were analyzed in the context of physical attacks, such as smudge attacks~\cite{aviv2010smudge} or shoulder surfing~\cite{schaub2012surfing,von2015easy,davin2017baseline,aviv2017towards}\\\\

    Occurrence Rates & Patterns were analyzed for how frequently they appear in the data, or how frequently subests of patterns (e.g., 3-grams) appear \\\\

        Guessability & Patterns were analyzed based on how guessable
        they are to an adversary trying to crack the password,
        typically reported in terms of the guessing
        entropy~\cite{kelley2012guess,bonneau2012science}\\\\

        Recall & Patterns were measured based on memorability and recall rates of the participants\\\\

        Demographic & Pattern properties were measured (e.g., using
        one of the above methods) and compared along demographic
        factors (e.g., handedness, male/female, location).\\\\

        \bottomrule
  \end{tabular}
  \caption{Overview of Analysis Method Techniques and their Descriptions}
  \label{tab:terms-analyze}
  \end{table}%
}

\newcommand{\tablecollected}[0]{%
\begin{sidewaystable*}
  \centering
  \scriptsize
\resizebox{0.9\linewidth}{!}{%

  \begin{tabular}{l l l | p{4cm} | p{4cm} |  c c | p{6cm}}
    \toprule
    {\bf Name}       & {\bf Reference}                & {\bf Year} & {\bf Collection Method}  & {\bf Analysis Method}  & {\bf \# Participants} & {\bf \# Patterns} & {\bf Protocol Description} \\
    \midrule
     ATOY & \cite{andriotis2013pilot} & 2013 & Local, Online, Mouse, Situational& Visual, Physical, Demographic & 144 & 144 & Recruited participants for a lab based study, asking them to generate one strong and one weak pattern using a web-based survey using a mouse. Performed visual and physical attack analysis on collected data.\\\midrule
     UDWH${}^\star$       & \cite{uellenbeck2013quantifying}        & 2013 & Adversarial, Local, On-device & Visual, Guessability, Recall & 113             & 106 / 573   & Recruited participants in university cafeteria to interact in an adversarial protocol of offensive and defensive patterns. Analyzed data based on guessability, visual features, common sub sequences, and recall rate. (\#patterns reported deffensive/offsenive)\\\midrule
    SWZ  & \cite{sun2014dissecting}       & 2014 & On-Device, Local & Visual, Meter  & 81  & 27 / 54 & Recruited participants at university to generate patterns with and without a password meter (\#patterns reported with-meter/without-meter)\\\midrule
    SCOKH      & \cite{song2015meteres}    & 2015 & Self-Reported, User-device, App & Visual, Meter, Physical  &101 & 49 / 52 & Recruited participants by distributing an application in the Android App Store called EnCloud that allowed participants to store data in the cloud protected with a pattern. Some participants used a password meter and other did not. Analysis used visual features, physical attacks, and guessability. (\#patterns reported with-meter/without-meter)\\\midrule
    ABK-MP-self${}^\star$   & \cite{aviv2015bigger} & 2015 & Self-reported, Online, User-device & Visual, Guessability, Demographics  & 757          & 443         & Recruitment via MTurk to either report their pattern or statistics about the pattern. Attention tests were used to checks for honesty, and participants must use Android Devices to complete survey.\\\midrule
    ABK-MP-adv${}^\star$        & \cite{aviv2015bigger} & 2015 & Adversarial, Local, Pen-and-paper & Visual, Guessability, Recall, Demographics  & 38          & 114 / 380   & Recruitment at university and group participants into small focus groups of size (8-to-20) to conduct adversarial protocol. (\#patterns reported defensive/offensive)\\\midrule
    SGSMA & \cite{siadati2015fortifying} & 2015 & Online, User-device, Situational & Visual, Guessability & 92 & 92 & Recruited via MTurk and asked participants to imagine receive new phone and generate a pattern for that phone (with some participants received  nudges to select strong patterns, see SGSMA-BLINK and SGSMA-EPSM)\\\midrule
    LDR${}^\star$   & \cite{loge2016user}      & 2016 & Online, User-device, Situational  & Visual, Guessability, Demographics & 802          & 841 / 842 / 838         & Recruited using snowball methods and asked participants to generate patterns under three application scenarios (\#patterns report for scenarios banking/shopping/unlock)\\    \midrule    

    ZEBOdLAH${}^\star$ & \cite{zezchwitz2016onquant} & 2016 & Online, User-device & Visual              & 506          & 506    & Recruitment via MTurk to generate patterns on participants own device (with some participants using background images or other nudges, see ZEBSOdLAH-bg) \\\\
    \midrule    
    CHCOSK & \cite{cho2017syspal} & 2017 & Online, User-Device & Recall, Guessability, Visual & 384 & 384 & Recruitment via MTurk for individuals who owned an Android device.  \\
    \midrule    
    TBLV & \cite{tupsamudre2017PassO} & 2017 & Online, Adversarial (variant), Mouse & Visual, Guessability & 11,960 & 34,548/35,249 & Recruitment within organization, asking them for 3 distinct patterns that are easy for them to remember but difficult for others to guess, and 3 distinct patterns which they think are being used by other participants.  Participants were able to choose between the standard $3\times 3$ arrangement and a circular arrangement (see TBLV-passO)\\

    \bottomrule
  \end{tabular}
}
  \caption{Overview of publications with collected datasets for the classical Android Unlock
    Pattern. The ${}^\star$ indicates datasets available to use for comparative analysis in Section~\ref{sec:analysis}.}
\label{tab:collected-data}
\end{sidewaystable*}
}

\newcommand{\tablealtered}[0]{%
  \begin{sidewaystable*}
    \centering
    \scriptsize
    \begin{tabular}{l l l | p{1.5cm} | p{1.5cm} |  c c | p{6cm}}
    \toprule
    {\bf Name}      & {\bf Reference}                & {\bf Year} & {\bf Collection Method}  & {\bf Analysis Method}   & {\bf \# Participants} & {\bf \# Patterns} & {\bf Protocol Description} \\
    \midrule
    UDWH-los  & \cite{uellenbeck2013quantifying}        & 2013 & Adversarial, Local, On-device, Altered & Guessability, Recall & 105             & 92 / 460          & Recruitment/protocol the same as UDWH but with altered contact point layout dubbed ``Left-out small'' with one point removed. (\#patterns reported defensive/offensive)\\\midrule
    UDWH-lol  & \cite{uellenbeck2013quantifying}        & 2013 & Adversarial, Local, On-device, Altered & Guessability, Recall & 99              & 87 / 439          &  Recruitment/protocol the same as UDWH but with altered contact point layout dubbed ``Left-out large'' with one point removed. (\#patterns reported defensive/offensive) \\\midrule
    UDWH-circ & \cite{uellenbeck2013quantifying}        & 2013 & Adversarial, Local, On-device, Altered & Guessability, Recall & 82${}^{\natural}$             & 86 / 429          & Recruitment/protocol the same as UDWH but with altered contact point layout in a ``circular'' arrangement. (\#patterns reported defensive/offensive) \\\midrule
    UDWH-rand & \cite{uellenbeck2013quantifying}        & 2013 & Adversarial, Local, On-device, Altered & Guessability, Recall & 80              & 80 / 399          & Recruitment/protocol the same as UDWH but with altered contact point layout in a ``random placement'' arrangement. (\#patterns reported defensive/offensive)\\\midrule
    ABK-MP-adv-4x4        & \cite{aviv2015bigger} & 2015 & Adversarial, Local, Pen-and-paper, Altered & Visual, Guessability, Recall & 40 & 119 / 382 & Protocol same as ABK-MP-adv except the grid-szize was expanded to 4x4. (\#patterns reported defensive/offensive)  \\\midrule
    SGSMA-EPSM & \cite{siadati2015fortifying} & 2015 & Online, User-Device, Situational, Altered & Visual Guessability & 72 & 72 & Protocol same as SGSMA except pattern entry was altered to include continuous meter feedback based on~\cite{andriotis2014complexity}.\\\midrule
    SGSMA-BLINK & \cite{siadati2015fortifying} & 2015 & Online, User-Device, Situational, Altered & Visual Guessability & 106 & 105 & Protocol same as SGSMA pattern entry was altered to include blinking contact points to suggest starting points.\\\midrule
    ZEBOdLAH-bg & \cite{zezchwitz2016onquant} & 2016 & Online, User-device, Mouse, Altered & Visual              & 496          & 496 / 496    & Recruitment via email lists and divided participants into two alterations one with static background image (girl's face) and one with dynamic background image (moving air bubbles (\#patterns reported for static/dynamic) \\
\midrule    
    CHCOSK-syspal & \cite{cho2017syspal} & 2017 & Online, User-Device & Recall, Guessability, Visual & 331/342/326/334 & 331/342/326/334 & Same as CHCOSK except that 1-point/2-point/3-point/Random-points were pre-selected that the participant must select patterns using those contact points\\
\midrule    
    TBLV-passO & \cite{tupsamudre2017PassO} & 2017 & Online, Adversarial (variant), Mouse, Altered & Visual, Guessability & 9,093 & 26,469/26,914 & Protocol same as TBLV.  Participants were able to choose between the standard $3\times 3$ arrangement and a circular arrangement.\\
    \bottomrule
  \end{tabular}
  \caption{Overview of collected datasets for variants of the  classical patterns ({\small ${}^{\natural}$ there apears to be a small inconsistency in the number of reported participants for this study})}
  \label{tab:collected-data-alternative}
\end{sidewaystable*}
}

\newcommand{\tablespatprop}{
\begin{table*}[t]
\centering
\small
\resizebox{\linewidth}{!}{
\begin{tabular}{r | c l | c l | c l || c l | c l | c l}
{\bf dataset} & {\bf length}  &   & {\bf slength} & & {\bf turns} & & {\bf crosses} & & {\bf kmoves} & & {\bf non-adj} & \\
\hline
UDWH-off       & 6.30 (1.68)  & a  & 5.88 (1.97)  & a & 2.64 (1.35)  & a & 0.08 (0.39)  & bc & 0.06 (0.30)  & a & 0.09 (0.32)  & bc\\
UDWH-def       & 6.55 (1.74)  & a  & 6.45 (2.08)  & a & 3.31 (1.40)  & b & 0.21 (0.58)  & a & 0.15 (0.43)  & b & 0.16 (0.42)  & ac\\
ABK-MP-adv-off & 6.34 (1.67)  & a  & 5.90 (1.75)  & a & 2.62 (1.15)  & a & 0.02 (0.18)  & c & 0.03 (0.23)  & a & 0.11 (0.32)  & abc\\
ABK-MP-adv-def & 6.11 (1.53)  & ab & 5.99 (1.89)  & ab & 2.94 (1.30)  & ab & 0.13 (0.59)  & ab & 0.09 (0.43)  & ab & 0.14 (0.42)  & abc\\
ABK-MP-self    & 6.05 (1.63)  & ab & 5.82 (2.08)  & ab & 2.65 (1.30)  & a & 0.14 (0.75)  & ab & 0.08 (0.44)  & ab & 0.16 (0.43)  & a\\
LDR-banking    & 5.92 (1.83)  & b  & 5.67 (2.44)  & b & 2.78 (1.58)  & a & 0.21 (1.08)  & a & 0.14 (0.66)  & b & 0.12 (0.43)  & abc\\
LDR-shopping   & 5.54 (1.70)  & c  & 5.05 (2.02)  & c & 2.28 (1.31)  & c & 0.08 (0.53)  & bc & 0.08 (0.41)  & ab & 0.07 (0.33)  & b\\
LDR-unlock     & 5.40 (1.57)  & c  & 4.92 (1.93)  & cd & 2.20 (1.23)  & c & 0.07 (0.48)  & bc & 0.06 (0.36)  & a & 0.07 (0.30)  & b\\
ZEBOdLAH       & 5.03 (1.44)  & d  & 4.66 (1.77)  & d & 2.22 (1.09)  & c & 0.08 (0.30)  & ab & 0.07 (0.32)  & ab & 0.08 (0.28)  & bc\\
\hline
 & \multicolumn{2}{c|}{$\chi^2=347.88,p=0.00$} & \multicolumn{2}{c|}{$\chi^2=320.02,p=0.00$} & \multicolumn{2}{c||}{$\chi^2=189.40,p=0.00$} & \multicolumn{2}{c|}{$\chi^2=69.11,p=0.00$} & \multicolumn{2}{c|}{$\chi^2=39.67,p=0.00$} & \multicolumn{2}{c}{$\chi^2=50.91,p=0.00$}\\
 & \multicolumn{2}{c|}{$\epsilon^2=0.075$} & \multicolumn{2}{c|}{$\epsilon^2=0.069$} & \multicolumn{2}{c|}{$\epsilon^2=0.041$} & \multicolumn{2}{c|}{$\epsilon^2=0.015$} & \multicolumn{2}{c|}{$\epsilon^2=0.009$} & \multicolumn{2}{c}{$\epsilon^2=0.011$}\\
\end{tabular}
}
\caption{Spatial features, displaying mean and std, inset. Significance testing conducted using Kruskal-Wallis rank sum test with post-hoc grouping using pairwise Mann-Whitney U-tests with a Bonferroni correction. Datasets with the same letter are {\em not} significantly different from each over. Effect size is calculated using the $\epsilon^2$ method, small effects are in the range $[0.01,0.08)$.}
\label{tab:vizfeatures}
\end{table*}
}

\newcommand{\tablestartend}{
\begin{table*}[t]
\centering
\scriptsize
\begin{tabular}{r | c l | c l || c l | c l}
{\bf dataset} & {\bf start-x} & & {\bf start-y} & & {\bf end-x} & & {\bf end-y} & \\
\hline
ABK-MP-adv-def & -0.34 (0.76)  & ab & 0.41 (0.81)  & ab & 0.15 (0.79)  & a & -0.20 (0.84)  & ab\\
ABK-MP-adv-off & -0.32 (0.80)  & a & 0.34 (0.83)  & a & 0.19 (0.86)  & a & -0.12 (0.87)  & a\\
ABK-MP-self & -0.49 (0.77)  & b & 0.39 (0.86)  & ab & 0.34 (0.79)  & a & -0.19 (0.84)  & ab\\
LDR-banking & -0.42 (0.82)  & ab & 0.40 (0.84)  & ab & 0.25 (0.81)  & a & -0.22 (0.81)  & ab\\
LDR-shopping & -0.42 (0.81)  & ab & 0.51 (0.79)  & b & 0.28 (0.80)  & a & -0.26 (0.82)  & ab\\
LDR-unlock & -0.43 (0.80)  & ab & 0.43 (0.83)  & ab & 0.34 (0.79)  & a & -0.32 (0.79)  & b\\
UDWH-def & -0.45 (0.83)  & ab & 0.28 (0.90)  & ab & 0.13 (0.81)  & a & -0.21 (0.79)  & ab\\
UDWH-off & -0.50 (0.79)  & b & 0.42 (0.85)  & ab & 0.21 (0.85)  & a & -0.26 (0.85)  & ab\\
ZEBOdLAH & -0.50 (0.73)  & b & 0.37 (0.81)  & a & 0.26 (0.80)  & a & -0.18 (0.79)  & a\\
\hline
 & \multicolumn{2}{c|}{$\chi^2=23.7$} & \multicolumn{2}{c||}{$\chi^2=23.9$} & \multicolumn{2}{c|}{$\chi^2=23.4$} & \multicolumn{2}{c}{$\chi^2=22.5$}\\
 & \multicolumn{2}{c|}{$p=0.00$}& \multicolumn{2}{c||}{$p=0.00$}& \multicolumn{2}{c|}{$p=0.00$}& \multicolumn{2}{c}{$p=0.00$} \\
 & \multicolumn{2}{c|}{$\epsilon^2<0.01$} & \multicolumn{2}{c||}{$\epsilon^2<0.01$} & \multicolumn{2}{c|}{$\epsilon^2<0.01$} & \multicolumn{2}{c}{$\epsilon^2<0.01$}\\
\end{tabular}
\caption{Start and end points when mapping (0,0) to the center point and using Gaussian distance (standard deviation inset): Used ANOVA to test significance difference and a post-hoc pairwise $t$-test with Bonferonni correction.  The $*$ indicates significant result from post-hoc analysis. Effect size is calculated using the $\epsilon^2$ method, small effects are in the range $[0.01,0.08)$, below which are considered trivial.}
\label{tab:startend}
\end{table*}
}

\newcommand{\tablemse}[0]{

\begin{table*}
  \centering
  \footnotesize
  \begin{minipage}{0.45\linewidth}
\resizebox{\linewidth}{!}{
\begin{tabular}{l | c c c c c c c c c  }
 &  \rotatebox[origin=r]{270}{\bf ABK-MP-adv-def } & \rotatebox[origin=r]{270}{\bf ABK-MP-adv-off } & \rotatebox[origin=r]{270}{\bf ABK-MP-self } & \rotatebox[origin=r]{270}{\bf LDR-banking } & \rotatebox[origin=r]{270}{\bf LDR-shopping } & \rotatebox[origin=r]{270}{\bf LDR-unlock } & \rotatebox[origin=r]{270}{\bf UDWH-def } & \rotatebox[origin=r]{270}{\bf UDWH-off } \\
\hline
{\bf ABK-MP-adv-off}  &  $5.71$ \\ 
{\bf ABK-MP-self}  &  $6.03$ & $5.07$ \\ 
{\bf LDR-banking}  &  $4.53$ & $5.15$ & $4.98$ \\ 
{\bf LDR-shopping}  &  $8.09$ & $8.38$ & $8.17$ & $5.61$ \\ 
{\bf LDR-unlock}  &  $8.47$ & $7.29$ & $7.01$ & $5.42$ & $4.17$ \\ 
{\bf UDWH-def}  &  $5.24$ & $6.24$ & $6.16$ & $4.39$ & $9.30$ & $9.75$ \\
{\bf UDWH-off}  &  $7.49$ & $5.58$ & $4.30$ & $7.43$ & $7.80$ & $7.12$ & $9.13$ \\ 
{\bf ZEBOdLAH}  &  $5.65$ & $6.54$ & $5.73$ & $3.18$ & $6.45$ & $6.84$ & $5.09$ & $8.12$ \\
\end{tabular}}
\caption{MSE of the frequency of patterns across datasets (all results $\times 10^{-8}$), smaller values indicate more similarity with 0.0 being identical}
  \label{tab:mse}
\end{minipage}
\hspace{0.04\linewidth}
  \begin{minipage}{0.45\linewidth}
\resizebox{\linewidth}{!}{
\begin{tabular}{l | c c c c c c c c c  }
 &  \rotatebox[origin=r]{270}{\bf ABK-MP-adv-def } & \rotatebox[origin=r]{270}{\bf ABK-MP-adv-off } & \rotatebox[origin=r]{270}{\bf ABK-MP-self } & \rotatebox[origin=r]{270}{\bf LDR-banking } & \rotatebox[origin=r]{270}{\bf LDR-shopping } & \rotatebox[origin=r]{270}{\bf LDR-unlock } & \rotatebox[origin=r]{270}{\bf UDWH-def } & \rotatebox[origin=r]{270}{\bf UDWH-off } \\
\hline
{\bf ABK-MP-adv-off}  &  $3.87$ \\
{\bf ABK-MP-self}  &  $2.03$ & $1.16$ \\
{\bf LDR-banking}  &  $2.57$ & $1.01$ & $0.56$ \\
{\bf LDR-shopping}  &  $4.87$ & $1.21$ & $1.44$ & $0.75$ \\
{\bf LDR-unlock}  &  $4.02$ & $1.20$ & $0.86$ & $0.52$ & $0.42$ \\
{\bf UDWH-def}  &  $0.17$ & $3.73$ & $1.93$ & $2.39$ & $4.72$ & $3.91$ \\
{\bf UDWH-off}  &  $6.60$ & $1.45$ & $1.95$ & $1.78$ & $1.37$ & $1.43$ & $6.24$ \\
{\bf ZEBOdLAH}  &  $0.81$ & $2.64$ & $0.81$ & $1.28$ & $2.75$ & $2.04$ & $0.75$ & $3.92$ \\
\end{tabular}}
\caption{MSE of the frequency of trigrams across datasets (all results $\times 10^{-3}$), smaller values indicate more similarity with 0.0 being identical}
  \label{tab:mse_gram}
\end{minipage}
\end{table*}
}

\newcommand{\tableguessing}[0]{
\begin{table}[t]
\centering
\scriptsize
\begin{tabular}{l | c | c | c | c}
& $\alpha=0.1$ & $\alpha=0.2$ & $\alpha=0.5$ & 20-guesses \\
\hline
UDWH-def & 10.96 & 11.60 & 12.04 & 1.52\% \\
ABK-MP-adv-def & 9.50 & 9.46 & 10.88 & 4.36\% \\
\hline
ABK-MP-adv-off & 6.63 & 6.75 & 8.60 & 19.20\% \\
ABK-MP-self & 5.81 & 6.59 & 9.06 & 20.10\% \\
ZEBOdLAH & 6.45 & 7.18 & 9.19 & 16.35\% \\
LDR-banking & 6.99 & 7.99 & 9.94 & 12.05\% \\
\hline
UDWH-off & 5.23 & 6.29 & 8.67 & 23.15\% \\
LDR-shopping & 5.40 & 6.18 & 8.51 & 21.85\% \\
LDR-unlock & 5.23 & 6.49 & 8.23 & 20.90\% \\
\end{tabular}
\caption{Partial guessing entropy for $\alpha=\{0.1, 0.2, 0.5\}$ of each of the data sets calculated using an average of 5-runs of 5-fold cross-validation, with the percentage of patterns guessed after the first 20 attempts.}
\end{table}
}

\newcommand{\tableguessingpure}[0]{
\begin{table}[t]
\centering
\scriptsize
\begin{tabular}{l | c | c | c | c}
& $\alpha=0.1$ & $\alpha=0.2$ & $\alpha=0.5$ & 20-guesses \\
\hline
UDWH-def & 10.02 & 11.51 & 13.28 & 4.76\% \\
ABK-MP-adv-def & 8.91 & 9.73 & 12.65 & 5.91\% \\
\hline
ABK-MP-adv-off & 6.85 & 7.70 & 9.40 & 13.47\% \\
ABK-MP-self & 6.00 & 7.23 & 9.31 & 14.62\% \\
ZEBOdLAH & 6.37 & 6.87 & 8.92 & 18.00\% \\
LDR-banking & 6.87 & 7.84 & 9.81 & 13.38\% \\
\hline
UDWH-off & 5.54 & 6.85 & 8.95 & 19.00\% \\
LDR-shopping & 5.50 & 6.74 & 8.35 & 18.88\% \\
LDR-unlock & 4.90 & 6.43 & 8.06 & 21.12\% \\
\end{tabular}
\caption{Partial guessing entropy for $\alpha=\{0.1, 0.2, 0.5\}$ of each of the data sets calculated using an average of 5-runs of 5-fold cross-validation, with the percentage of patterns guessed after the first 20 attempts.}
\label{fig:entropy}
\label{tab:strength-medium}
\end{table}
}



\section{Introduction}

Studies examining password systems are an important part of computer and usable
security. The challenge with password systems, generally, is that humans choose
passwords as opposed to using randomly generated or mnemonic
passwords/pass-phrases~\cite{kuo2006mneonic, shay2012correct,
  morris1979password, mazurek2013measuring}.  With the advent of mobile
computing on smartphones and tablets, unlock authentication choices used to
locking and unlocking mobile device plays a significant role in information
security. 

The two prominent platforms, Apple's iPhone running iOS and Android OS devices,
dominate the market and both provide ingrained mechanisms for protecting the
unlock procedure. While biometrics, e.g., fingerprint- or face-identification,
are second\-ary/cont\-inuous authentication options, passcode/knowledge-based
authentication is still the primary means for securing a mobile device. 

The most common knowledge-based passcode authentication is PINs, where the
iPhone requires users to create and enter a passcode consisting of (at-least) a
4-digit PIN (later updates may require a 6-digit PIN). Android, from the time of
its initial launch, offers a wider variety of unlock authentication techniques,
including PIN, text-based passwords, face identification, and, most relevant to
this paper, the graphical password pattern. 



Since its launch, the Android unlock pattern has been widely studied in many
contexts. Foremost, studies of choice of unlock authentication
type~\cite{egelman2014readytolock, harbach2014sa, vanbruggem2013modifying,
  harbach2016anatomy,harbach2016keep, vonZezschwitz2013wild} has shown Android
patterns remains relatively popular as an authentication choice.  Patterns has
also been studied with respect to side channels, such as smudge
attacks~\cite{aviv2010smudge}, sensor attacks~\cite{aviv2012practicality}, video
attacks~\cite{ye2017cracking}, shoulder
surfing~\cite{schaub2012surfing,von2015easy,davin2017baseline,aviv2017towards} or eye
tracing~\cite{chen2018eyetell}.  Other studies have examined the impact of
guessing attacks~\cite{uellenbeck2013quantifying, aviv2015bigger}. 

Most relevant to this paper, there have also been a number of
studies~\cite{song2015meteres, sun2014dissecting, heidt2016refining,
  uellenbeck2013quantifying, loge2016user, aviv2015bigger, andriotis2013pilot,
  zezchwitz2016onquant} analyzing user choice of Android graphical password
patterns, variations
therein~\cite{aviv2015bigger,uellenbeck2013quantifying,tupsamudre2017PassO,cho2017syspal}
(such as adding or removing contact points), methods to affect
choice~\cite{song2015meteres, sun2014dissecting, andriotis2014complexity} (such
as password meters), and demographic factors in
selection~\cite{aviv2016analyzing,loge2016user}. What is missing from the
literature is a systemized comparison of the related work in this space that
compares both the methodology and the results.


In this paper, we perform such analysis by first performing an extensive
classification and categorization across the prior studies in a taxonomy of
collection methods and analysis techniques used. Second, we were able to obtain
9 distinct datasets from 4 prior
publications~\cite{uellenbeck2013quantifying,aviv2015bigger,loge2016user,zezchwitz2016onquant}
and perform an extensive comparison of both the visual features of the patterns
and the relative security. From this analysis and our classifications we make
the following contributions:

\begin{itemize}
\item Collection methods can minimally affect spatial features, such as the length of the
  patterns, but for other visual properties, e.g., crosses, start, and end,
  there does not appear to be correlations with collection method.

\item Collection methods can affect the comparative guessing strength. We find
  that there exists three regimes, a high, medium and low strength regime; we
  conjecture that the medium strength regime likely is the most realistic as it
  contains the highest clustering, including a data set with self reported
  patterns.

\item Patterns in the high and low strength regimes are related to adversarial
  collection methods, whereby participants both select patterns they do not
  want others to guess and generate patterns to guess others. The so-called
  ``defensive'' patterns are significantly harder to guess compared to the
  medium strength regime, and the ``offensive'' patterns are significantly
  easier to guess.

\item Collection whereby participants are encouraged to generate
  patterns that match a scenario, e.g., are asked to generate a pattern to secure a
  bank account, can provide both medium and low strength datasets, dependent on
  the strength of the motivator in the scenario.

\item While pattern collection on devices is likely preferred, we do not see
  significant differences between data sets collected using a device or
  pen-and-paper.

\end{itemize}



\figpatternsample{}

\section{Related Work and Background}

\paragraph{Mobile Authentication Unlock Choices}
There are three basic choices for knowledge-based unlock authentication on
mobile devices, depending on platform and model.
\begin{itemize}

\item {\em PIN} based authentication, sometimes referred to as  {\em
  passcodes} on iPhones: where a user is asked to recall a PIN of a
  length at least four digits (newer iPhones, however, will require a
  6-digit passcode~\cite{iphone6digit}).
\item {\em Pattern} based authentication: where a user is asked to
  recall a gesture that interconnects a set of 3x3 contact points. On the Android OS, four or more points should be selected.  The user cannot jump over points not visited before. We
  will focus almost exclusively on this authentication method in this
  survey.
\item {\em Password} based authentication, sometimes called an {\em
  alpha-numeric passcode}: where a user is asked to recall a standard
  text-based password (entered using a soft-keyboard) to unlock the
  device.
\end{itemize}
On more recent mobile devices, users also have the option to select a biometric,
such as a fingerprint or face recognition, but as users are {\em also} required
to select a PIN or pattern in this setting, we consider the choices above the
primary choices. {\em Moreover, an attacker attacking an authentication scheme will likely focus on
  knowledge based attacks---something that can be guessed or enumerated---which
  make user choice of authentication secrets still quite relevant. }

Beyond the graphical password pattern, there have been a number of studies
examining user choice for
PINs~\cite{bonneau2012birthday,kim2012pin,datagenetics} and
passwords~\cite{morris1979password, mazurek2013measuring, kelley2012guess,
  wash2016understanding, ur2016users}. Following this section, we will present
the relevant work in pattern based data collection and analysis.



\paragraph{Mobile Device Locking Practices}
There have been a number of studies of mobile device unlocking
behavior~\cite{egelman2014readytolock, harbach2014sa, vanbruggem2013modifying,
  harbach2016anatomy, harbach2016keep, harbach2016keep,vanbruggem2013modifying}
dating back from 2013 through 2016. From these studies we know that  users
perceive patterns  as more secure and less error-prone (as opposed to PINs) in
entry~\cite{vonZezschwitz2013wild}; however the opposite is often true in
practice. Generally, these studies suggest that it is important to continue to
study Android's graphical password system. Despite the fact that other
methods exist, it remains a popular choice that many users (perhaps
wrongly) see as a more secure choice for protecting their device from
unauthorized access. 



\paragraph{Background on Android Graphical Pattern}
The Android graphical password pattern belongs to a family of draw-metric
graphical passwords systems like Draw-A-Secret~\cite{jermyn}, Qualitative
Draw-A-Secret~\cite{lin2007graphical}, PassPoints~\cite{wiedenbeck},
Pass-Go~\cite{tao2008passgo}, PassShapes~\cite{weiss} and
PassDoodles~\cite{varenhorst2004passdoodles} to name a few.  These solutions
require the user to draw a predefined shape or an arbitrary representation.

The Android graphical password pattern presents the user with a grid of 3x3
contact points on which the password pattern is ``drawn'' (see Figure~\ref{fig:patternsample}). If the pattern is
successfully recreated, entry to the device is granted.  The drawing of a
pattern is constrained such that (1) a pattern must contain at least 4 contact
points, (2) a contact point may only be used once, (3) a pattern must be entered
without lifting, and (4) a user may not avoid a previously un-selected contact
point. In Figure~\ref{fig:patternsample}, an example of the allowable strokes
originating from the upper left corner are shown.

\section{Limitations}

It is relevant to also note the unique challenges associated with measuring and
comparing data related to the Android unlock patterns and unlock authentication
methods, generally. Foremost is that, unlike text-based passwords (and PINs),
pattern authentication is not used in other contexts beyond mobile devices and
is also not used for remote authentication. This impacts data collection
because, for example, much of the common knowledge of the strength and
variations in text based passwords comes from the leaks of large password
datasets from remote authentication services. Android pattern unlock, however,
are primarily used as a local authentication process, so obtaining real,
user-worn graphical passwords through leaks or other means is highly unlikely.

The limits of not having known real data to analyze challenges testing the
veracity of other collection methods because there does not exist a clear ground
truth for comparison. Much of the data collected is only {\em
  self-confirming} in that researchers look to find consistency between the
datasets with respect to commonalities (e.g., most frequent patterns) in the
user generated/reported patterns and the strength of those patterns. Here, we
attempt to cluster relevant data sets together based on features to determine if
there is broader consistency and if these consistencies can be associated with a
given collection method.

We are additionally limited in that the size of the datasets analyzed are not as
rich as one would want from, say, a leaked password dataset. Again, this is due
to the fact that leaks of real pattern data is not available, and thus all the
available datasets are smaller, user-study collected. Fortunately, we can use
alpha-guesswork~\cite{bonneau2012science}, which provides comparable results
even for smaller datasets, but we acknowledge that this is a limitation on our
findings.

Finally, the survey of data was completed by reviewing relevant conferences and
journal in the space of human computer interaction, computer security, and
usable security. We did not use a formal system of review, such as searching a
known research database, and we acknowledge this limitations. But, with
extensive background in this space and a relatively limited number of
publications in the area, we are confident that we have broad coverage of the
space.

\tableterms{}

\section{A Taxonomy of Collection Methods}

Here, we provide a survey and taxonomy of the relevant work regarding data collection for Android
unlock patterns, as identified by reviewing top conferences and
journals in the area of computer security, usable security, and computer human
interaction.  This taxonomy is organized based on the terms found in
Table~\ref{tab:terms} and is applied to the relevant research in
Table~\ref{tab:collected-data}. Studies are named based on the first letters of the authors' surnames in which the studies are first presented, and for
simplicity we only focus on the portions of study that collect patterns
directly, without modifications to the scheme such as in
SysPal~\cite{cho2017syspal} or 4x4 patterns~\cite{aviv2015bigger}.

\subsection{Protocol Scenarios}
A study must incentivize participants to act in realistic ways when generating
patterns, and we observed three major forms of scenario building to incentivize
participants: adversarial/game-theoretic settings, self-reporting, and
situational settings.

\paragraph{Adversarial Incentives}
One method to provide incentives to generate patterns that have a security
connection is to design an {\em Adversarial} protocol (first proposed by
Uellenbeck et. al~\cite{uellenbeck2013quantifying} and later used in a
pen-and-paper study by Aviv et. al~\cite{aviv2015bigger} as well as by
Tupsamudre et. al~\cite{tupsamudre2017PassO}). The core idea for this protocol
is to have participants create a set of patterns that they wish to protect, so
called {\em defensive} patterns, and a set of patterns that they believe others
may have chosen, so called {\em offensive} patterns. If a participant is able to
generate a defensive pattern that others did not guess with their offensive
patterns, or come up with offensive patterns that others selected as their
defensive pattern, the participant earns a reward. 

This was primary method used in UDWH~\cite{uellenbeck2013quantifying}, labeled
with *-off and *-def, collected on a mobile device. Additionally, the
ABK-MP-adv~\cite{aviv2015bigger} used a similar method, but collected using
pen-and-paper. For collecting the TBLV dataset, Tupsamudre
et. al~\cite{tupsamudre2017PassO} ran an organization-wide contest allowing each
participant to choose three defensive and three offensive patterns. The
challenge with adversarial technique, as we will show, is that it may both over
incentivize participants in the defensive setting to create unrealistic,
overly-strong patterns and similar so for the offensive setting, but instead
weaker patterns.

\tablecollected{}
\tablealtered{}



\paragraph{Self Reporting}
Another collection method is to directly ask participants to {\em Self-Report}
the patterns they use. The challenges with this approach are obvious: what is to
stop participants from lying? Additionally, it is counter to security practices
for participants to release this information. These limitations are
non-negligible, but researchers can build in checks to exclude participants who
abuse the survey and provide assurances to incentivize participation. For
example, the survey can have participants enter the pattern multiple times,
conduct the survey on the participants own Android device, or ask the
participant directly (without penalty) if they provided truthful
answers. Additionally, as patterns are only useful with access to the physical
device, participants may feel comfortable revealing this information.

These methods were used to collect the ABK-MP-self~\cite{aviv2015bigger} dataset
which applied a self-report method by recruiting participants from
MTurk. Participants were given the option to report their pattern (on their own
device) or answer statistics (e.g., length, start point, common sequences) about
their pattern. Attention tests were used to ensure that participants answered
honestly. 

\paragraph{Situational Representations}
Another technique to increase the variation of patterns is to place
participants in situational scenarios with reduced or increased risk
to data. The idea being that increased risk scenarios may illicit
participants to report stronger passwords as compared to the reduced
risk scenarios. These techniques have been used in text-based password
research~\cite{ur2015observing}.

The LDR~\cite{loge2016user} data set used this technique in its protocol to
increase the diversity of reported patterns. In particular, researchers asked
participants to generate a pattern for use as a bank password, as a shopping web
site password, and as normal device unlock password.  One of their results shows
that patterns created for the bank are significantly stronger than those created
for online shopping and device unlock.  In a similar, but less situational
setting, ATOY dataset was collected with a protocol asking participants to
provide one easy to remember password and one hard to guess
password~\cite{andriotis2013pilot}. SGSMA also provided a situation setting in
their dataset collection directing participants to consider setting up a new
phone~\cite{siadati2015fortifying}. 

\subsection{Participant Recruitment}

\paragraph{Local vs. On-line}
Collecting Android unlock patterns can be conducted in two main ways, locally or
remotely. A {\em local} study occurs when the researchers and participants are
co-located, for example, in the form of a lab study where participant are asked
to come into a setting and generate patterns under the supervision of the
researcher(s). The benefits of such an approach relate to the experimenter's
ability to control extraneous and independent variables.  However, logistics of
bringing in a large sample of participants within a short time period is not
without its challenges. Additionally, the presence of a research can lead to
Hawethorn effect whereby participants behave differently due to being directly
observed or having an expectation of a given outcome.

In contrast, a remote study is conducted without the direct supervision of the
researcher, and is generally conducted {\em on-line} via a web-based platform
accessed through the browser. They can be distributed using services like Amazon
Mechanical Turk, the marketplace, or using a snowball technique.  Users may feel
more comfortable interacting with technologies within their own environment
during remote studies, although the risk of interruptions in an uncontrolled
environment is higher than in an in-lab study. A challenge with the on-line setting
is verifying that participants act in ways that are conducive to the study,
which can be directly observed in the local setting.

Good examples of local studies in our survey are the data sets ATOY, UDWH,
SGSMA, SWZ, and ABK-MP-adv~\cite{uellenbeck2013quantifying, andriotis2013pilot,
  sun2014dissecting, aviv2015bigger,siadati2015fortifying}. In each of these
studies the researchers were present during data collection; in ATOY, this took
the form of a lab computer where participants were asked to enter patterns that
were easy and hard; in UDWH, this took the form of research recruiting
participants in the institutions cafeteria and instructing them in the
procedures of the protocol; and in ABK-MP-adv participants were divided into
study groups and operated under the supervision of the researchers to complete
the protocol.

Good examples of online studies are ABK-MP-self, LDR, ZEBOdLAH, CHKCOSK, and
TBLV~\cite{aviv2015bigger,loge2016user,zezchwitz2016onquant,cho2017syspal,tupsamudre2017PassO}. In
each of these studies, the participants were recruited either via a system like
Amazon Mechanical Turk~\cite{aviv2015bigger,zezchwitz2016onquant,cho2017syspal},
snowball efforts~\cite{loge2016user}, or company mailing
lists~\cite{tupsamudre2017PassO} to complete a survey hosted on-line. The
researcher and the participant are separated from each other, as opposed to a
local study. 




\paragraph{Application Based}
A separate, but related form of on-line surveys recruitment, is by distributing
a specialized application (i.e., an App) developed using Android's development
tools and posting the App via one of the Android Application marketplaces. We
describe this technique as {\em App} in Table~\ref{tab:terms} and can be an
effective way to get known, real Android users to participate in the study. The
challenges with this technique comes down to ethics; users may interpret the
application as non-research and use the application in ways that are unintended
by the researchers. However, proper informed consent can account for this.

A good example of this methodology is the SCOKH
study~\cite{song2015meteres}. The researchers released an application into the
market place called {\em EnCloud} which purported to be a encrypted cloud
storage service where data was secured using an Android unlock pattern. The
participants were then allowed to choose patterns under different conditions,
e.g., having a meter or not, to measure the effectiveness of their choices.

\subsection{User Input Methods}

There are also varied forms for the ways in which participants generate a
pattern for study. For example, as the Android unlock pattern is used on a
touchscreen device, it is most natural to conduct the survey where users enter
patterns via touch gestures on a smartphone or tablet. This natural setting is
described as {\em On-device}, and if it occurred on participants' own devices,
we describe this as {\em User-device}. Both App based studies and on-line
studies can utilize on-device/user-device methods as well as local collections.

A good example of the distinction between on-device (but not user-device) is the
UDWH data set where participants were recruited at a school cafeteria and were
directed to use a device with the survey pre-built
in~\cite{uellenbeck2013quantifying}. ABK-MP-self, LDR, ZEBOdLAH, SGSMA and
SCOKH, in contrast, were conducted on-line and allowed participants to use their
own device to complete the survey~\cite{siadati2015fortifying, aviv2015bigger,
  loge2016user, song2015meteres, zezchwitz2016onquant} often requiring
participants to use an Android device (checked via the user-agent field) or
because the survey was distributed through the App marketplace.

For some online studies, though, it could be the case that some
participants used a {\em Mouse} to enter patterns (as in ATOY), which
may affect the kinds of patterns that are deemed ``usable'' as the
mouse allows for finer motor control as compared to
gestures~\cite{andriotis2013pilot}.

Additionally, as the Android unlock patterns are ``drawn'' it is also
possible to conduct studies using {\em pen-and-paper} where participants
literally draw patterns using a pen on a sheet of paper with printed
contact points, as was the case in ABK-MP-adv. Such a non-technical
data collection method may seem counter intuitive for this space, but
the sensory aspect of drawing translates well between touching and
writing. There is also evidence that this technique does produce
realistic results~\cite{aviv2016analyzing}, but some spatial aspects
of patterns (e.g., the starting points) may be slightly changed as
compared to other collection methods.

\tabletermsanalyze{}
\section{Analysis and Comparison of Data\-sets}
\label{sec:analysis}

Next, we turn our attention to the actual collected data, performing analysis
over that data using the most common techniques. A list of analysis methods used
in the prior work are provided in Table~\ref{tab:terms-analyze}; we focus on
occurrence rates, visual features, and guessability.  We do not consider ad-hoc
methods of pattern strength, as used in password
meters~\cite{song2015meteres,sun2014dissecting} as these are not grounded and
are typically based on visual features anyway. We also do not consider recall
rates as they are not germane to comparing datasets.

Unfortunately, we were not able to acquire all relevant datasets, either due to
ethical reasons or institutional oversight, but we were able to access 9
distinct datasets presented in four prior publications: ABK-MP-self, ABK-MP-adv-def,
ABK-MP-adv-off~\cite{aviv2015bigger}; LDR-banking,
LDR-shopping, LDR-unlock~\cite{loge2016user}; UDWH-def,
UDWH-off~\cite{uellenbeck2013quantifying}; and
ZEBOdLAH~\cite{zezchwitz2016onquant}.


\figfreqpatterns{}

\paragraph{Occurrence Rates}
The most basic analysis one can perform to compare datasets is to look at the
most commonly used patterns. One can visualize the most frequent patterns (such
a visual is provided in Figure~\ref{fig:freqpatterns}), and a casual
observations provides some intuition that there exists a number of high
occurring patterns that are consistent across datasets, such as a the {\bf Z} or
{\bf L} shape (in various symmetries).

We avoid analysis of the frequency of individual patterns, which has been
covered in prior work related to these datasets, and instead concern ourselves
with comparing the datasets with respect to how frequent repeated patterns occur
as well as the overall uniqueness of the data. Both of these metrics are related
to the strength of the dataset; the less unique (or more repetitions) a dataset
is (or has) the easier it is to predict the patterns and thus guess the
patterns.

Repetitions are calculated by counting the number of patterns that occur two ore more times
in the dataset, which we can then normalize by the size of the dataset because
larger datasets should have proportionally larger number of repetitions. The
uniqueness of the dataset measures the proportion of unique patterns to the
total number of data in the dataset. For example, a 50\% unique dataset implies
that of the 100 patterns in the dataset, 50 of them are unique and the other 50
are repeats.  These results are presented in Table~\ref{tab:rep}.

Using a $\chi^2$ test, we can test for significant differences in the
proportions of both repetitions and uniqueness, and in both cases, we find that
there are significant differences ($\chi^2=20.31,p<0.01$ for repetitions, and
$\chi^2=213.03,p=0.0$ for uniqueness). Further investigation into the
differences using a post-hoc, pairwise test using a false discover rate
correction, showed that there are distinct groupings of similar datasets in each
category. The groupings are represented in Table~\ref{tab:rep} using letters,
such that datasets with the same letter are {\em not} significantly different.

For the proportion of repetitions, the defensive datasets stand out, but
particularly the UDWH-def dataset which has significantly fewer repetitions than
the other datasets and is only similar to the other defensive dataset. This
suggests that the adversarial collection method tends to encourage participants
to select more unique patterns for the defensive collection, and as these were
both collected in-person, the effect of having other participants and
researchers present may further encourage this behavior.

When measuring dataset uniqueness, the post-hoc analysis reveals other
relationships, breaking down into three basic regimes. The highest proportion of
unique patterns is, again, the defensive datasets ({\em a} grouping). There is
also a low regime with the lowest proportion of unique patterns which contains
the LDR-shopping and LDR-unlock ({\em d} grouping). In the middle, bottom-half,
are the two offensive datasets ABK-MP-adv-off, UDWH-off (overlapping in {\em b})
and the top-half contains the remaining datasets of ABK-MP-self, ZEBOdLAH, and
LDR-banking (overlapping at {\em c} groupings).

This suggests that the adversarial model produces similar results in both the
higher regime defensive and middle regime offensive. The scenario driven data
collection with the weakest motivator, shopping and, surprisingly, unlocking,
have some of the least unique sets of patterns. Finally, the remaining datasets,
which contain collection with stronger scenarios and self-reporting, on device,
have a close consistent groupings. We will see similar groupings when analyzing
the strengths of the datasets as uniqueness properties are related to
guessability strength metrics.

\begin{table*}[t]
\centering
\scriptsize
\begin{tabular}{r | c | c | c c || c c || c | c}
{\bf dataset } & {\bf size } & {\bf rep. } & {\bf norm. rep. } & & {\bf \%unique } & & {\bf avg } & {\bf std } \\
\hline
UDWH-def       & 106 & 3 & 0.028 & a& 97.2\% & a & 0.029 & 0.168 \\
ABK-MP-adv-def & 113 & 8 & 0.071 & ab & 92.9\% & a & 0.076 & 0.265 \\
UDWH-off       & 573 & 78 & 0.136 & b & 59.2\% & b & 0.690 & 2.196 \\
ABK-MP-self    & 440 & 61 & 0.139 & b & 67.7\% & c & 0.477 & 1.484 \\
LDR-shopping   & 841 & 119 & 0.141 & b & 52.6\% & d & 0.903 & 2.694 \\
ZEBOdLAH       & 507 & 72 & 0.142 & b & 69.0\% & c & 0.449 & 1.297 \\
ABK-MP-adv-off & 378 & 57 & 0.151 & b & 65.6\% & bc & 0.524 & 1.292 \\
LDR-banking    & 838 & 128 & 0.153 & b & 66.2\% & c & 0.510 & 1.482 \\
LDR-unlock     & 842 & 138 & 0.164 & b & 49.8\% & d & 1.010 & 2.604 \\
\end{tabular}
\caption{The size and number of repeated (rep.) patterns in each of the dataset, sorted by the normalized repetitions (norm. rep.). Using a $\chi^2$ test, there is significant differences in the proportions of repetitions ($\chi^2=20.31,p<0.01$) and uniqueness ($\chi^2=213.03,p=0.0$). Pairwise posthoc analysis with a false discovery rate (fdr) correction was used to determined letter based grouping whereby datasets with the same letter are {\em not significantly} different. Also provided is the average number of repetitions (avg) and the standard deviation (std).}
\label{tab:rep}
\end{table*}


\figvizfeatures

\figlengths

\paragraph{Spatial Features and Visual Properties}
We analyzed the provided datasets using three {\em spatial features} and three
{\em visual properties}. Each of these pattern classification methods
have been applied in prior work as a way to understand the shapes of
patterns~\cite{cho2017syspal,loge2016user,siadati2015fortifying,aviv2015bigger, aviv2016analyzing,andriotis2013pilot,uellenbeck2013quantifying}, or as way to estimate the strength of a pattern, as in
a password meter~\cite{song2015meteres,sun2014dissecting}. For a pictorial, refer to
Figure~\ref{fig:vizfeatures}, as each is explained.
\begin{itemize}
  \item Spatial Features: 
    \begin{itemize}
    \item {\em Length} (length): The number of contact points used in the pattern. For
      example, the {\bf Z}-shaped pattern used 7 contact points.
    \item {\em Stroke Length} (s-length): The length of the gesture lines in the
      pattern by using the Gaussian distance measure with the center contact
      point mapped to (0,0) and each contact point being one unit apart. For
      example, the {\bf Z}-shaped pattern has a stroke length 6.82, with the top
      and bottom line worth 2 units, and the diagonal is $2\cdot\sqrt{2}$.
    \item {\em Turns} (turns): The number of direction changes in a pattern. For
      example, the {\bf Z}-shaped pattern makes two direction changes, thus has 2
      turns.
    \end{itemize}
  \item Visual Properties: 
    \begin{itemize}
    \item {\em Crosses} (crosses): When two lines overlap and cross over in any
      shape.
    \item {\em Knight Moves} (kmoves): When a line connects two contact points
      over 1 and down 2 (or in symmetries), like a knight moves in chess.
    \item {\em Non Adjacency} (non-adj): When a previous contacted point is
      traced over to connect two non-adjacent contact points. In the example in
      Figure~\ref{fig:vizfeatures}, the middle contact point is first, then
      right, then over the middle to the left, connecting two non-adjacent
      contact points.
    \end{itemize}
\end{itemize}

The results of calculating these features and properties across the datasets are
presented in Table~\ref{tab:vizfeatures}. Significance testing was done using
Kruskal-Wallis as there was unequal variance or non-normal distributions in the
data. As this is a conservative test, and the datasets are relatively large, we
found that there was significant differences among the dataset for each of the
tested features. Using pairwise Mann-Whitney U-tests with a Bonferroni
correction, we sought to identify groups of similar datasets, these are labeled
with letter codes in Table~\ref{tab:vizfeatures}. As before, datasets, under the category,
with the same letter are {\em not} significantly different.

\tablespatprop

\tablestartend{}

For length and slength, there is a clear grouping which suggests that collection
methods may have had an effect on pattern length features. In particular, both
adversarial method's datasets are much longer than the scenario-driven ones, with
the self-reporting in the middle. As length is correlated with slength and
turns, similar groupings exist there. For visual properties, e.g., crosses,
kmoves, non-adj, we see that these groupings by collection do not persists, and
there are is much overlap across different datasets.

These results suggest that the collection method can impact some of the
features, namely those based on the length of the pattern, but the effect is
small ($\epsilon^2 < 0.08$). Adversarial (e.g, *-off,*-def) collection methods
encourage slightly longer patterns, while scenario-driven (ZEBOdLAH, LDR-*)
collection tends towards slightly shorter. While other visual features vary
across datasets, the overlapping similarities do not appear to correlate across
collection methods and there is very small, perhaps trivial, effects.

\paragraph{Start- and End-conditions}
A similar analysis is used to understand the start and end conditions of
patterns, that is, which contact points in the grid do users tend to select for
initiating and terminating a pattern.  We compute averages by assigning
the middle point of the pattern to the (0,0) in the Cartesian plain, as before,
and calculating the average x and y shift for the start and end locations when
mapped into the plain. The analysis is found in Table~\ref{tab:startend}.

Using Kruskal-Wallis, we find that there does exist significant differences
between the data sets, but these differences are small, and in post-hoc analysis
using pairwise Mann-Whitney U-test with Bonferroni corrections, there are large
groupings of similarities between the datasets that appear uncorrelated with
dataset or collection method. Additionally, the effect size of
$\epsilon^2 < 0.01$ is trivial, all suggesting that collection methods likely
plays little role in spatial shifting of the start and end points. Users seems
consistent in preferring to start in the upper-left and end in the lower-right.




\begin{table}
  \centering
  \scriptsize
  \begin{tabular}{l | rrrrr}
                 &    min  &1st quart.&  median  &3rd quart.&    max\\
    \hline
       UDWH-def  &   0.52  &    5.68  &    8.83  &   13.13 &   19.93\\
 ABK-MP-adv-def  &   0.52  &    4.87  &    7.64  &   11.76 &   19.93\\
    \hline
 ABK-MP-adv-off  &   0.52  &    3.86  &    6.46  &    9.13 &   19.93\\
    ABK-MP-self  &   0.52  &    3.91  &    6.00  &    9.59 &   19.93\\
     ZEBOdLAH    &   0.52  &    4.29  &    6.38  &    9.43 &   19.93\\
     LDR-banking &   0.52  &    4.21  &    6.48  &    9.39 &   19.93\\
    \hline
    UDWH-off     &   0.52  &    3.60  &    5.50  &    9.04 &   19.93\\
    LDR-shopping &   0.52  &    3.65  &    5.27  &    8.42 &   19.93\\
       LDR-unlock &   0.52  &    3.76  &    5.27  &    7.94 &   19.93\\
  \end{tabular}
\caption{Basic statistical information about pattern strength per data set. Data sets are grouped based on similarity of collection protocols and results. The data is divided into a {\em high}, {\em medium}, and {\em low} strength regime based on the median pattern strength.}
  \label{tab:strength-median}
\end{table}








\paragraph{Strength of individual patterns}
Pattern strength is inversely related to the frequency and probability of a
pattern being used in real life, as a rational attacker will try to guess more
frequent/likely patterns first to maximize his success for a limited number of
attempts. While the frequency of patterns is known within the sample sets, it is
imperative to estimate the likelihood of patterns, generally.  

Different approaches have been used to do this estimation. We use Markov models
to calculate the probabilities of users selecting a given pattern which can then
be used as an approximations for the pattern strength.  The technique we apply
follows closely that of Aviv et al.~\cite{aviv2015bigger}, which expands on the
Markov model of Uellenbeck et al.~\cite{uellenbeck2013quantifying} by adding in
additional predictions related to the start points, length, and finish points of
the pattern.


Markov models are based on the observation that subsequent tokens, such as
letters in normal text or contact points in the pattern, are rarely independently
chosen by humans, but can often be quite accurately modeled based on a short
history of tokens.  For example, in English texts, the letter following a
\texttt{t} is more likely to be an \texttt{h} than a \texttt{q}, and for Android
pattern unlock, a similar process occurs by which users are more likely to
choose some transitions over others based on prior state, e.g., starting in the
upper left, a user will more likely move to the right as opposed to the center
contact point.

Using these facts, one can construct an $n$-gram Markov model that estimates the
likelihood of a pattern by considering the probability of a transition from the
previous $n-1$ contact points to the next contact point. For a given sequence of
contact points for a pattern $p=\{c_1,\ldots,c_m\}$, an $n$-gram Markov model
estimates its probability as
\begin{eqnarray}
  \label{eq:markov}
  &&P(c_1,\ldots,c_m)\\ 
  && = P(c_1,\ldots,c_{n-1}) \cdot
  \prod_{i=n}^m  P(c_i|c_{i-n+1},\ldots,c_{i-1}). \nonumber
\end{eqnarray}

The required \emph{initial probabilities} $P(c_1,\ldots,c_{n-1})$ and
\emph{transition probabilities} $P(c_n|c_{1},\ldots,c_{n-1})$ can be
determined empirically from the relative frequencies from training data. 
%
To handle the varying length of patterns we additionally multiply with a length
term $P(|p| = m)$. Also, as suggested by Aviv et
al~\cite{aviv2015bigger}, we consider special (non-contact) start transitions
and end transitions to capture start and end conditions for transitions less
than length $n$ as they occur at the beginning and end of a pattern.

One commonly applies further post-processing to the raw frequencies: So-called
\emph{smoothing} tries to even out statistical effects in the transition matrix
and initial probabilities. In particular, smoothing avoids relative frequencies
of $0$, as these would yield an overall probability of $0$ regardless of the
remaining probabilities.
We use smoothing based on a uniform probability drawn from the set of {\em all}
389,112 possible patterns, assuming each are equally likely.


In our analysis, to determine the strength of an individual pattern, we trained a
3-gram Markov model using all patterns in our collected dataset
with one instance of the pattern being measured left out. This process ensures
independence of training and evaluation sets while preserving information about
pattern frequencies. For example, the pattern may be very frequent, but
only a single instance is removed.
The estimated probabilities for patterns, $\hat p$, could be used directly as a
strength measurement, but as probabilities over a large set of possibilities,
they are quite small.  A more readable measure is $-\log(\hat p+\epsilon)$,
where logarithms are to basis $2$, and we add a small constant
$\epsilon=10^{-6}$ to cap strength scores at $19.9315 \approx 20$.  We use this
strength measure in the rest of this work as a representation of a likelihood of
pattern as estimated by the Markov model.

We compare the strength scores for the individual datasets and present the
results in Table~\ref{tab:strength-medium}. A Shapiro-Wilk test suggest that the
data is not normally distributed ($W=0.90,p < 2.2e-16$), and so we compare the
data using a Kruskal-Wallis to find that that there are significant differences
across the datasets ($\chi^2=122.23, p=0.00$) The results from the post-hoc test
(pairwise Mann-Whitney U with Holm correction) are shown in
Table~\ref{tab:sig-diff-median}.  The table shows the corrected $p$-values from
the pairwise comparison, and marks significant differences in bold.

These results directly inform the grouping of our datasets based on similarity
of pattern strength, and are similar to the groupings found in the dataset
uniqueness analysis as the Markov model probabilities are sensitive to that
metric. This grouping into a low, medium, and high strenght regime informs our
analysis going forward.

\tableguessingpure{}



\begin{table*}[tp]
  \centering
  \small
  \resizebox{\textwidth}{!}{
  \begin{tabular}{l | c c | c c c c | c c c}
               & UDWH-def    & ABK-MP-adv-def &  ABK-MP-adv-off & ABK-MP-self & ZEBOdLAH    & LDR-banking & UDWH-off    & LDR-shopping& LDR-unlock   \\
\hline
UDWH-def       & -           & \it 0.55084    & \bf 3.1e-06     & \bf 1.3e-05 & \bf 0.00011 & \bf 3.6e-06 & \bf 3.1e-08 & \bf 5.0e-11 & \bf 4.0e-12  \\
ABK-MP-adv-def & \it 0.55084 & -              &  \bf 0.01777    & \bf 0.04427 & \it 0.22124 & \it 0.06071 & \bf 0.00063 & \bf 4.5e-06 & \bf 1.2e-06  \\    
\hline                                                                                                                 
ABK-MP-adv-off & \bf 3.1e-06 & \bf 0.01777    &  -              & \it 1.00000 & \it 1.00000 & \it 1.00000 & \it 1.00000 & \it 0.05711 & \bf 0.01500  \\    
ABK-MP-self    & \bf 1.3e-05 & \bf 0.04427    & \it 1.00000     & -           & \it 1.00000 & \it 1.00000 & \it 0.57343 & \bf 0.01500 & \bf 0.00363  \\  
ZEBOdLAH       & \bf 0.00011 & \it 0.22124    & \it 1.00000     & \it 1.00000 & -           & \it 1.00000 & \bf 0.01500 & \bf 7.9e-06 & \bf 8.0e-07  \\ 
LDR-banking    & \bf 3.6e-06 & \it 0.06071    & \it 1.00000     & \it 1.00000 & \it 1.00000 & -           & \bf 0.02543 & \bf 2.8e-06 & \bf 1.2e-07  \\ 
\hline                                                                                                                 
UDWH-off       & \bf 3.1e-08 & \bf 0.00063    & \it 1.00000     & \it 0.57343 & \bf 0.01500 & \bf 0.02543 & -           & \it 1.00000 & \it 0.87558  \\   
LDR-shopping   & \bf 5.0e-11 & \bf 4.5e-06    & \it 0.05711     & \bf 0.01500 & \bf 7.9e-06 & \bf 2.8e-06 & \it 1.00000 & -           & \it 1.00000  \\ 
LDR-unlock     & \bf 4.0e-12 & \bf 1.2e-06    & \bf 0.01500     & \bf 0.00363 & \bf 8.0e-07 & \bf 1.2e-07 & \it 0.87558 & \it 1.00000 &  -           \\ 
  \end{tabular}}
  \caption{Results from post-hoc test (pairwise Mann-Whitney U test with Holm correction) on  strength.  Given are the (corrected) $p$-values, where significant differences are marked bold.  Grouping is loosely based on the differences in strength.}
  \label{tab:sig-diff-median}
\end{table*}

\paragraph{(Partial) Guessing Entropy based on Markov Model}
(Partial) guessing entropy~\cite{massey-1994-guessing-entropy,bonneau2012science} is a measure of the strength of
human-chosen authentication secrets, such as pattern sequences, with a
strong theoretical foundation.  Intuitively, it models an ``ideal''
attacker who knows the probabilities of all patterns, and guesses
those patterns in order of decreasing likelihood.
This has the advantage that it does not depend on a specific guessing
algorithm, but has the disadvantage that it assumes an ideal (and
possibly unrealistic) adversary.
%
We use the logarithmic variant of this measure, $\tilde G_\alpha (X)$, which
gives ``bits of information'' (see~\cite{bonneau2012science} for definitions and
details).

Computing $\tilde G_\alpha$ requires an accurate approximation of the
distribution $X$, which normally requires large sample sizes beyond
the collected data.  In line with previous work~\cite{uellenbeck2013quantifying} we use an approximation based on
Markov models to compute $\tilde G_\alpha$.  Thus the entropy results
are relative to the chosen Markov model, and typically constitute an
upper bound only.
We report (partial) entropy estimates for the levels $\alpha \in \{0.1, 0.2, 0.5\}$,
i.e., for an attacker trying to break 10\%, 20\%, or 50\% of the accounts, respectively.

To do the guessing, we performed a 5-fold cross-validation on each dataset by
which we trained the Markov model using 4 of the folds and guessed on the
remaining fold. We guessed based on the absolute ordering of all 389,112
patterns based on the likelihood measures as computed by the Markov model. More
advanced and aggressive guessing strategies could be used (as described by Aviv
et. al~\cite{aviv2015bigger}), but this strategy produces the most repeatable
results. 

The results are presented in Table~\ref{fig:entropy} which are an average of 3
runs of the cross-fold validation. The results show similar groupings as in
Table~\ref{tab:sig-diff-median}. We also note that these entropy measures are
not precisely as reported in prior work~\cite{aviv2015bigger,uellenbeck2013quantifying} and is
likely due to small variations (or over-fitting) in modeling for the Markov
model; however, general trends remain.

\section{Discussion of Collection Methods}


\paragraph{Adversarial Setting}
Two datasets are collected under the adversarial settings: The UDHW and the
ABK-MP-adv datasets.  Tables~\ref{tab:strength-median} and~\ref{fig:entropy}
shows that the two defensive datasets are the strongest datasets available to
us, and in fact the post-hoc tests show that the two defensive sets are
significantly stronger than almost every other dataset (the only two exceptions
are ABK-MP-adv-def/ZEBOdLAH and ABK-MP-adv-def/LDR-banking). 

It seems most likely that these differences can be explained by the collection
method. This is evident due to the fact that the corresponding {\em offensive}
data is collected during the same experiment from the same user base and shows
significant differences in strength --- namely, they are some of the weakest in
terms of strength --- but while significant differences exist in spatial
features, there are reactively small magnitude differences. We also see
similarity between the offensive and defensive (respectively) for repetitions
and uniqueness qualities.  This all suggests that the adversarial setting
encourages participants to draw the same patterns based on visual indicators,
but somehow are incentivized to choose less common, harder to guess
patterns. Second, all the other datasets behave much more similar (in terms of
strength and guessability) compared to the defensive sets. For example, the
ABK-MP-adv-def and UDWH-def are 3 and 4 bits higher in partial guessing entropy
compared to the next highest data set.



\paragraph{Situational Settings}
On the weak end of the spectrum, three datasets are generally weaker than the
other datasets, the LDR-shopping and LDR-unlock datasets, as well as the
UDWH-off. The three LDR datasets were collected using the same situational
protocol, but it was observed in the original publication~\cite{loge2016user}
that the three datasets behave differently and that the banking scenario
prompted participants to choose stronger patterns. The three datasets are
significantly weaker than any other dataset, with three exceptions
(UDWH-off/ABK-MP-adv-off, UDWH-off/ABK-MP-self, and
LDR-shopping/ABK-MP-adv-off). From these results, we speculate and would
recommend that researchers use strong situational motivation. 

\paragraph{Medium strength datasets}
The four remaining datasets, i.e., ABK-MP-adv-off, ABK-MP-self, ZEBOdLAH, and
LDR-banking, have no significant differences in strength amongst each
other. They form a more or less consistent subset with very similar strength and
guessability characteristics. We conjecture that this group of datasets may most
likely mimic reality of pattern strength as they include the only self-reported
dataset (ABK-self) and there is such strong similarity.

\paragraph{Participant recruitment}
Both UDWH and ABK-MP-adv performed local recruitment, while most other datasets
in our evaluation performed online recruitment.  In principle, this could
provide an alternate explanation for the strength of the defensive datasets, e.g., some form of a Hawthorn effect whereby the presence of the researcher (or other participants) affected the results.
However, the differences in collection protocol, specifically the strong
external incentives from other participants attacking the pattern, seem a much
more likely explanation for the observed strength differences.

There is some argument to be made that application-based data collection may
provide the most reliable data from a population using their phone with an
installed application with real-world accounts. Unfortunately, we do not have
access to a dataset collected under this protocol. 



\paragraph{Input methods}
Almost all datasets available to us were collected on a mobile device (either
user device or provided device), but it doesn't seem likely that this difference
influences the strength of patterns.  (The only exception is the ABK-MP-adv
dataset which was collected using pen-and-paper.) In most cases it seems
advisable to have an input method as close to the original input as possible,
and thus perform a study on a provided or the user's own device.  Sometimes,
e.g., to address privacy concerns, data collection using pen-and-paper can be
preferable. This is consistent with the results previously reported by Aviv
et. al~\cite{aviv2014understanding} that there were minimal differences in the
pen-and-paper sample. As such, we woudl recommend that studies be conducted
on-device when possible, but pen-and-paper can be used as substitute in an
adversarial setting with only considering the offensive data set.



\section{Conclusion}

We present the first cross dataset comparison of Android unlock
patterns, as well as a systemized review of the collection and analysis methods
used in this space. From this work, we find that there exist significant
differences, both in terms of pattern strength (as measured using guessability
metrics) and visual/spatial properties. However, the impact of spatial
properties is relatively small, albiet significant, but collection method has a
large effect on pattern strength. Adversarial settings both over and under
incentivize strong and weak patterns, and weak situational motivators can also
have a negative effect. We find that there is a strong, consistent midrange of
data that we conjecture that most likely contains the most realistic
patterns. Additionally, the input method and participant recruitment settings
seem to have limitted or no effect. 

Based on this survey and analysis, we would recommend that researchers carefully
consider the collection method to ensure that there are clear motivators. For
example, use a variety of data collection methods to assess a range of possible
values when performing strength. Additionally, researchers should use grounded
strength metrics, based on guessability, as there is not enough differences in
spatial and visual features to make strong claims.

\bibliographystyle{abbrv}
\bibliography{ref}

\counterwithin{figure}{section}
\counterwithin{table}{section}





\end{document}